\documentclass[final,1p,times,twocolumn]{elsarticle}

\usepackage{amssymb}

\usepackage{lineno}

\usepackage{subfigure}

\journal{Nuclear Instrument and Methods in Physics Research Section A}

\begin{document}

\begin{frontmatter}

\title{Investigation of Hamamatsu H8500 phototubes as single photon detectors. }

\author[a]{R.A. Montgomery}
\author[b]{M. Hoek}
\author[a]{V. Lucherini}
\author[a]{M. Mirazita}
\author[a]{A. Orlandi}
\author[a]{S. Anefalos Pereira}
\author[a]{S. Pisano}
\author[a,c]{P. Rossi}
\author[a]{A. Viticchi\`{e}}
\author[d]{A. Witchger\footnote{Presently at: Department of Physics, North Carolina State University, 2401 Stinson Drive, Raleigh, NC 27695-8202, USA}}

\address[a]{INFN Laboratori Nazionali di Frascati, Via Enrico Fermi, 40, 00044 Frascati, Italy}
\address[b]{Institut f{\"u}r Kernphysik, Johannes Gutenberg-Universit{\"a}t Mainz, Johann-Joachim-Becher-Weg 45, D 55128 Mainz, Germany}
\address[c]{Jefferson Laboratory, Thomas Jefferson National Accelerator Facility, 12000 Jefferson Avenue, Newport News, VA 23606, USA}
\address[d]{Department of Physics, Duquesne University, 317 Fisher Hall, Pittsburgh, PA 15282, USA}

\begin{abstract}
We have investigated the response of a significant sample of Hamamatsu H8500 MultiAnode PhotoMultiplier Tubes (MAPMTs) as single photon detectors, in view of their use in a ring imaging Cherenkov counter for the CLAS12 spectrometer at the Thomas Jefferson National Accelerator Facility.  For this, a laser working at 407.2\,nm wavelength was employed. The sample is divided equally into standard window type, with a spectral response in the visible light region, and UV-enhanced window type MAPMTs. The studies confirm the suitability of these MAPMTs for single photon detection in such a Cherenkov imaging application.
\end{abstract}


\end{frontmatter}
\section{Introduction}
\label{sec:Introduction}

The CEBAF electron accelerator of the Thomas Jefferson National Accelerator Facility (TJNAF), in Newport News (VA, USA),  is currently undergoing an upgrade to increase its maximum energy from 6 to 12\,GeV~\cite{CEBAF12}. In the experimental Hall-B, the CLAS spectrometer~\cite{CLAS} is being modified and upgraded to CLAS12~\cite{CLAS12}, in order to operate in the new experimental conditions. The major focus of the Hall-B physics program at 12\,GeV will be the study of the internal dynamics and 3-dimensional imaging of the nucleon, quark hadronisation processes, and kaon versus pion production in hard exclusive and semi-inclusive scattering, to provide access to the flavour decomposition of the non-perturbative distribution functions.

The main features of CLAS12 are its capability to operate at high luminosities (on the order of $10^{35}$\,$cm^{-2}s^{-1}$), and operation of highly polarized beam and nucleon targets. The design of the CLAS12 spectrometer is described in \cite{CLAS12}. The spectrometer in its present configuration does not provide an efficient kaon identification at large momenta, from 3\,-\,8\,GeV/c. At such momenta, the semi-inclusive kaon yield is one order of magnitude smaller than the pion yield. The required rejection factor for pions is then around 1\,:\,500, corresponding to a 4\,$\sigma$ pion-kaon separation for a contamination in the kaon sample of a few percent. Moreover the baseline spectrometer detectors do not allow the separation of positive kaons from protons in the 5\,-\,8\,GeV/c momentum interval.

Improved particle identification and event reconstruction can be achieved in this momentum range by replacing the foreseen low-threshold Cherenkov counter with a Ring Imaging CHerenkov (RICH) detector, without any impact on the baseline design of CLAS12. A study for a RICH detector to be implemented within the CLAS12 spectrometer is reported in \cite{prossi}. It is shown that, using aerogel radiators, it is possible to achieve the required identification of kaons up to 8\,GeV/c momentum if the photon detector employed has a pad size less than than 1\,cm$^2$. In this respect, a possible choice is that of Multi-Anode PhotoMultiplier Tubes (MAPMTs) having single photon detection capabilities.

This paper describes a set of experimental tests which were performed at the Laboratori Nazionali di Frascati (LNF) to measure the response of Hamamatsu H8500 MAPMTs to photons in the blue wavelength region, in order to assess their performance as single photon detectors to be used in a RICH counter. This wavelength was selected to match the peak intensity of useful Cherenkov radiation produced in aerogel (UV-visible region). For the first time, a large sample of H8500 MAPMTs has been tested, and the variation between their responses assessed. Moreover, two different H8500 MAPMT types have been tested: both the standard and the UV-enhanced window types (14\,units of each). The Hamamatsu H8500 MAPMT has previously been investigated by Matsumoto \emph{et al.}, through beam test studies, to be a suitable MAPMT for use in a RICH detector incorporating aerogel as the radiator~\cite{Matsumoto}.

\section{The Hamamatsu H8500 }
\label{sec:H8500}
\begin{figure}[h!]
\begin{center}
\includegraphics[width=.4\textwidth]{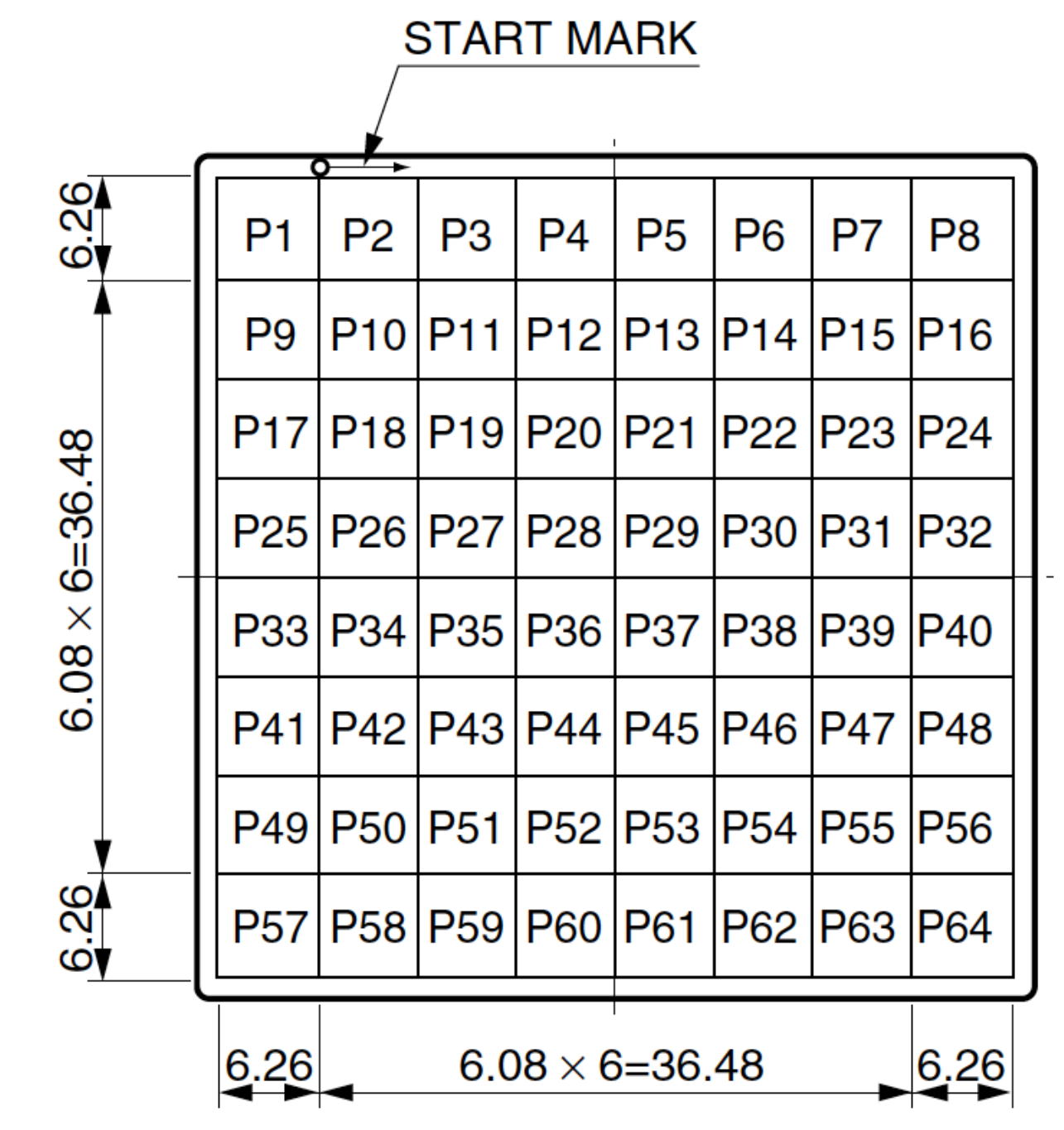}
\caption{\small \sf Numbering of the 64 pixels of an H8500 MAPMT, as seen by looking towards the entrance window (front view). Image taken from~\cite{H8500DataSheet}. The dimensions of the MAPMT are also marked.
}
\label{fig:PixelMap}
\end{center}
\end{figure}

The Hamamatsu H8500 is a multianode bialkali photomultiplier tube, comprised of  8\,$\times$\,8 pixels and having a 12-stage amplification. We show in Fig.~\ref{fig:PixelMap} the numbering scheme from Hamamatsu for the 64 pixels of the H8500, as seen by looking at the entrance window of the MAPMT.
The external size is 52\,$\times$\,52\,mm$^2$, with an active area of 49\,$\times$\,49\,mm$^2$ allowing a high packing fraction of $89\,\%$. The pixels basically have a square cross-section with side lengths of 5.8\,mm for centre pixels and 5.98\,mm for edge pixels. The device has a pixel\,-\,to\,-\,pixel pitch of 6.08\,mm in the centre and 6.26\,mm for edge pixels, as shown in Fig.~\ref{fig:PixelMap}. This provides an imaging plane with small dead-space. Two versions of such MAPMT are available:  the H8500C type, with spectral response in the visible light range from 300 to 650\,nm wavelengths; and the H8500C-03 type, with enhanced response in the UV region, from 185 to 650\,nm. For both MAPMTs, the peak efficiency is at a wavelength of about 400\,nm, in the blue light region, where the typical quantum efficiency obtained from the Hamamatsu datasheet is around $27\,\%$~\cite{H8500DataSheet}. The High Voltage (HV) supply range recommended by Hamamatsu is between -900\,V and -1100\,V, with a reference value of -1000\,V~\cite{H8500DataSheet}. Hamamatsu measured various relevant parameters of the phototubes at this reference voltage, some of which are reported in Tab.~\ref{tab:pmt_sheets} for the 14 H8500C and 14 H8500C-03 MAPMTs that we tested.
\begin{table}[h!]
\begin{center}
\caption{\small \sf Selected characteristics of the 28 H8500 MAPMTs used in the tests, according to the Hamamatsu datasheets and as measured at HV= -1000\,V. The expected dark count rates of each MAPMT, as calculated from the dark current values measured by Hamamatsu, are also given. The reference area for the dark count rates corresponds to the MAPMT size itself.}
\label{tab:pmt_sheets}
\begin{tabular}{||c|c|c||c|c|c||} \hline
ID & type & Serial no. & Dark current (nA) & Dark count rate (kHz) & Gain ($10^6$) \\ \hline
\hline
1 & H8500C & CA4655 & 4.23 & 9.01 & 2.93 \\ \hline
2 & & CA4658 & 0.16 & 0.53 & 1.89 \\ \hline
3 & & CA4667 & 0.09 & 0.39 & 1.45 \\ \hline
4 & & CA4683 & 0.24 & 0.89 & 1.69 \\ \hline
5 & & CA4686 & 1.20 & 2.87 & 2.61 \\ \hline
6 & & CA5342 & 0.27 & 0.60 & 2.81 \\ \hline
7 & & CA5348 & 0.12 & 0.27 & 2.73 \\ \hline
8 & & CA5525 & 0.29 & 0.69 & 2.64 \\ \hline
9 & & CA5562 & 0.24 & 0.70 & 2.15 \\ \hline
10 & & CA5575 & 0.50 & 1.12 & 2.70 \\ \hline
11 & & CA5577 & 0.25 & 0.38 & 4.10 \\ \hline
12 & & CA5675 & 1.44 & 3.26 & 2.76 \\ \hline
13 & & CA5683 & 0.20 & 0.45 & 2.78 \\ \hline
14 & & CA5687 & 0.91 & 3.30 & 1.72 \\ \hline
\hline
15 & H8500C-03 & DA0168 & 0.16 & 0.36 & 2.78 \\ \hline
16 & & DA0172 & 0.17 & 0.77 & 1.37 \\ \hline
17 & & DA0174 & 0.73 & 2.16 & 2.11 \\ \hline
18 & & DA0179 & 0.24 & 0.83 & 1.81 \\ \hline
19 & & DA0181 & 0.12 & 0.44 & 1.71 \\ \hline
20 & & DA0348 & 0.23 & 0.71 & 2.01 \\ \hline
21 & & DA0349 & 0.47 & 0.98 & 2.99 \\ \hline
22 & & DA0353 & 0.17 & 0.70 & 1.52 \\ \hline
23 & & DA0355 & 0.16 & 0.66 & 1.52 \\ \hline
24 & & DA0356 & 0.12 & 0.26 & 2.84 \\ \hline
25 & & DA0357 & 0.17 & 0.35 & 3.03 \\ \hline
26 & & DA0359 & 0.14 & 0.38 & 2.29 \\ \hline
27 & & DA0360 & 0.29 & 0.48 & 3.81 \\ \hline
28 & & DA0361 & 0.39 & 0.67 & 3.64 \\ \hline
\end{tabular}
\end{center}
\end{table}

The MAPMTs are very low noise devices, with small dark currents, and the expected dark count rates ($R_{dc}$) for each of the MAPMTs, as calculated from the dark currents measured by Hamamatsu ($I_{dc}$), are also given in Tab.~\ref{tab:pmt_sheets}. The dark currents measured by Hamamatsu are quoted globally per MAPMT, and therefore the reference area for the dark count rates given in Tab.~\ref{tab:pmt_sheets} corresponds to the MAPMT size itself. Assuming that the majority of dark count signals correspond to a single PhotoElectron (p.e.) level, the expected dark count rates were calculated as shown in Equation~\ref{eq:DarkCountRate}, where: $dt$ is the selected time interval and was set to 1\,s; $G$ is the gain of the MAPMT; $e$ is the elementary charge.
\begin{equation}
R_{dc} = \frac{I_{dc}\times dt}{G \times e}
\label{eq:DarkCountRate}
\end{equation}
These dark count rates were used to calculate the expected background arising from dark noise in the MAPMTs of the CLAS12 RICH detector. They were concluded to be negligible, with a mean probability of $\sim$\,3$\times$10$^{-4}$ dark counts expected from each MAPMT per CLAS12 detector data acquisition time window of 250\,ns. The dark noise behaviour of the MAPMTs is therefore not considered to cause any concern for their application in the CLAS12 RICH detector and so we do not discuss this topic any further.

\section{The Test Set-up}
\label{sec:setup}
\begin{figure}\begin{center}
\includegraphics[width=.75\textwidth]{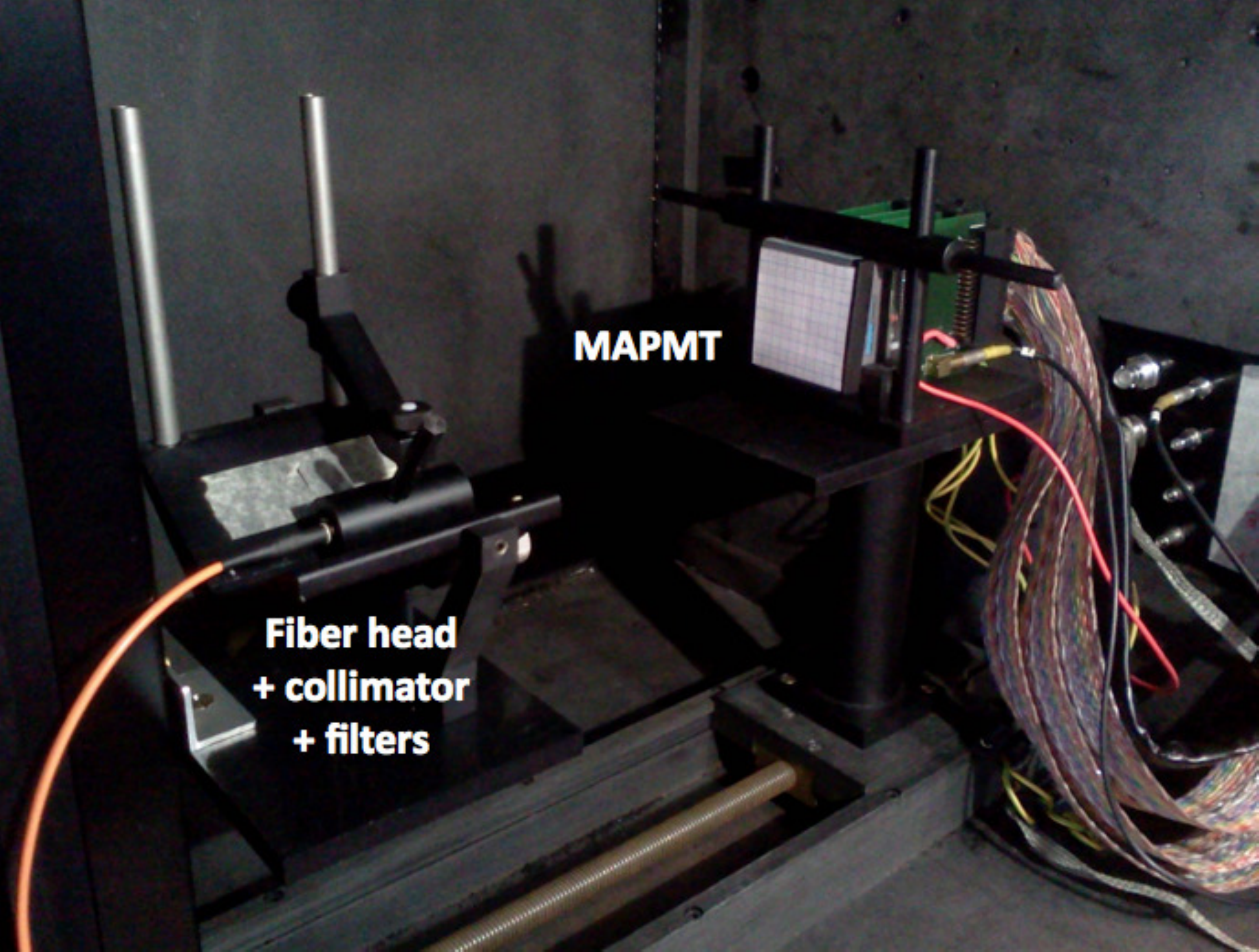}
\caption{\small \sf The set-up used for tests of the Hamamatsu H8500 MAPMTs, inside the light-tight box. During measurements, the distance between the fiber head/collimator/filters and the MAPMT was as small as possible (roughly 1mm), and is larger in the image for display purposes only.
}
\label{fig:set_up}
\end{center}
\end{figure}

We tested the response of the H8500 MAPMTs reported in Tab.~\ref{tab:pmt_sheets} to a low intensity laser beam, at the LNF. The test set-up included a laser emitting blue light (in the region close to where the Hamamatsu H8500 MAPMT has its maximal quantum efficiency) and standard VME electronics for the readout. Two motorised step motors were also employed to remotely move the laser beam across the MAPMT entrance window. During measurements, the distance between the end of the collimator and the MAPMT surface was as small as possible (roughly 1\,mm), in order that the MAPMT surface was positioned in the focal plane of the laser beam. The set-up was installed inside a light-tight box to isolate the MAPMT from any background light, as shown in Fig.~\ref{fig:set_up}.

\subsection{The Laser and Motors}
For the light source we used a PiLas EIG1000D picosecond injection laser~\cite{PiLas}, a compact system based on a laser diode, emitting monochromatic light at the wavelength of 407.2\,nm. The light was transported from the laser diode to inside the light-tight box via a fiber-optical cable. A 1\,mm diameter, 50\,mm long black PVC custom-made mechanical collimator was attached to the lens collimator connected to the end of the fiber-optical cable and then mounted onto a mechanical support which was driven by two high precision NRT150 motors from Thorlabs~\cite{motors} (one for each of the horizontal and vertical movements). The motors were remotely controlled so that the support could be moved across the entire surface of the MAPMT entrance window and illuminate one by one of all its 64 pixels. With this set-up it was possible to perform automised scans of the response of all pixels of each MAPMT. For the scans, with the collimator as close as possible to the MAPMT, the diameter of the laser spot on the MAPMT surface was 0.9\,mm. This value was obtained through the use of a CCD (Hamamatsu C9260-901 board camera) to image the laser spot at the MAPMT position. The intensity of the laser beam was reduced down to the level of a few photons per pulse by using Edmund Optics UV-VIS neutral density filters~\cite{filters}, which were mounted after the collimator. Finer adjustments were possible by changing the settings of the laser tune (relative intensity) on its controller unit, allowing to vary the photon intensity within the linear range of its performance. The tests were done typically with the filters having a total attenuation factor on the order of $\approx$\,10$^{6}$ and a laser tune setting of $\approx$\,25\,\%. The pulse repetition rate was driven by and controlled by the Data AcQuisition (DAQ) readout program and was set to 100\,Hz for the duration of the tests.

\subsection{Electronics}
We used standard VME electronics to read the 64 anodes of one H8500 at a time. A CAEN V2718 VME Bridge, connected to a computer via an optical fiber, was used to generate the {\it Trigger In} signal for the PiLas. After arrival of this {\it Trigger In} signal and before emitting the light pulses, the PiLas controller generated a TTL {\it Trigger Out} signal that, reshaped through a CAEN N89 NIM/TTL converter and stretched to a length of 40\,ns through a CAEN N93B Dual Timer, was used as the gate signal for the readout of two CAEN V792 QDCs. The CAEN V792 QDC is a fast, 32 channel charge\,-\,to\,-\,digital converter with a dynamical range of 400\,pC and a resolution of $\approx$\,100\,fC. The analogue signals of all the 64 pixels of an MAPMT were extracted through custom designed readout boards, which were designed to provide one common ground line amongst all pixels. The signals were then sent directly to the V792 QDCs using four 16-channel twisted-pair flat-ribbon cables.

\begin{figure}\begin{center}
\includegraphics[width=.65\textwidth]{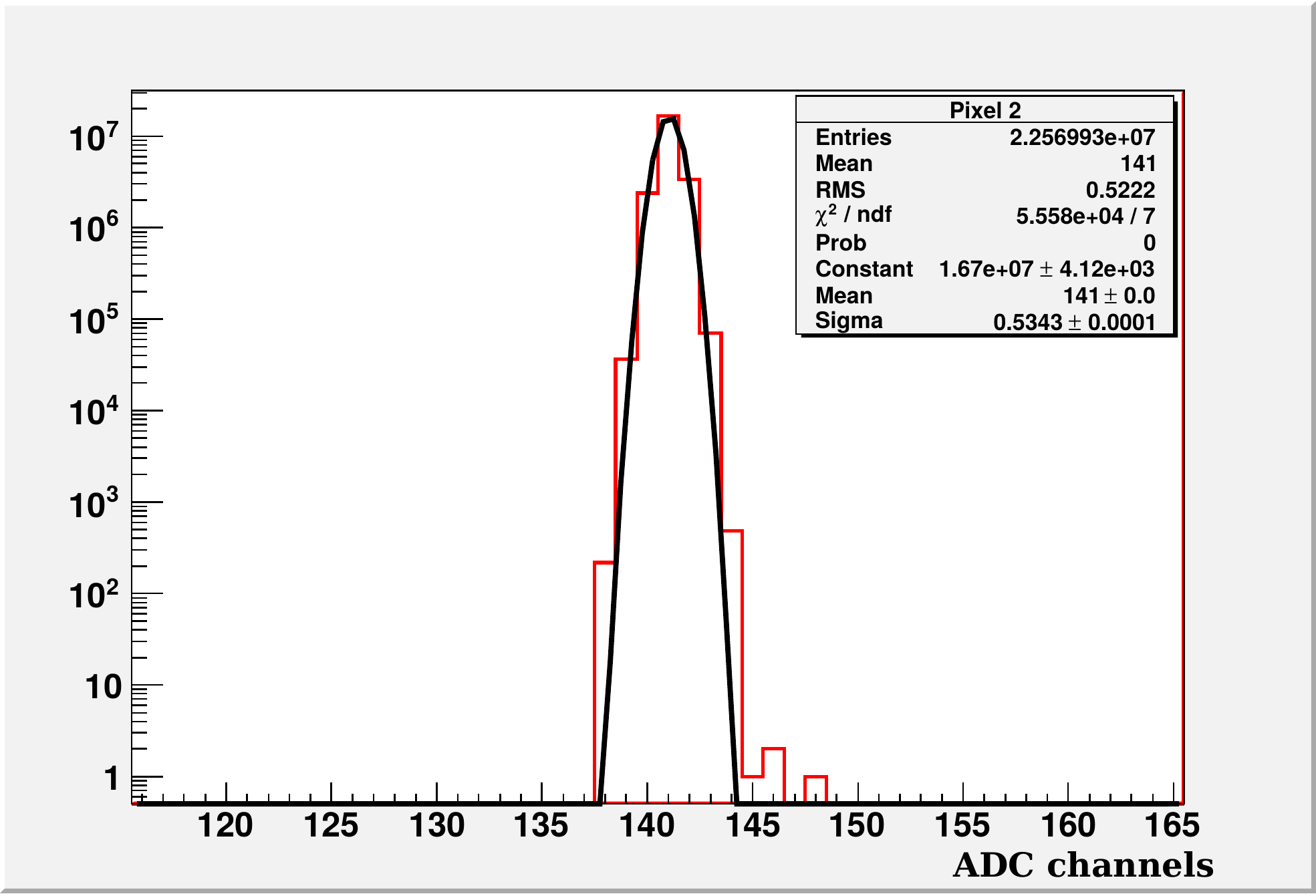}
\caption{\small \sf Pedestal measurement, both the data (histogram) and a Gaussian fit, for a representative pixel of one MAPMT after several hours of continuous data taking at HV = -1000\,V. The MAPMT entrance window was covered with a black cap and the set-up put inside the light-tight box.
}
\label{fig:DAQ_noise}
\end{center}
\end{figure}

Having the goal of measuring QDC spectra in the few p.e. regime, it was extremely important to check the noise level of the electronics. For this, we ran for several hours with the entire set-up in its operating conditions, but with the surface of the MAPMT covered by a black cap, so that no light could reach the photocathode. The result for a representative pixel of one of the tested MAPMTs is shown in Fig. \ref{fig:DAQ_noise}.
We find for this pedestal a Gaussian distribution with a width of 0.53\,QDC channels, i.e. $\sim$53\,fC. For the recorded events (more than $22.5\times10^{6}$) we find an excess of about 250 entries beyond a 5\,$\sigma$ cut, corresponding to a probability of $1.1\times10^{-5}$ for finding noise hits above the 5\,$\sigma$ threshold. Similar results were found for all the MAPMTs. The behaviour also reflects the low dark-noise characteristics of the MAPMTs, as listed in Tab.~\ref{tab:pmt_sheets}, and confirms that background noise from the MAPMTs is not expected to be an issue for their application in the CLAS12 RICH detector. Overall, the width of the Gaussian distributions fitted to the pedestals were about 0.5\,QDC channels, corresponding to 50\,fC of integrated charge. The pedestal noise matched exactly the resolution of the QDCs used, indicating a low noise set-up suitable for the single photon tests.

\subsection{MAPMT Scan Method}
Before starting the systematic study of the response of all MAPMTs, a single MAPMT was selected to perform a finer scan of some typical pixels with a laser spot of diameter 90\,$\mu$m and step size of 0.1\,mm in both directions, in order to verify the uniformity of the response across its pixels' surfaces~\cite{Montgomery}. The response across the pixel active area was found to be sufficiently homogeneous and uniform for the CLAS12 RICH application. Furthermore, the MAPMT pixel size and dead-space regions measured by the fine scans were as expected, fulfilling the position resolution requirement of the RICH detector. From these tests it was concluded that a coarser resolution scan could be employed to test the significant sample of 28 MAPMTs.

The tests of the 28 MAPMTs have been performed using an automated procedure, in which a pedestal run (with laser off) was performed first. Then the laser head was moved to illuminate the centre of pixel 1 (see Fig.~\ref{fig:PixelMap}), the laser was switched on and 10\,k events were taken with all 64 pixels read out simultaneously. The laser then was moved to the centre of the next pixel and the measurement was repeated. With this procedure for the laser measurements, we were able to study the entire MAPMT response when a single pixel was illuminated by the laser and also when the pixel was not illuminated, the latter condition providing useful information on the cross talk between channels. After sequencing through all 64 pixels, the laser was switched off and another pedestal run was taken, to check the stability of the electronics.

\begin{figure}[h!]
\begin{center}
\mbox{
\subfigure[Pixel 25]{\scalebox{1.0}{\includegraphics[width=0.48\linewidth]{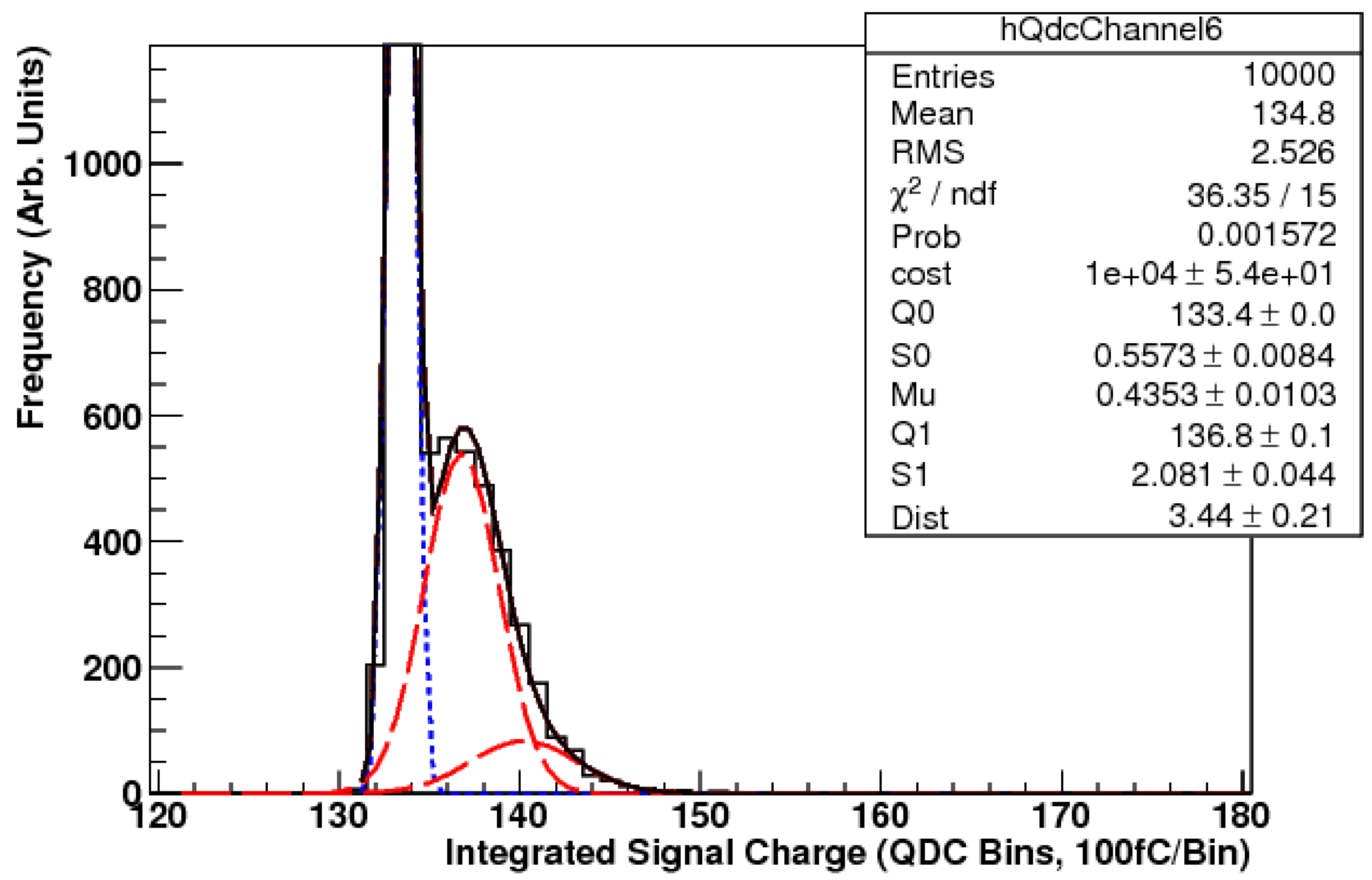}}} \quad
\subfigure[Pixel 36]{\scalebox{1.0}{\includegraphics[width=0.475\linewidth]{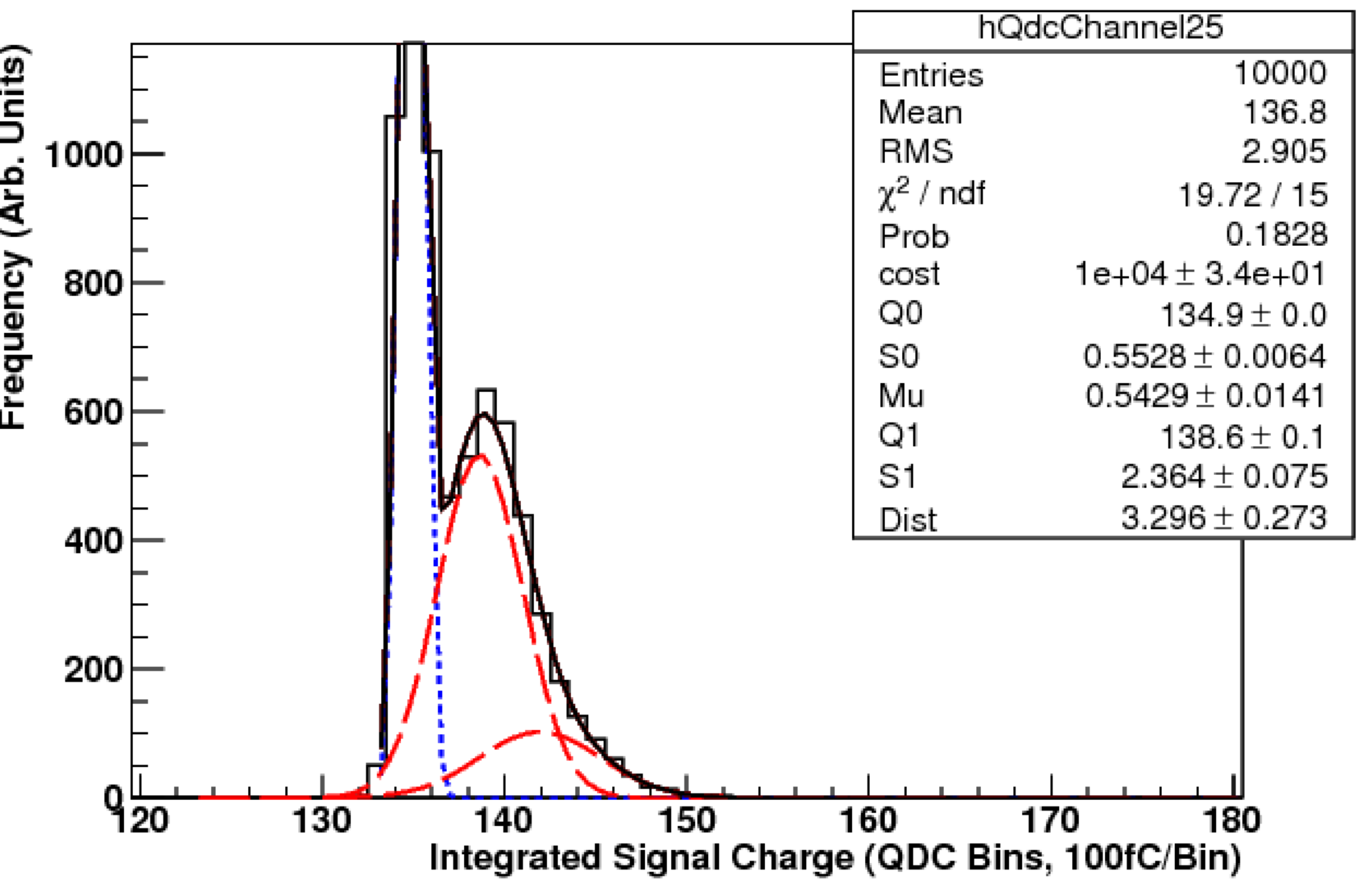}}}
}
\caption{\small \sf Two representative QDC spectra for pixels 25 and 36 of the CA4658 MAPMT at HV=-1000\,V (the pixel numbers mapped to electronic QDC readout channels 6 and 25 respectively). The curves show the total fit (solid black curve), the pedestal (dotted blue curve) and p.e. (dashed red curve) contributions. The results for the free parameters of the fits may also be read in the plots, where: $cost$ is the normalisation constant ($A$); $Q0$ and $S0$ are the mean and width of the pedestal Gaussian distribution; $Mu$ is $\mu$, the average number of detected p.e.; $Q1$ and $S1$ are the mean and width of the first p.e. Gaussian distribution; $Dist$ is $d$, the distance between the p.e. peaks.}
\label{fig:ca4658_1000_QDC}
\end{center}
\end{figure}

\subsection{QDC Spectra Analysis}
The analysis of the data have been performed following the method described in~\cite{perrino01}. For this, the QDC spectrum measured by each pixel has been fitted using Equation~\ref{eq:QDCfit}, where $q$ is the QDC channel readout and $P(q)$ and $G_k(q)$ are functions (normalised to 1) describing the shape of the pedestal and of the $k^{th}$ p.e. peak, respectively.
\begin{equation}
f(q) = A \left[ e^{-\mu} P(q) + \sum_{k=1}^{N}{ \frac{\mu^k e^{-\mu} }{k!} G_k(q) }  \right]
\label{eq:QDCfit}
\end{equation}
Each term in the sum is weighted by the Poisson distribution having $k=0,1,2,...,N$ p.e., with $\mu$ being the average number of detected p.e.. The parameter $A$ represents a normalisation constant term that, since all the factors in the sum are normalised to one, gives the total number of recorded events. The mean position $Q_k$ and width $S_k$ of the $k^{th}$ p.e. peak, for $k>1$, are computed as $Q_k = Q_1 + (k-1)\times d$ and $S_k = \sqrt{k} S_1$, where: $Q_1$ is the mean position of the first p.e. peak; $d$ is the distance between successive p.e. peaks; $S_1$ is the width of the first p.e. peak. Typically, the average number of detected p.e. was $\mu<1$, so that the total number of p.e. peaks considered in the fit was chosen to be 5. The fit description used was slightly simplified with respect to the full possible model given in~\cite{perrino01}, which may include contributions from different background effects, such as photon conversions on the first dynode stage of the MAPMT for example. This was due mostly to the relatively low resolution of the QDC spectra, in conjunction with the fact that the relation given in Equation~\ref{eq:QDCfit} was found to describe the spectra well, whilst also providing the most successful automisation and convergence of the fits. For the narrow pedestals measured (see Fig.~\ref{fig:DAQ_noise}) a simple Gaussian model was used, while for the p.e. peaks several different parametrisations have been tried. It was found that the best results (in terms of convergence of the fits and of uniformity of the function parameters) were obtained using Gaussian distributions as well, even though other shapes could provide in some cases slightly better values of $\chi^2$.

\section{Results at the Nominal Supply Voltage}
\label{sec:1000V}
Our first test was to check the average gain values quoted by Hamamatsu at the reference supply voltage of HV = -1000\,V (see Tab.~\ref{tab:pmt_sheets}). Two representative QDC spectra are shown  in Fig.~\ref{fig:ca4658_1000_QDC} for the CA4658 MAPMT, measured with the laser illuminating the selected pixels.
In each of the plots of Figs.~\ref{fig:ca4658_1000_QDC}\,(a) and (b) the overall results of the fits are shown (solid black curve), together with the individual contributions (pedestal in dotted blue and the p.e. peaks in dashed red curves). Since the average number of detected p.e. is $\mu \approx 0.5$, the Gaussian contributions for $k > 2$ p.e. are relatively small.

\begin{figure}\begin{center}
\includegraphics[width=.7\textwidth]{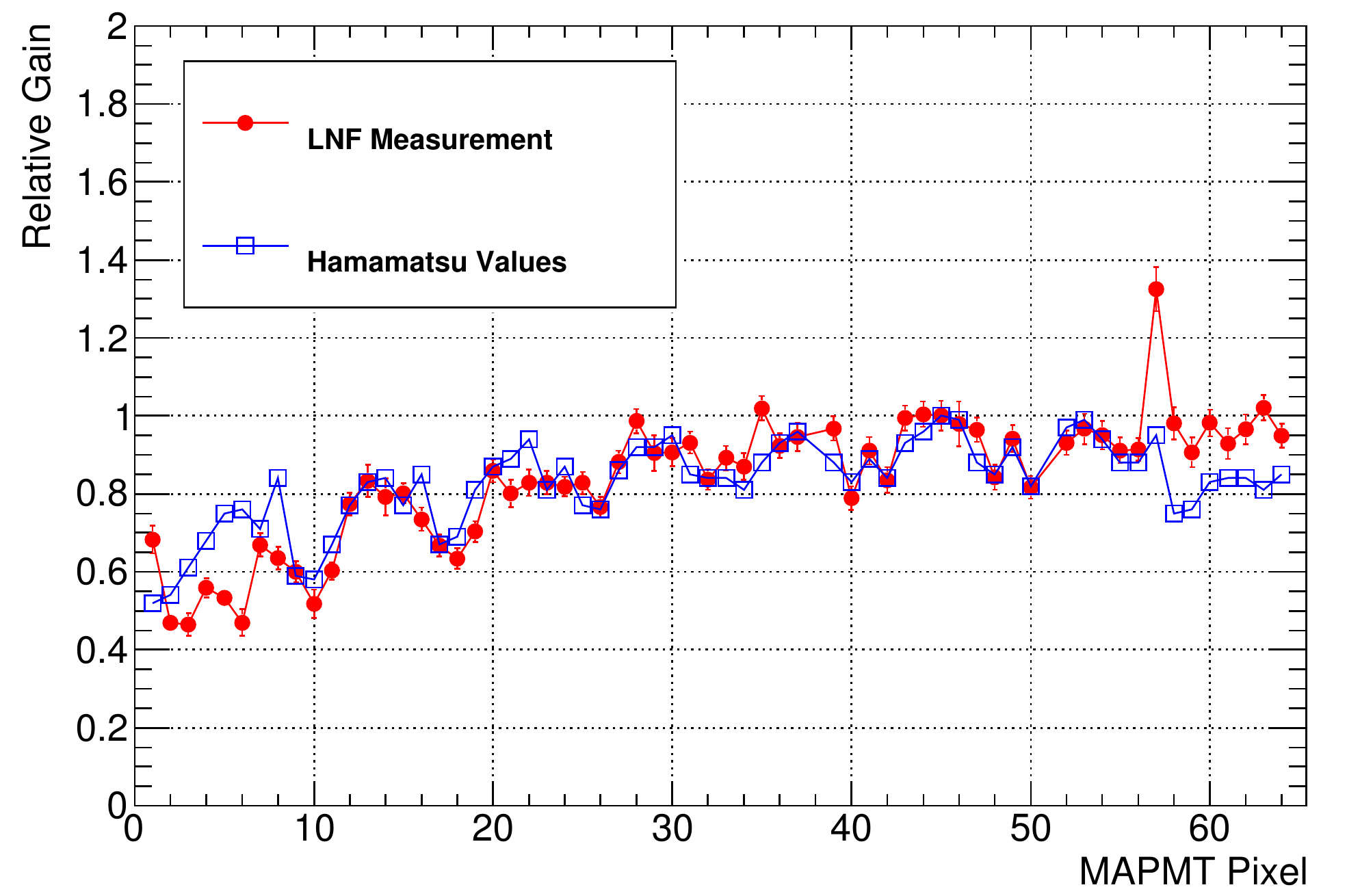}
\caption{\small \sf Comparison between the gains of all 64 pixels of the CA4658 MAPMT at HV=-1000\,V as given by Hamamatsu (blue squares) and as extracted from fits ($Q_{1}$) to the measured spectra (red circles). In both cases, the values have been normalised to the highest gain of the Hamamatsu measurement.
}
\label{fig:CmpRelGains_ca4658}
\end{center}
\end{figure}

According to the Hamamatsu datasheet, pixel 36 (Fig.~\ref{fig:ca4658_1000_QDC}\,(b)) has a gain about 20\,\% higher than that of pixel 25 (Fig.~\ref{fig:ca4658_1000_QDC}\,(a)), and in fact we see a significantly improved separation of the first p.e. peak from the pedestal for that pixel.
Furthermore, the gains extracted from the fits to the spectra of all 64 pixels of the CA4658 MAPMT are compared to those quoted by Hamamatsu in Fig.~\ref{fig:CmpRelGains_ca4658} (two channels are missing due to faulty electronic lines). In both cases, the values obtained from either the data or the Hamamatsu test sheet results have been calculated relative to the values for highest gain pixel determined by the Hamamatsu measurements.
There is a very strong agreement between the two relative gain distributions for the pixels. For the top and bottom rows of the MAPMT (pixels 1\,-\,8 and 57\,-\,64 respectively), the curves match less closely with each other and the mismatch for these two rows is not well understood, however the agreement for these rows remains within $\sim$20\,\%.

\begin{figure}[h!]
\begin{center}
\mbox{
\subfigure[Normalisation constant, A.]{\scalebox{1.0}{\includegraphics[width=0.48\linewidth]{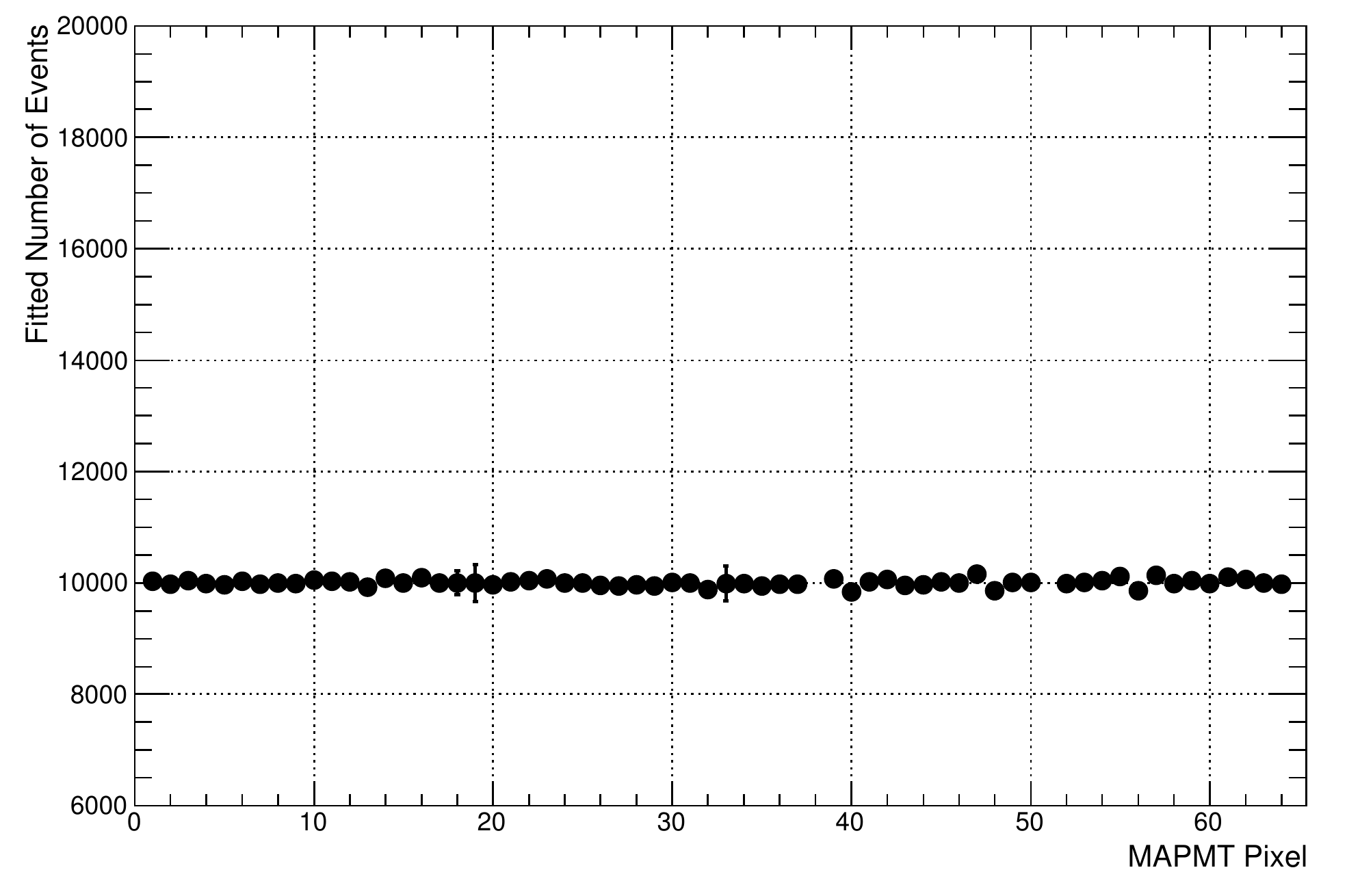}}} \quad
\subfigure[Average number of detected p.e., $\mu$.]{\scalebox{1.0}{\includegraphics[width=0.48\linewidth]{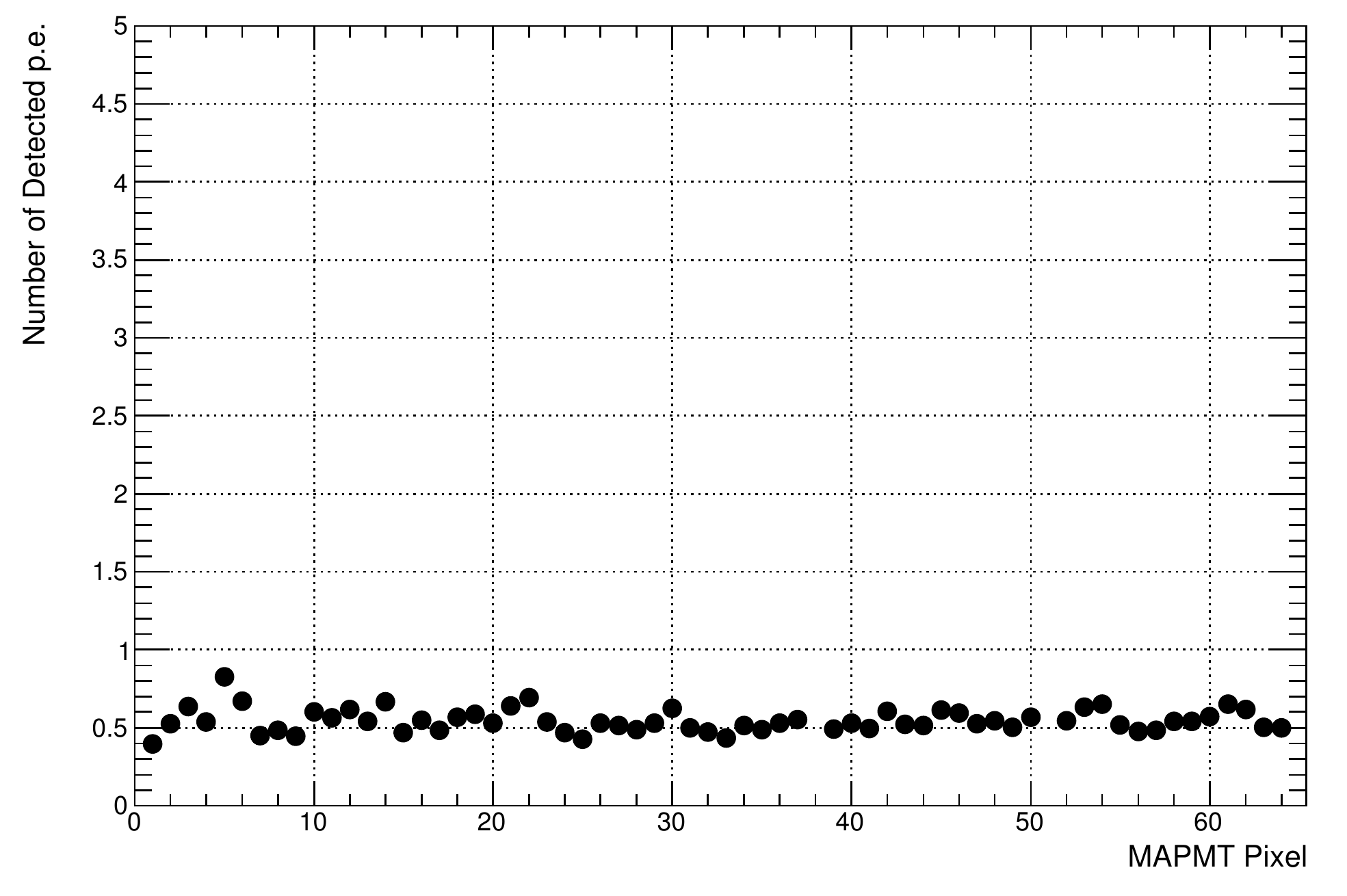}}}
}\\
\mbox{
\subfigure[Width of the first p.e. peak, $S_{1}$.]{\scalebox{1.0}{\includegraphics[width=0.48\linewidth]{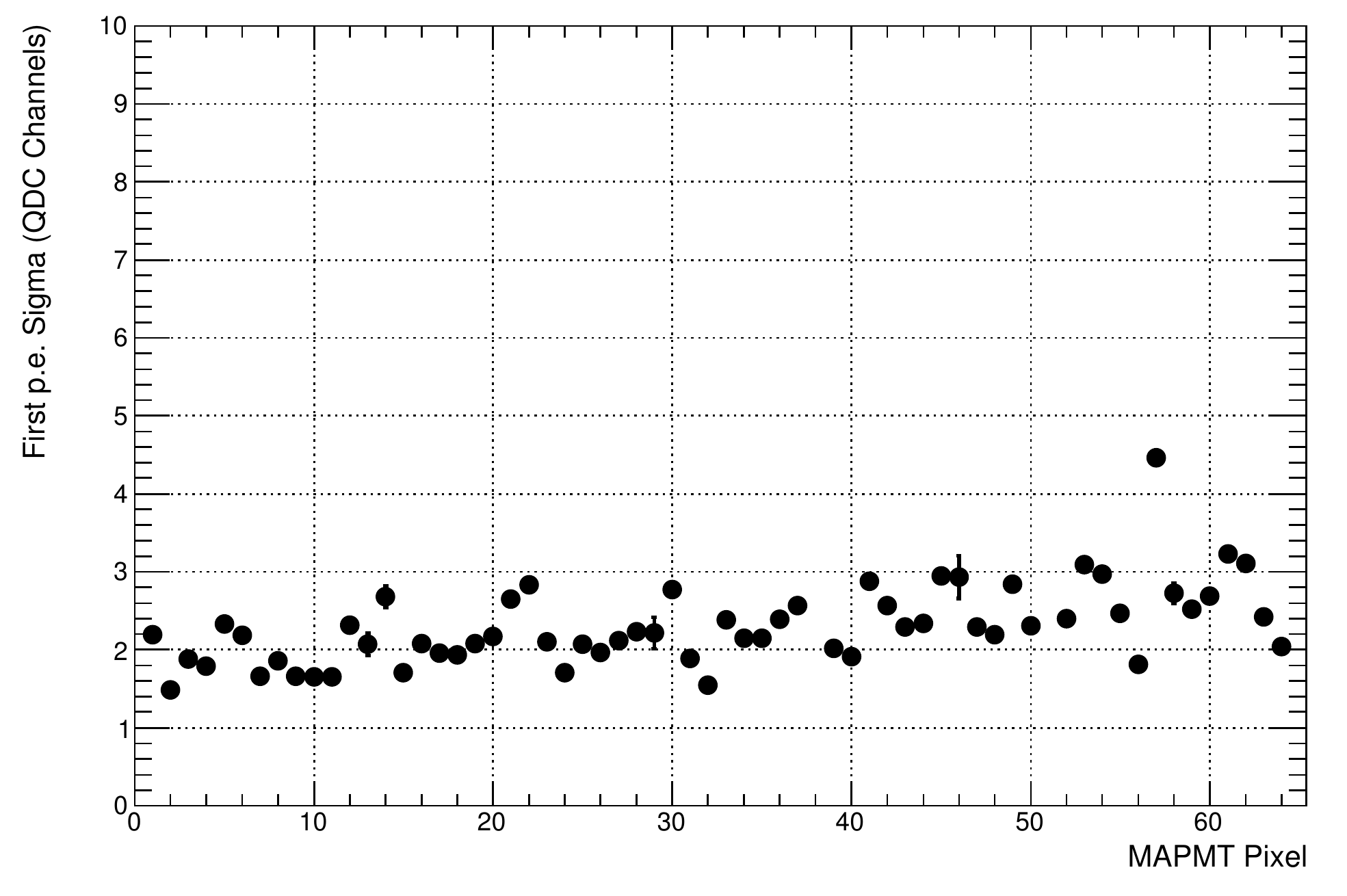}}} \quad
\subfigure[Gain of the first p.e. peak, $Q_{1}-Q_{0}$.]{\scalebox{1.0}{\includegraphics[width=0.48\linewidth]{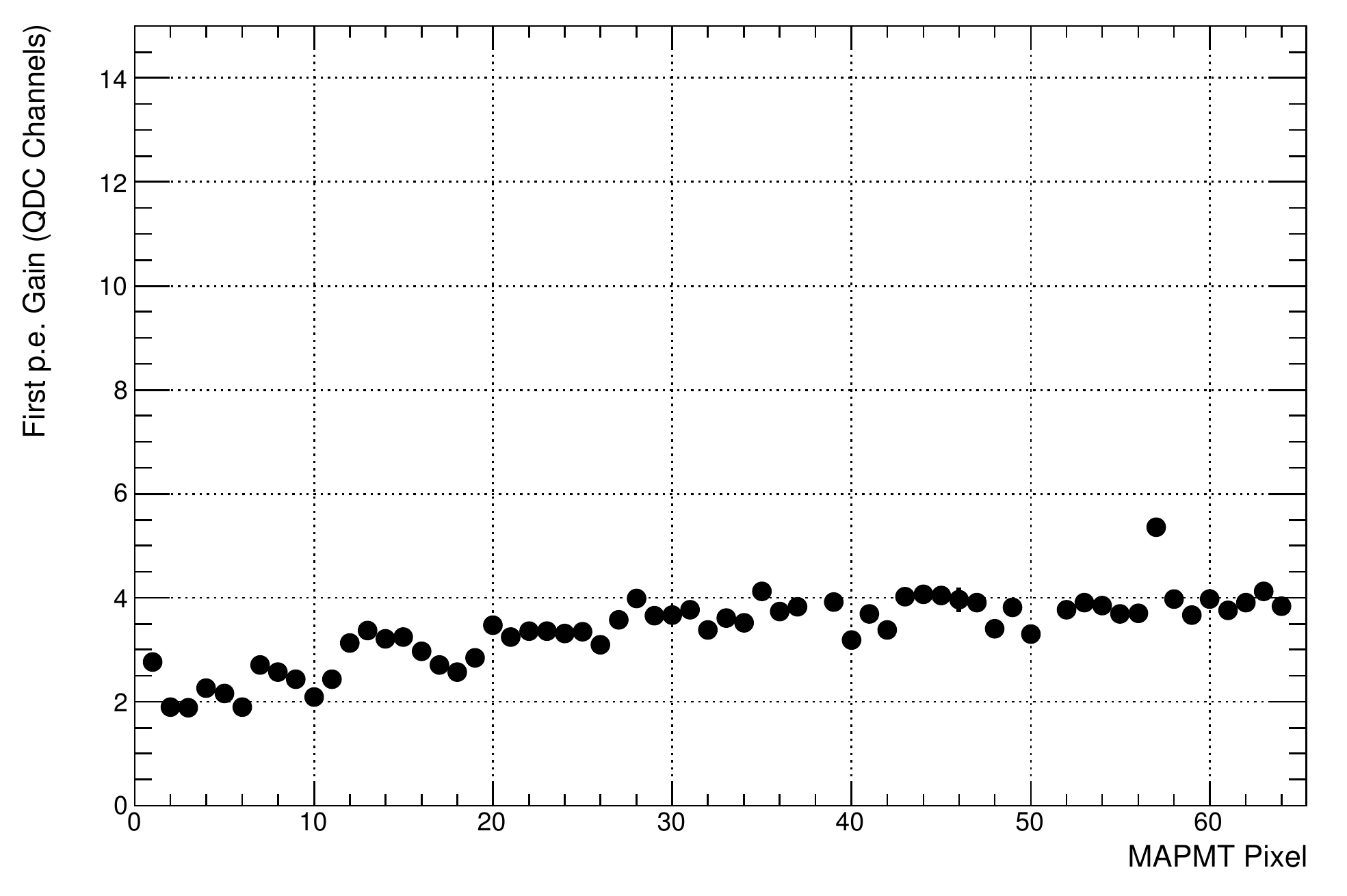}}}
}
\caption{\small \sf Pixel dependence of some fitted parameters for the CA4658 MAPMT spectra at HV=-1000\,V.}
\label{fig:ca4658_1000_pars}
\end{center}
\end{figure}

The results of the fitted parameters for all pixels of the MAPMT CA4658 are shown in Figs.~\ref{fig:ca4658_1000_pars}\,(a), (b), (c) and (d), where, from top left to bottom right, we can see:
\begin{itemize}
\item{the normalisation constant A, which should correspond to the recorded events per pixel;}
\item{the average number of detected p.e., $\mu$;}
\item{the width of the Gaussian of the first p.e. peak, $S_1$, in QDC channels;}
\item{the gain of the first p.e. peak, in QDC channels, defined as the difference between the means of the first p.e. and the pedestal distributions ($Q_{1}-Q_{0}$).}
\end{itemize}
For all pixels we found that, in general, the normalisation constant is equal to the total number of events ($10^4$) and that the average number of p.e. is roughly the same. These were both used as indications of the reliability of the fitting procedure, in addition to the $\chi^2$ values returned. In most cases the $\chi^2$ values were not exceeding 2. There do exist a small subset of channels for which the fit at 1000\,V is less reliable, even if the $\chi^2$ of the fits is good. These channels correspond to the pixels with smaller gains and thus with the first p.e. peak closer to the pedestal. The slight increase of the measured gain ($Q_{1}-Q_{0}$) as the pixel number increases is in agreement with the gain map provided by Hamamatsu for this MAPMT (see Fig.~\ref{fig:CmpRelGains_ca4658}).

\begin{figure}\begin{center}
\includegraphics[width=.6\textwidth]{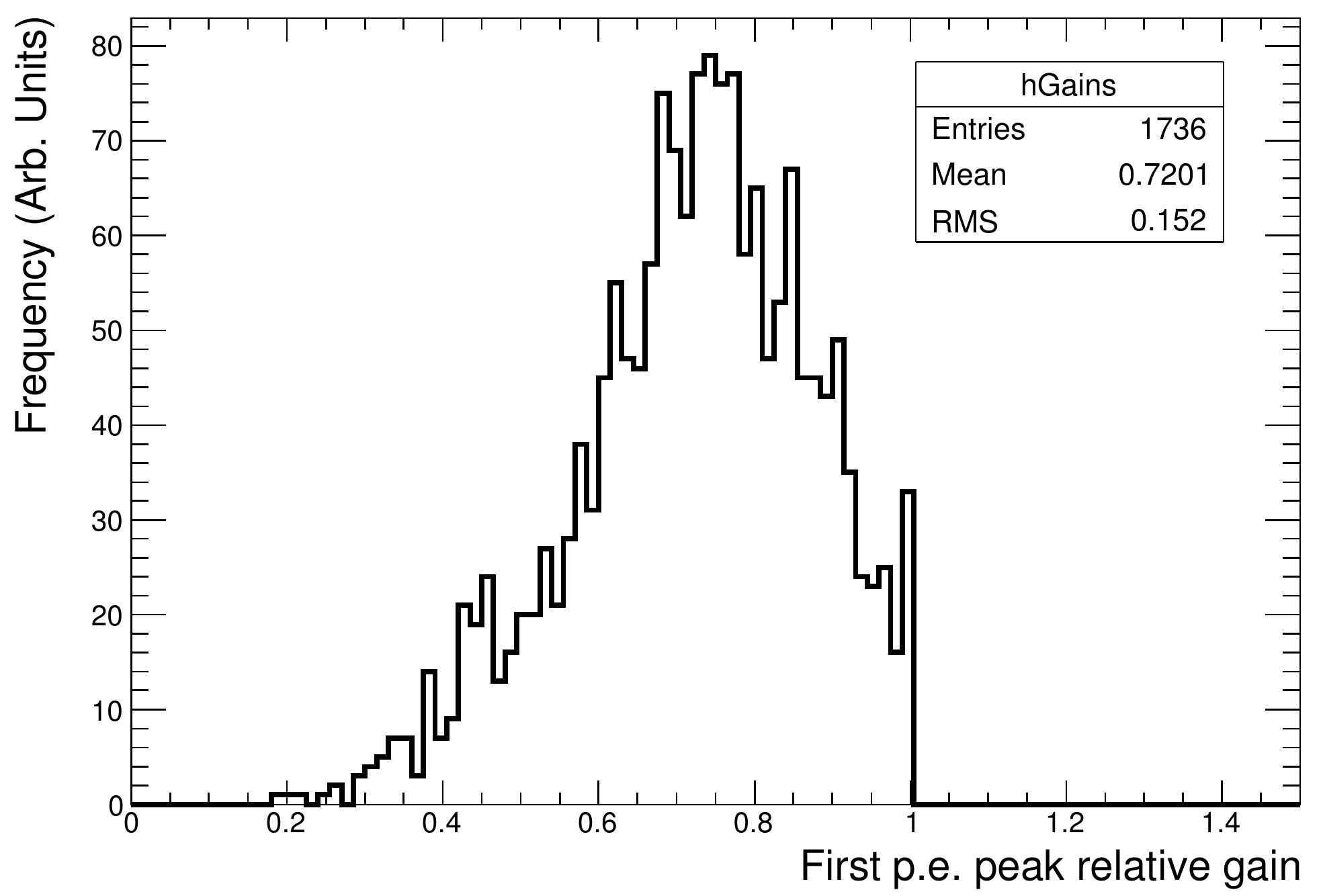}
\caption{\small \sf Fitted gain parameter $Q_{1}-Q_{0}$ extracted for each pixel of the 28 MAPMTs, normalised to the corresponding highest gain pixels.
}
\label{fig:gains_allpixels_1000}
\end{center}
\end{figure}

\begin{figure}[h!]
\begin{center}
\mbox{
\subfigure[Width of first p.e. peak, $S_{1}$.]{\scalebox{1.0}{\includegraphics[width=0.48\linewidth]{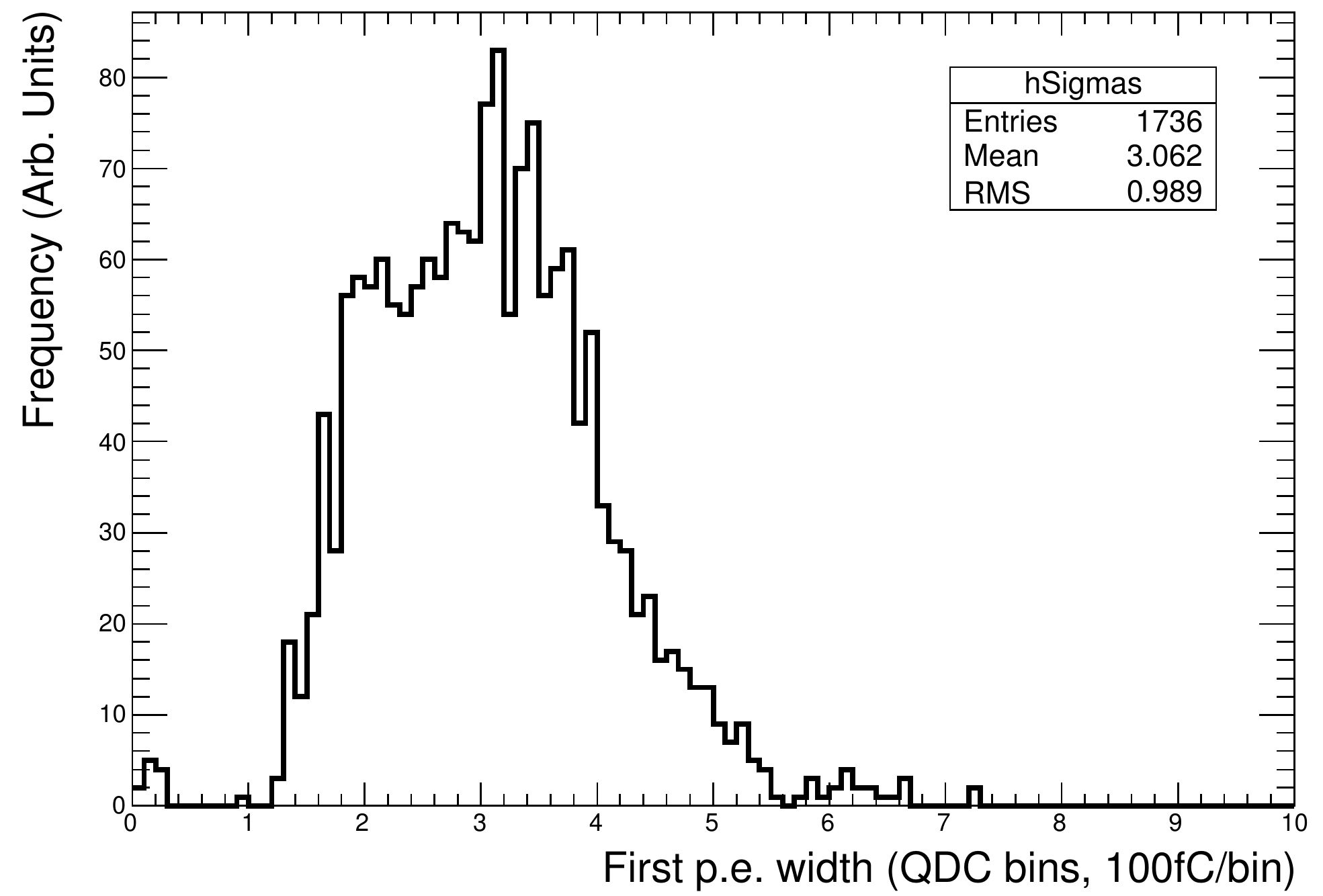}}} \quad
\subfigure[Width of first p.e. peak, $S_{1}$, divided by its gain, $Q_{1}-Q_{0}$.]{\scalebox{1.0}{\includegraphics[width=0.48\linewidth]{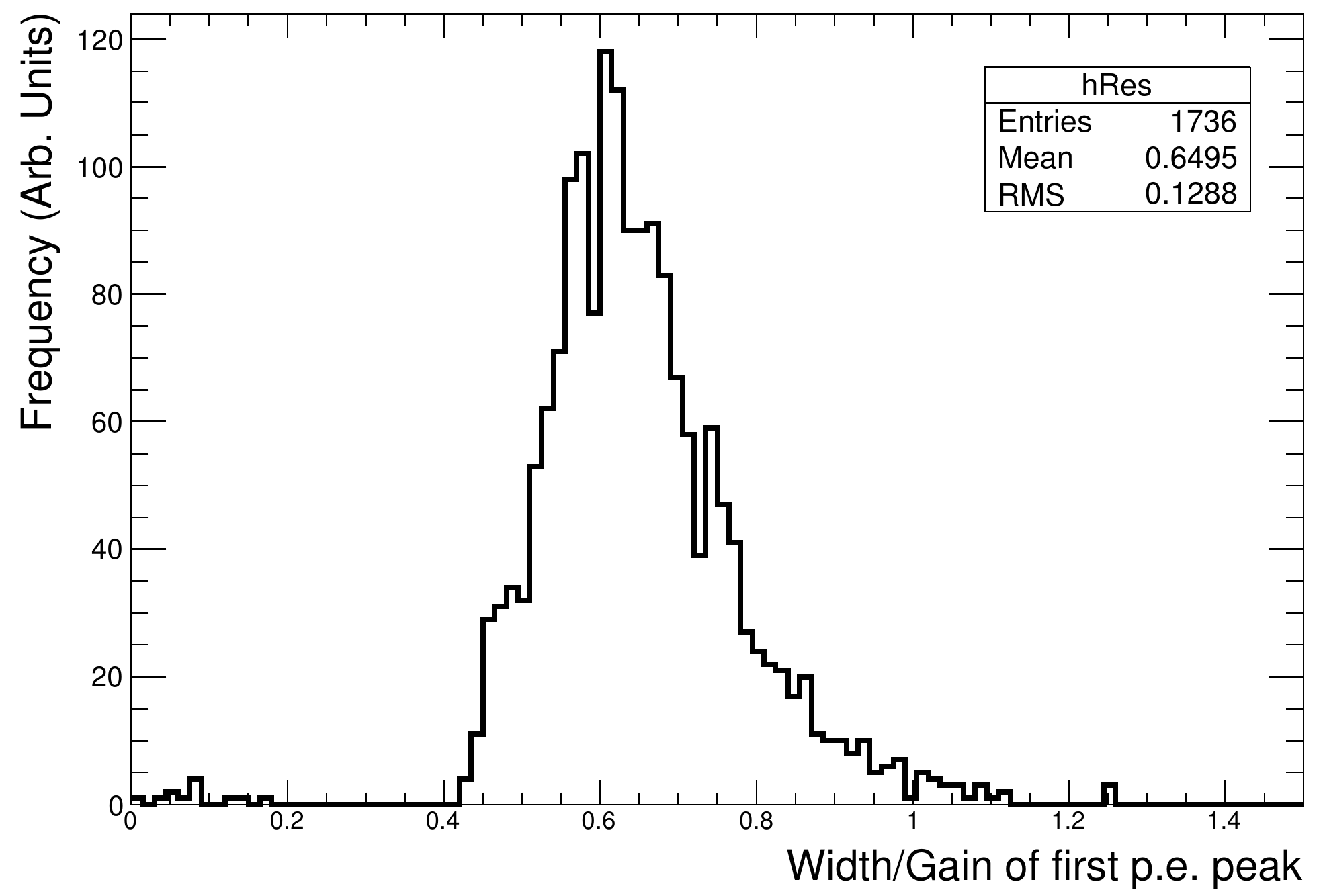}}}
}
\caption{\small \sf Width of first p.e. peak and ratio of this width to the gain as returned by the fit, for all pixels of the MAPMTs.}
\label{fig:SigmaRes_allpixels_1000}
\end{center}
\end{figure}

To compare the uniformity of all 28 MAPMTs, the fitted gains $Q_{1}-Q_{0}$ extracted from each pixel of each MAPMT are shown in the distribution of Fig.~\ref{fig:gains_allpixels_1000}. Each pixel's gain has been calculated relative to the highest gain pixel of each MAPMT, and electronic readout channels which were non-functional have been removed.
The distribution is narrowly peaked, with an RMS of only $\sim$15\,\%, which confirms an adequately small spread in the gains of the MAPMTs. If necessary such a gain dispersion may be easily compensated by standard front end electronics pre-amplifiers. The distribution of the fitted single p.e. width, $S_{1}$, is shown in Fig.~\ref{fig:SigmaRes_allpixels_1000}\,(a). An RMS of $\sim$1\,QDC unit is obtained, again indicating a consistent shape of the single p.e. gain, $Q_{1}-Q_{0}$.
Fig.~\ref{fig:SigmaRes_allpixels_1000}\,(b) gives the ratio of single p.e. width, $S_{1}$, divided by the single p.e. gain, $Q_{1}-Q_{0}$. This indicates the spread how the first p.e. peak is resolved amongst the MAPMTs. The distribution is narrowly centred around 65\,\%.

\begin{figure}\begin{center}
\includegraphics[width=.6\textwidth]{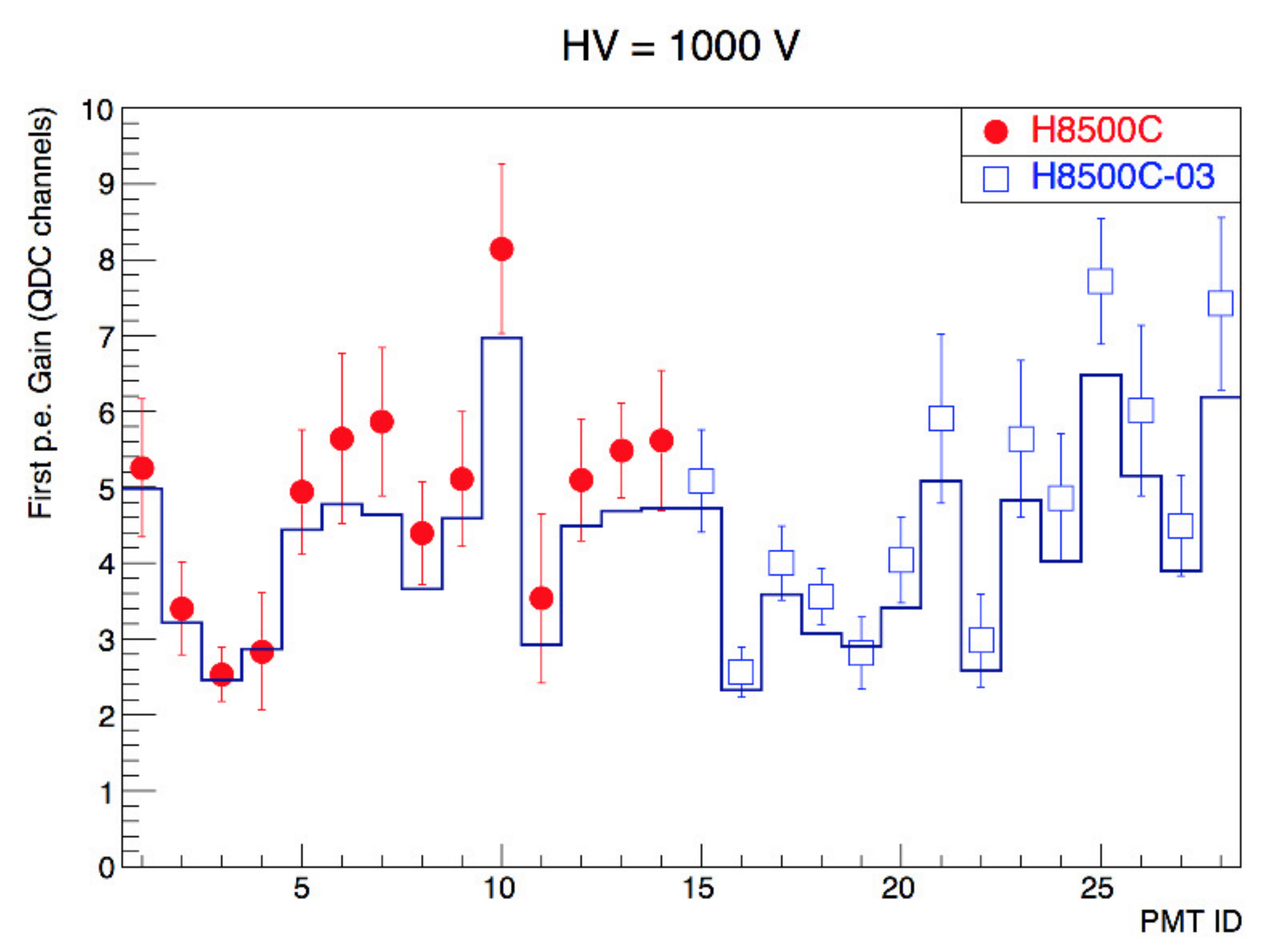}
\caption{\small \sf Average measured absolute gain $Q_{1}-Q_{0}$ at HV = -1000\,V for the 14 H8500C MAPMTs (red circles) and the 14 H8500C~-03 MAPMTs (empty squares) compared with the Hamamatsu test sheet values (histogram). The error bars represent the RMS values of the distributions of results for the 64 pixels of each MAPMT.
}
\label{fig:h8500_1000}
\end{center}
\end{figure}

To further summarise the results and compare the responses amongst different MAPMTs, we calculated the averages of the fitted parameters over the 64 pixels of each MAPMT at HV = -1000\,V. In Fig.~\ref{fig:h8500_1000} we show, for instance, the average absolute gains $Q_{1}-Q_{0}$ for the 28 MAPMTs under test (red circles for the 14 H8500C MAPMTs and empty blue squares for the 14 H8500C-03 MAPMTs), compared with the absolute gains obtained from the Hamamatsu test sheets. As seen, the agreement is very good. The measured average absolute gains are systematically higher than those obtained by Hamamatsu, and is likely due to the different method used by Hamamatsu to measure the gains of the MAPMTs. In the Hamamatsu measurements the whole MAPMT is illuminated simultaneously, whereas we only illuminated a spot of diameter 0.9\,mm in the centre of each pixel. The Hamamatsu measurement averages the gain over the entire area of each pixel (including its borders), whereas we compare this value with a smaller representative area of each pixel.

\section{Results at Higher Supply Voltages}
\label{sec:highHV}
The measurements performed at the reference supply voltage of -1000\,V have shown that the small separation between the pedestal and the first p.e. peak (typically a few QDC channels, see Fig. \ref{fig:ca4658_1000_pars}\,(d)) makes the separation of the signal from the background difficult in some cases. Thus, we repeated all the measurements at different HV values: -1040\,V and -1075\,V, in order to study the behaviour of the MAPMT response as a function of the supply voltage. Measurements were not performed at the maximum recommended HV value of -1100\,V, although the MAPMTs were each operated at this HV for a brief.

\begin{figure}[h!]
\begin{center}
\mbox{
\subfigure[1000\,V.]{\scalebox{1.0}{\includegraphics[width=0.48\linewidth]{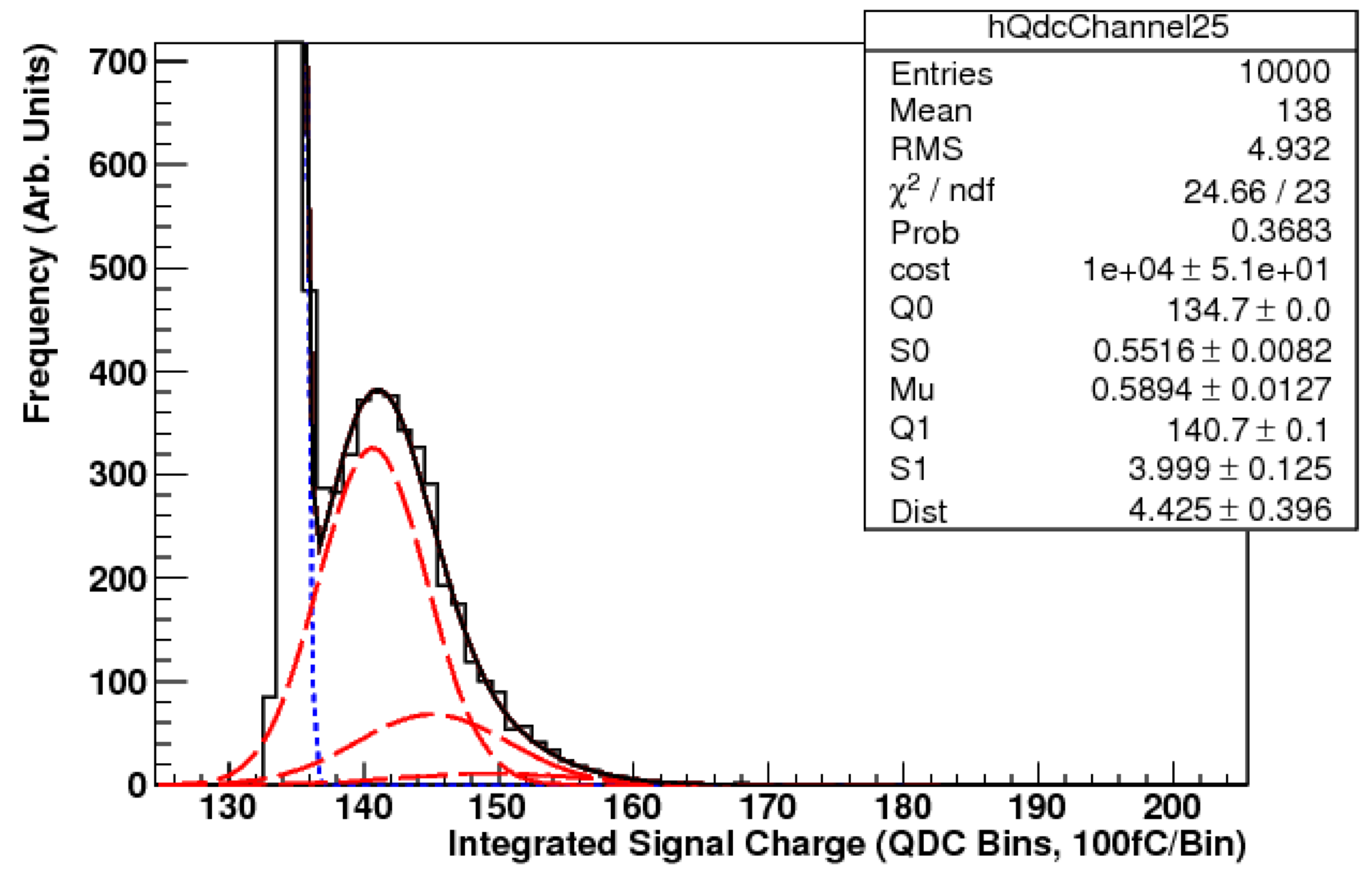}}}\quad
\subfigure[1040\,V]{\scalebox{1.0}{\includegraphics[width=0.49\linewidth]{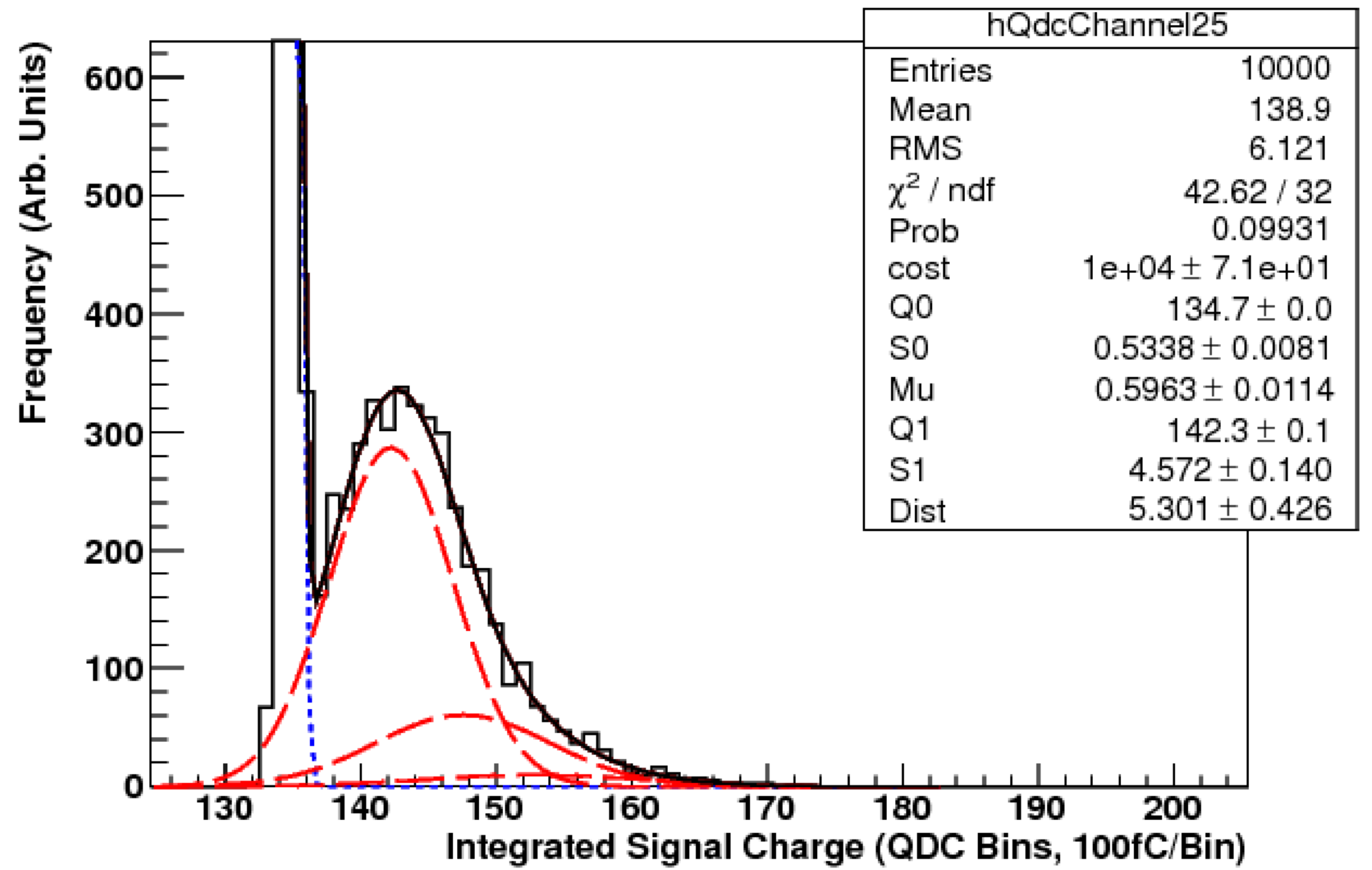}}}
}
\mbox{
\subfigure[1075\,V]{\scalebox{1.0}{\includegraphics[width=0.48\linewidth]{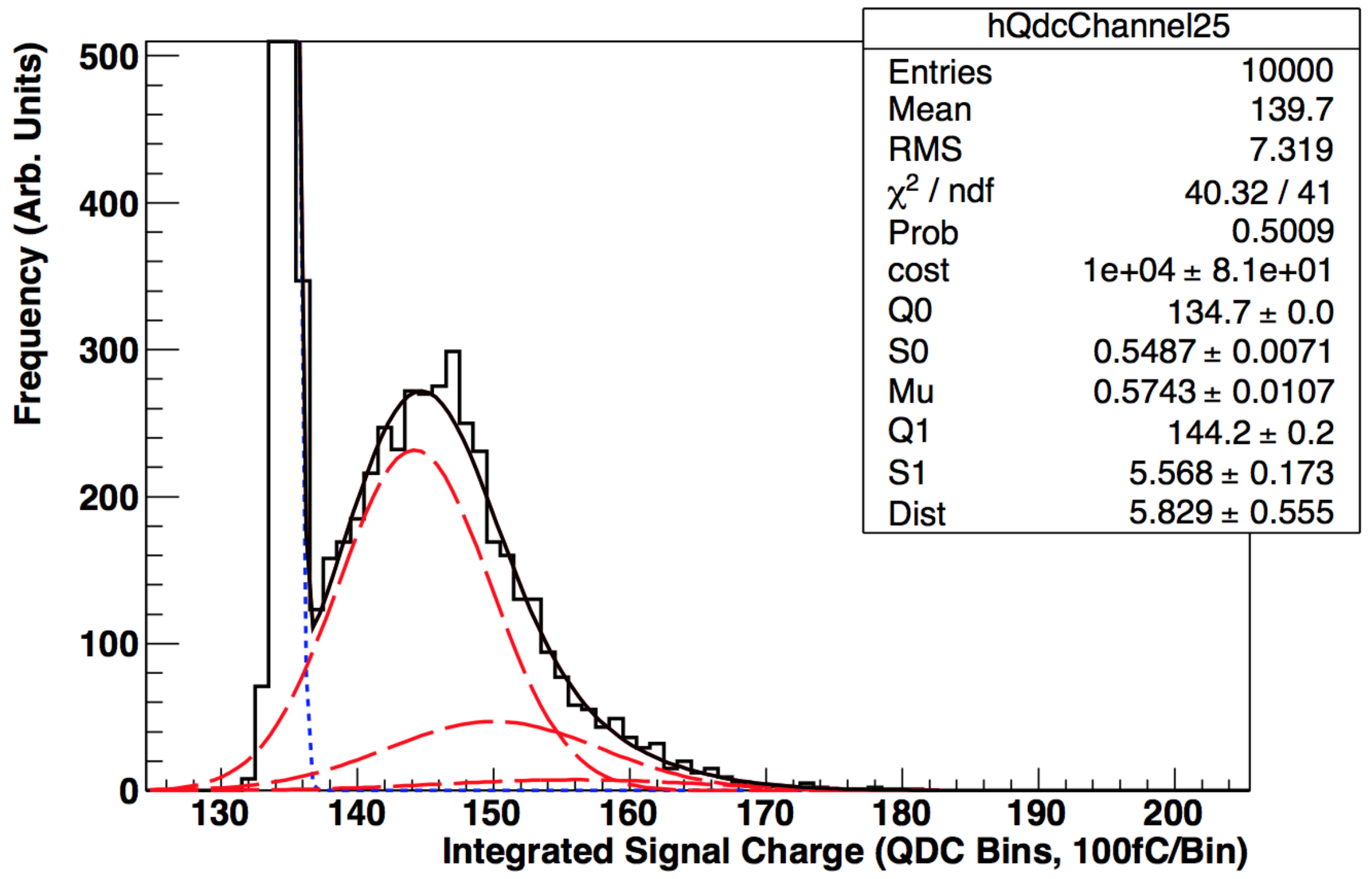}}}
}
\caption{\small \sf QDC spectra for pixel 36 (QDC channel 25) of the CA4655 MAPMT for HV=-1000, -1040 and -1075 V. The curves are the results of the fits, as in Fig.~\ref{fig:ca4658_1000_QDC}.}
\label{fig:ca4658_25_hv_QDC}
\end{center}
\end{figure}

\begin{figure}\begin{center}
\includegraphics[width=.6\textwidth]{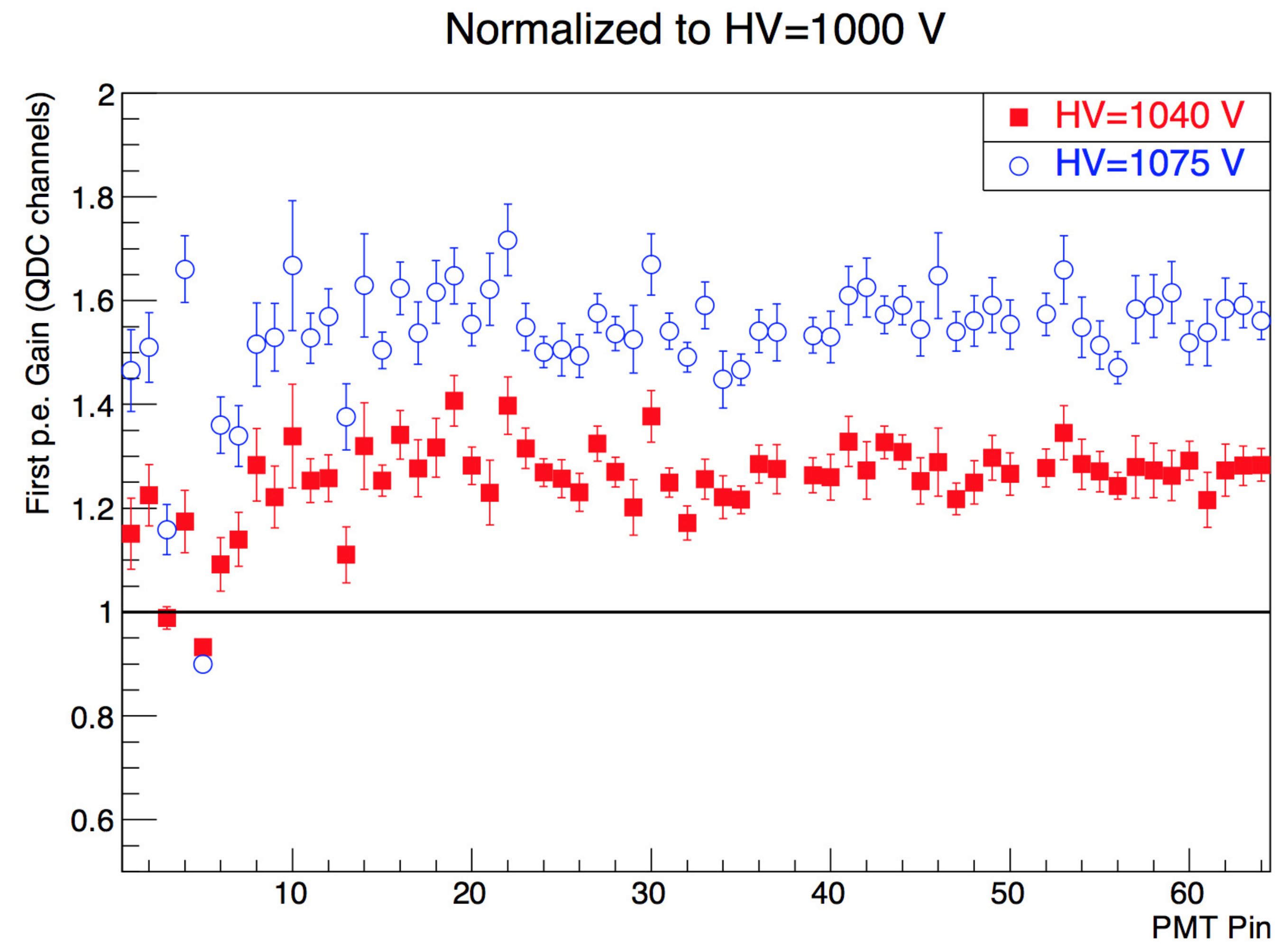}
\caption{\small \sf Relative gain ($Q_{1}-Q_{0}$) for the 64 pixels of the MAPMT CA4655 at -1040\,V (red squares) and -1075\,V (empty circles) with respect to the measurements at -1000\,V (horizontal line at value 1).
}
\label{fig:ca4658_gain_hv}
\end{center}
\end{figure}

In Fig.~\ref{fig:ca4658_25_hv_QDC}, we compare the measured QDC distributions of pixel 36 (mapping to QDC channel 25) of the CA4658 MAPMT for the HV values of -1000, -1040 and -1075\,V.
The improvement in the separation of the signal from the pedestal is clearly visible as the HV, or gain, increases. The separation increases from 6\,QDC channels at -1000\,V to 9.5\,QDC channels at -1075\,V. As a further example, the relative increase in the gain ($Q_{1}-Q_{0}$) at -1040\,V and -1075\,V with respect to the -1000\,V measurements, for all the 64 pixels of the CA4658 MAPMT, is shown in Fig.~\ref{fig:ca4658_gain_hv}. For the few results which are equal to or less than 1, this is due to a either faulty electronic channels or low gain pixels, where the fitting procedure was not as successful.
On average, for all the 28 MAPMTs tested, the gain increase ($Q_{1}-Q_{0}$) found at -1075\,V was between $50\%$ and $80\%$ with respect to that at -1000\,V.

\begin{figure}[h!]
\begin{center}
\mbox{
\subfigure[1000\,V.]{\scalebox{1.0}{\includegraphics[width=0.325\linewidth]{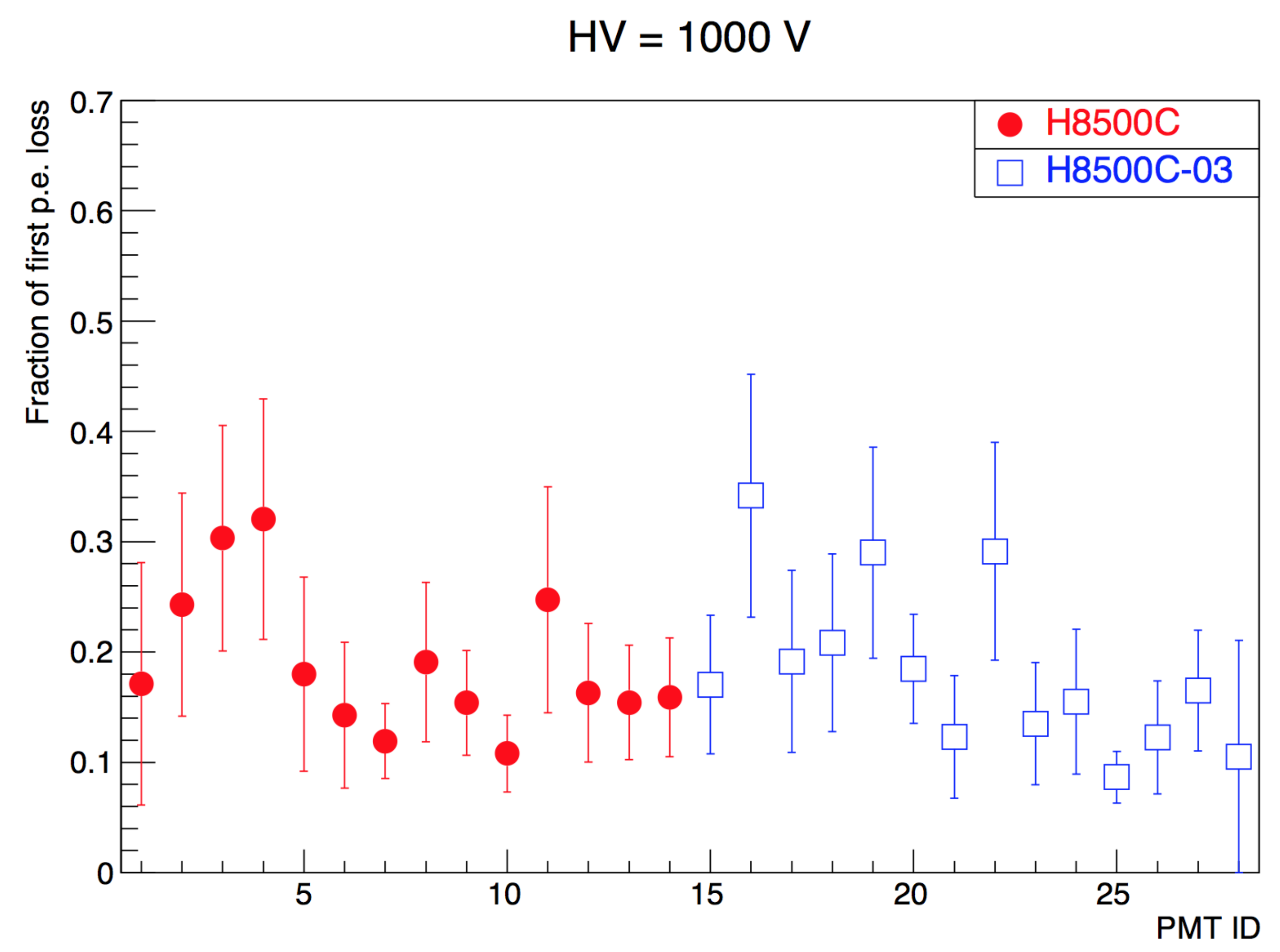}}}
\subfigure[1040\,V]{\scalebox{1.0}{\includegraphics[width=0.325\linewidth]{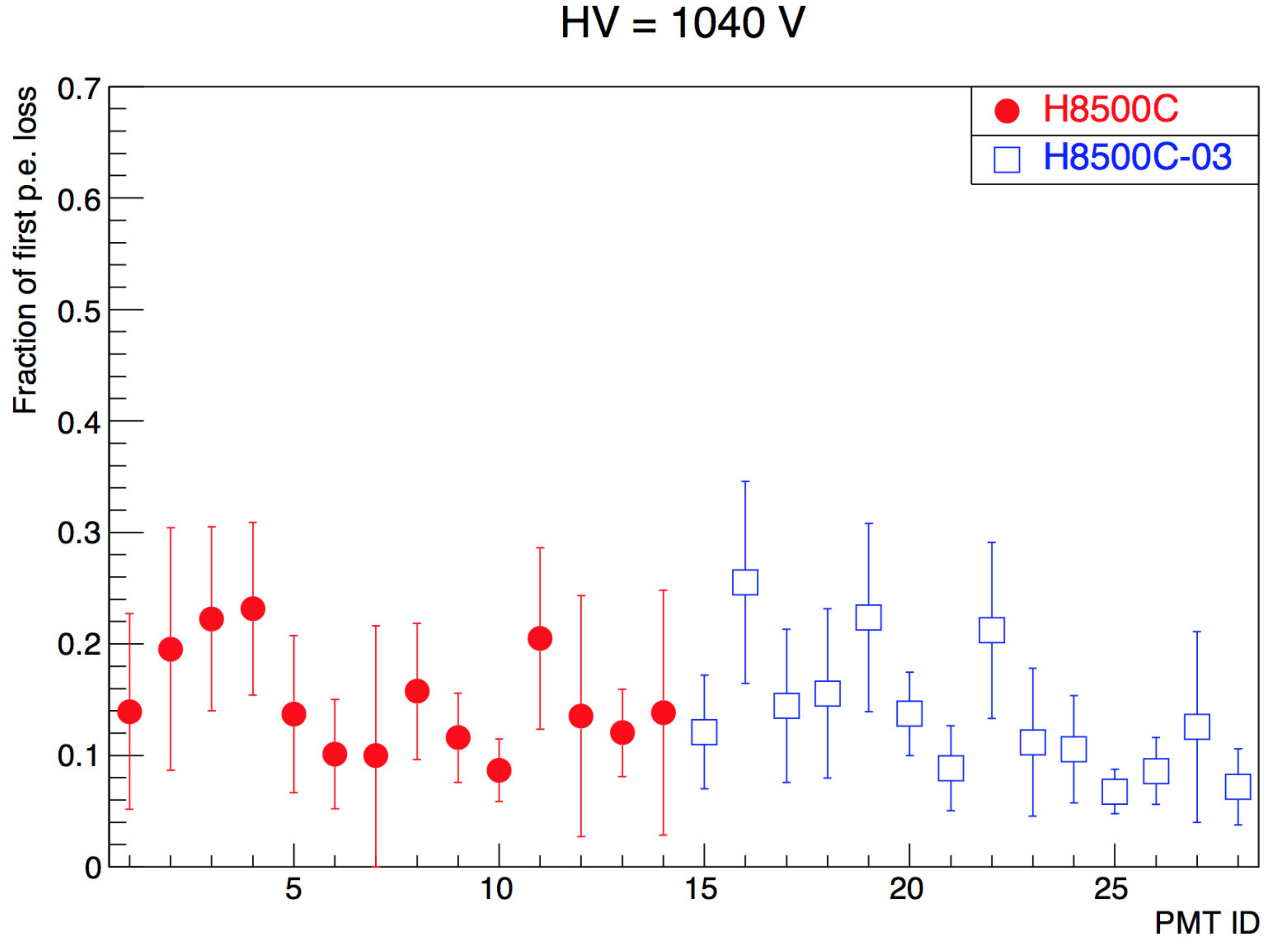}}}
\subfigure[1075\,V]{\scalebox{1.0}{\includegraphics[width=0.325\linewidth]{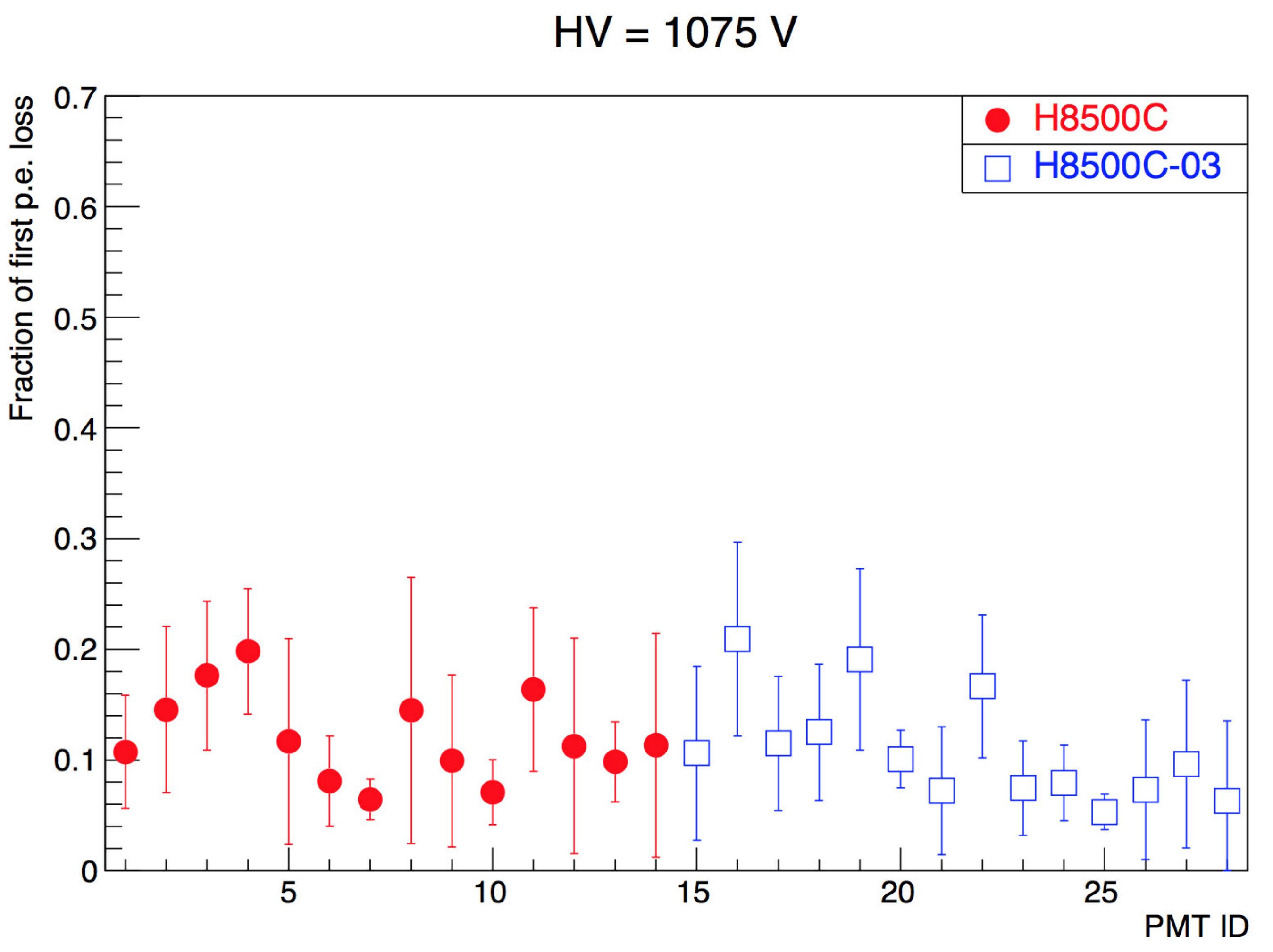}}}
}
\caption{\small \sf Fractional loss of the single p.e. signal below the 3\,$\sigma$ cut above the pedestal, estimated from the integration of the Gaussian fit to the first p.e. peak, at HV=-1000\,V, -1040\,V and -1075\,V. Average values of all 64 pixels for each of the 14 H8500C MAPMTs (red circles) and the 14 H8500-C03 MAPMTs (empty squares) are given. The error bars represent the RMS of the distributions for the 64 pixels of each MAPMT.}
\label{fig:h8500_loss}
\end{center}
\end{figure}

The separation between the first p.e. signal from the pedestal is the crucial point for a single photon counting application such as the CLAS12 RICH detector. In fact, in the RICH detector, the occupancy will be no more than one photon per pixel, thus it is important that the fraction of the single p.e. spectrum below the pedestal peak is as small as possible. One can estimate this loss of events by integrating the Gaussian fit of the one p.e. peak below a suitable cut selected to remove the QDC pedestal. Assuming a cut at 3\,$\sigma$ above the pedestal peak, we obtained the results shown in Figs.~\ref{fig:h8500_loss}\,(a), (b) and (c), for the three supply voltages of -1000\,V, -1040\,V and -1075\,V.
We see that, at the highest voltage, except for one MAPMT, the average loss is below 20\,\% and that, within one RMS, almost all of the MAPMTs have a loss fraction below 30\,\%.

\section{Cross Talk Analysis}
\label{sec:Xtalk}
\begin{figure}\begin{center}
\includegraphics[width=.75\textwidth]{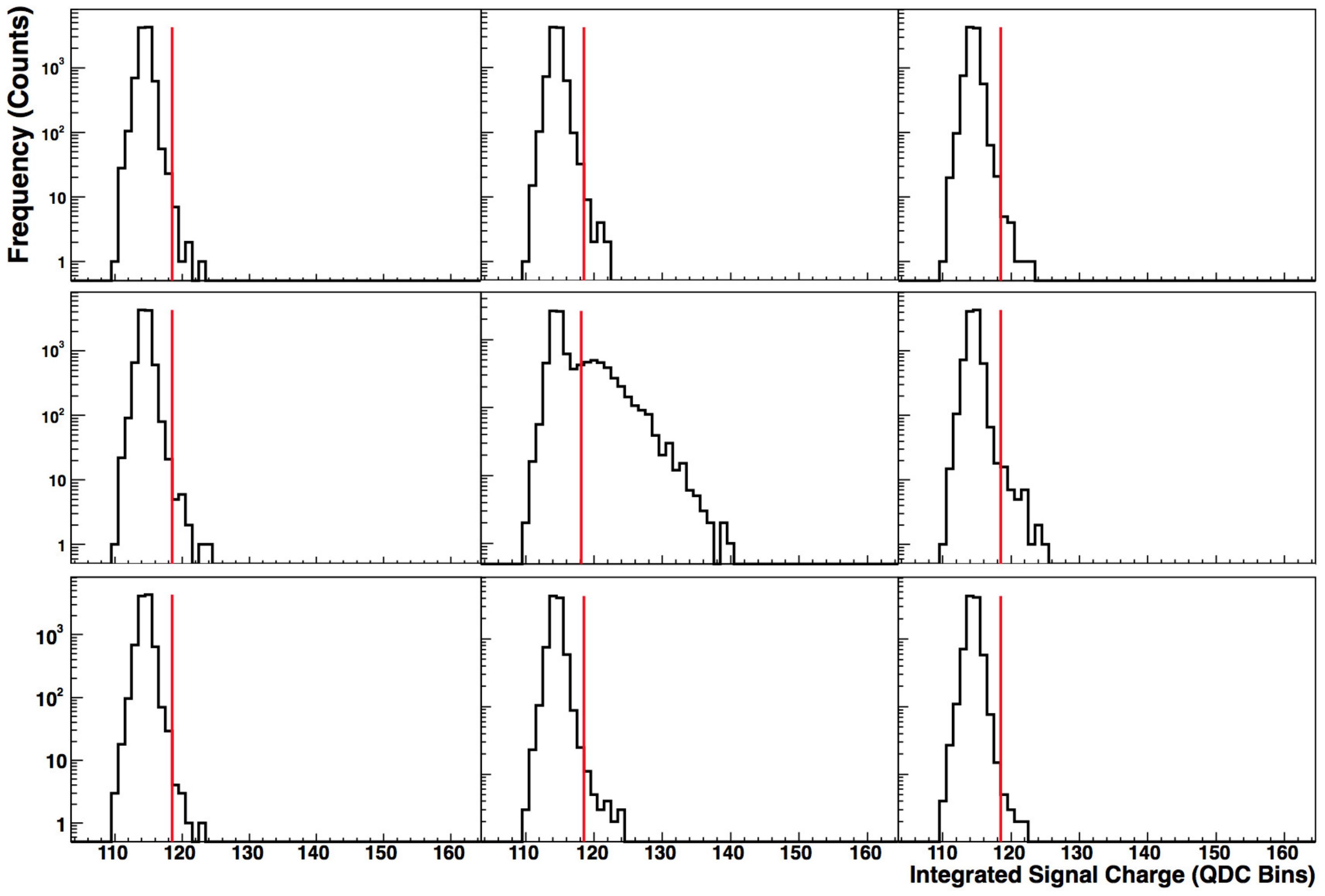}
\caption{\small \sf QDC distributions for one pixel of the CA4658 MAPMT when it is illuminated (central histogram) and when the laser strikes its eight neighbouring pixels, i.e. when it is not illuminated (surrounding histograms). The red vertical lines indicate the pedestal threshold cuts used. The y-axes denote the frequency (counts) and the x-axes denote the integrated signal charge (QDC bins, 100fC/bin).
}
\label{fig:ca4658_Xtalk_34_1075}
\end{center}
\end{figure}

As previously described, in each measurement of an MAPMT and at a given supply voltage, only one pixel was illuminated by the laser at any time. The data from all 64 pixels, however, were simultaneously readout by the DAQ during each measurement, thus allowing an analysis of the cross talk between pixels to be made. The procedure is shown in Fig.~\ref{fig:ca4658_Xtalk_34_1075}. The cross talk levels shown here are higher than those intrinsic to the MAPMT alone since they include contributions from stray laser photons within the set-up. A subsequent study with a mask allowing only the illumination of a single pixel revealed a significant reduction of the measured cross talk. Furthermore, we used the measurements taken at the HV setting of -1075\,V to provide an upper limit for the measurements, since the cross talk magnitudes are expected to increase with HV when determined with a fixed threshold.

\begin{figure}[h!]
\begin{center}
\mbox{
\subfigure[Side-sharing neighbours.]{\scalebox{1.0}{\includegraphics[width=0.48\linewidth]{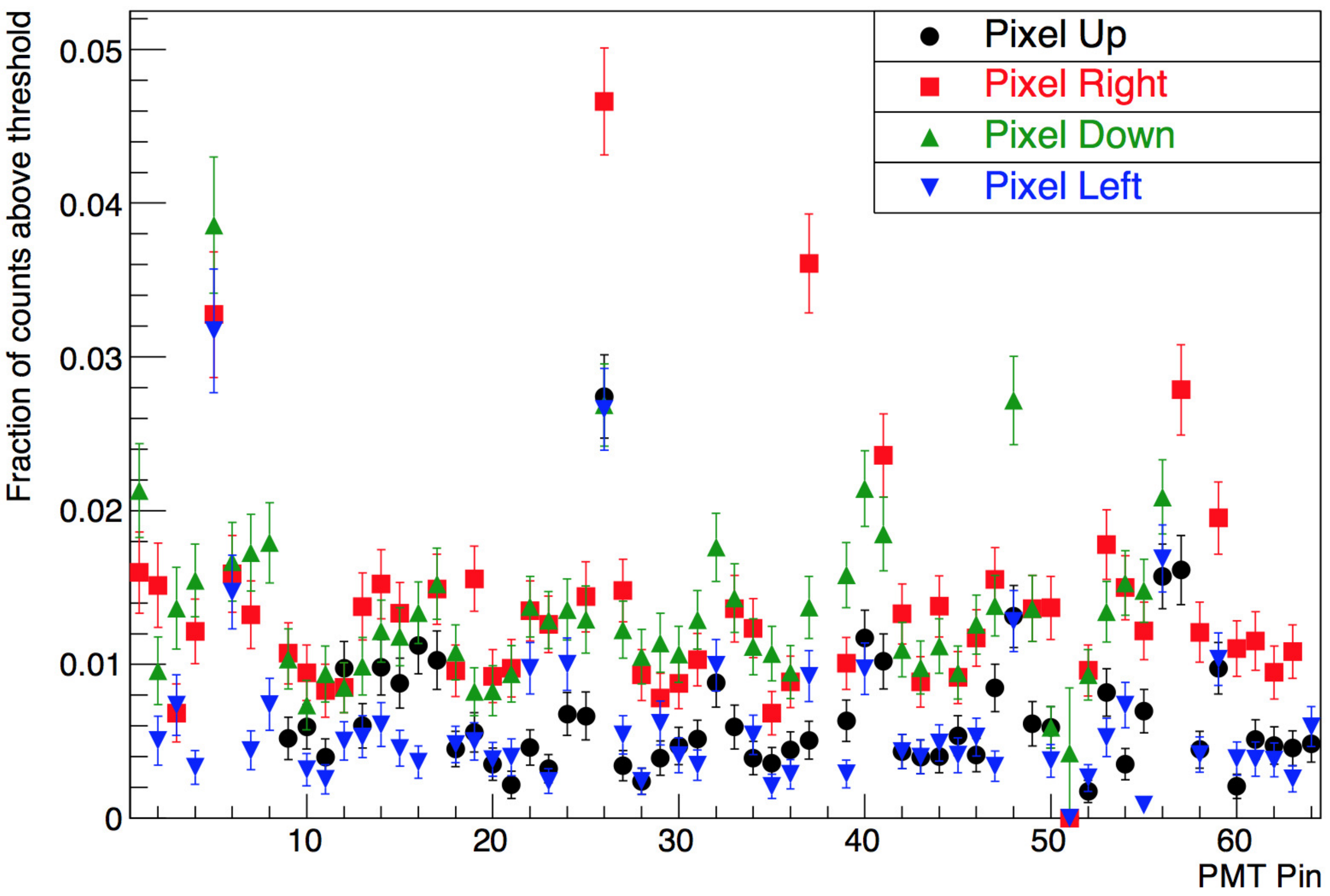}}}
\subfigure[Cornering neighbours.]{\scalebox{1.0}{\includegraphics[width=0.48\linewidth]{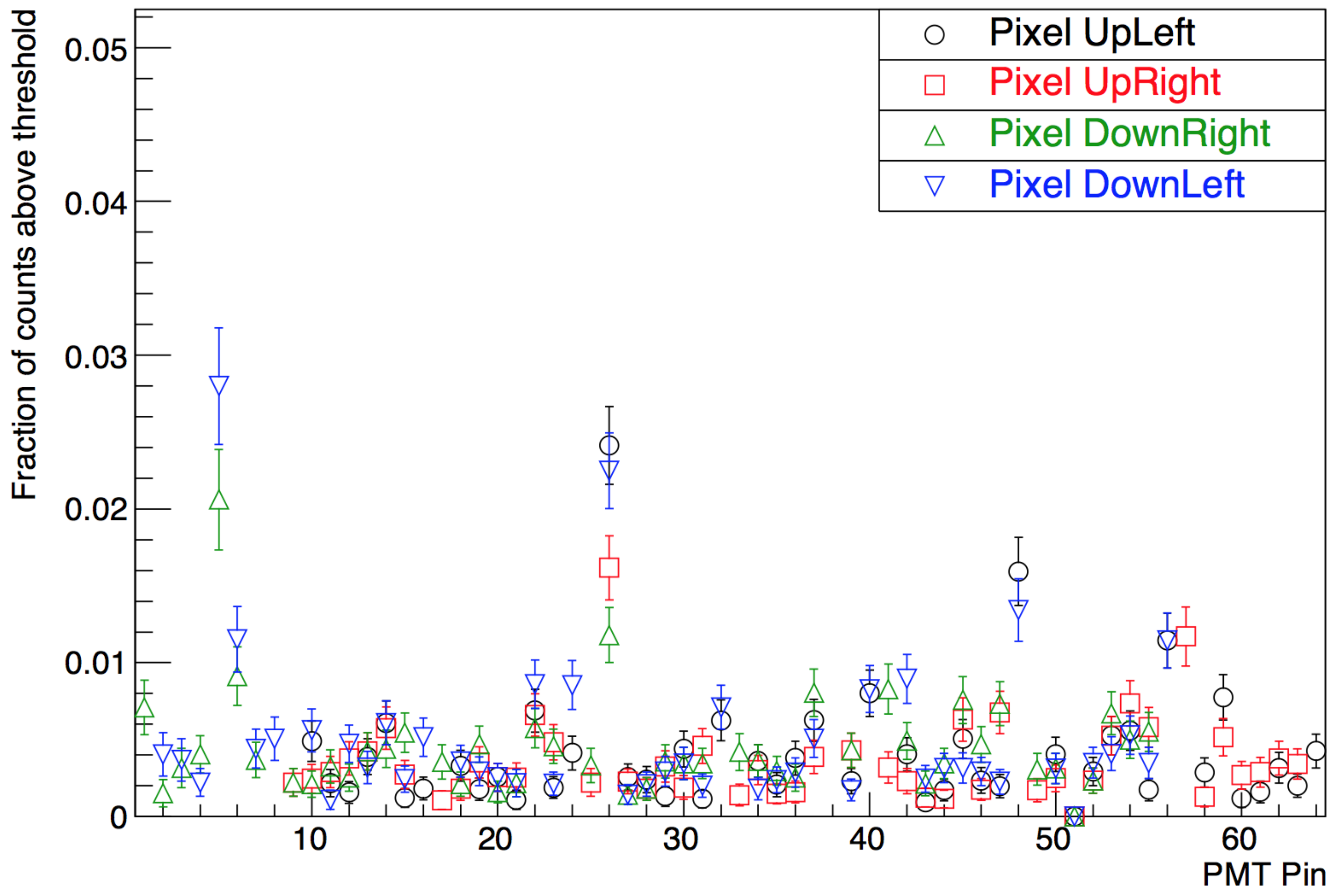}}}
}
\caption{\small \sf Fraction of cross talk hits for the 64 pixels of the CA4658 MAPMT when each of the eight adjacent pixels ((a) up, down, left and right side-sharing neighbours and (b) the 4 cornering neighbours) are illuminated.}
\label{fig:ca4658_Xtalk_1075}
\end{center}
\end{figure}

\begin{figure}[h!]
\begin{center}
\mbox{
\subfigure[]{\scalebox{1.0}{\includegraphics[width=0.75\linewidth]{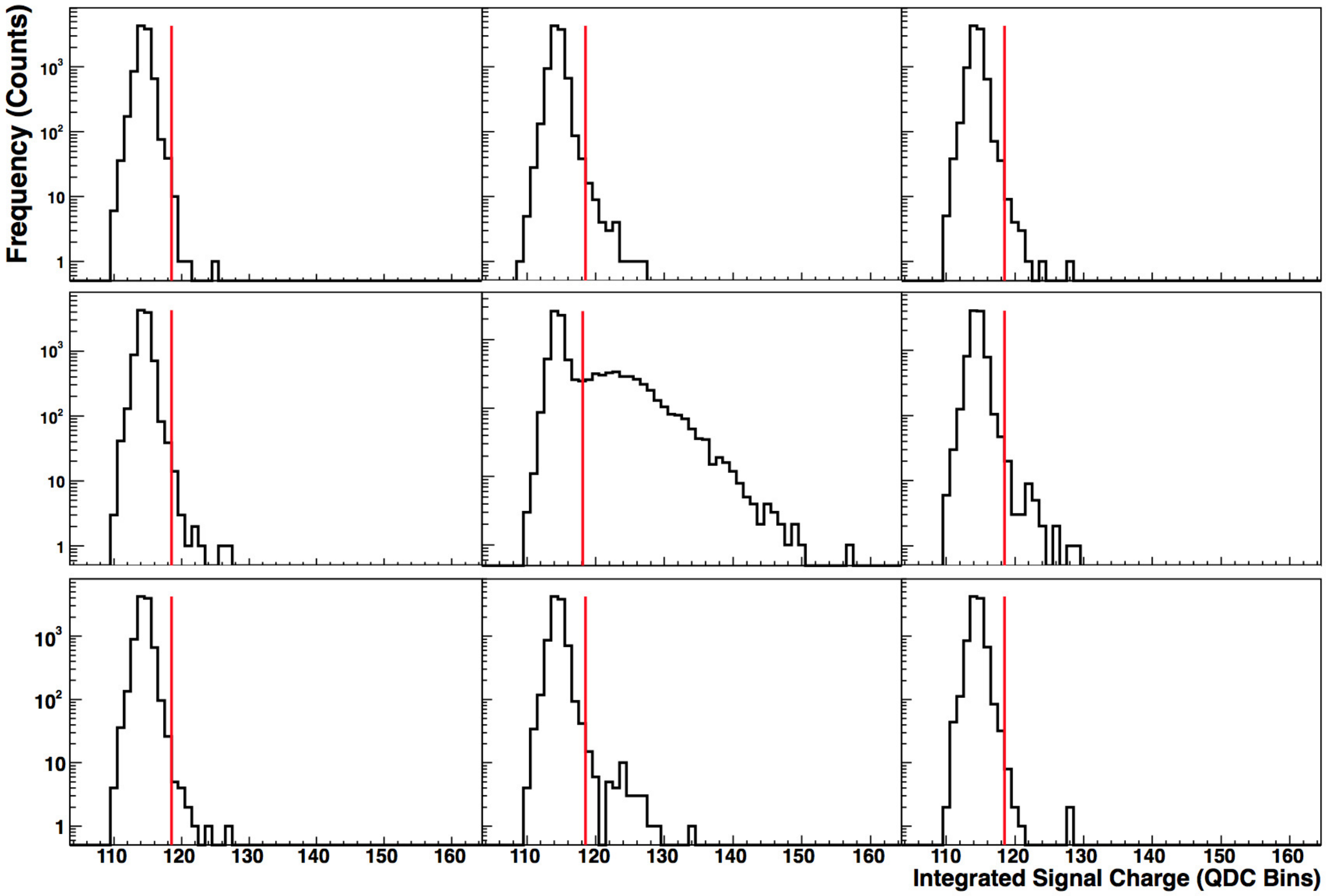}}} }
\\
\mbox{
\subfigure[]{\scalebox{1.0}{\includegraphics[width=0.75\linewidth]{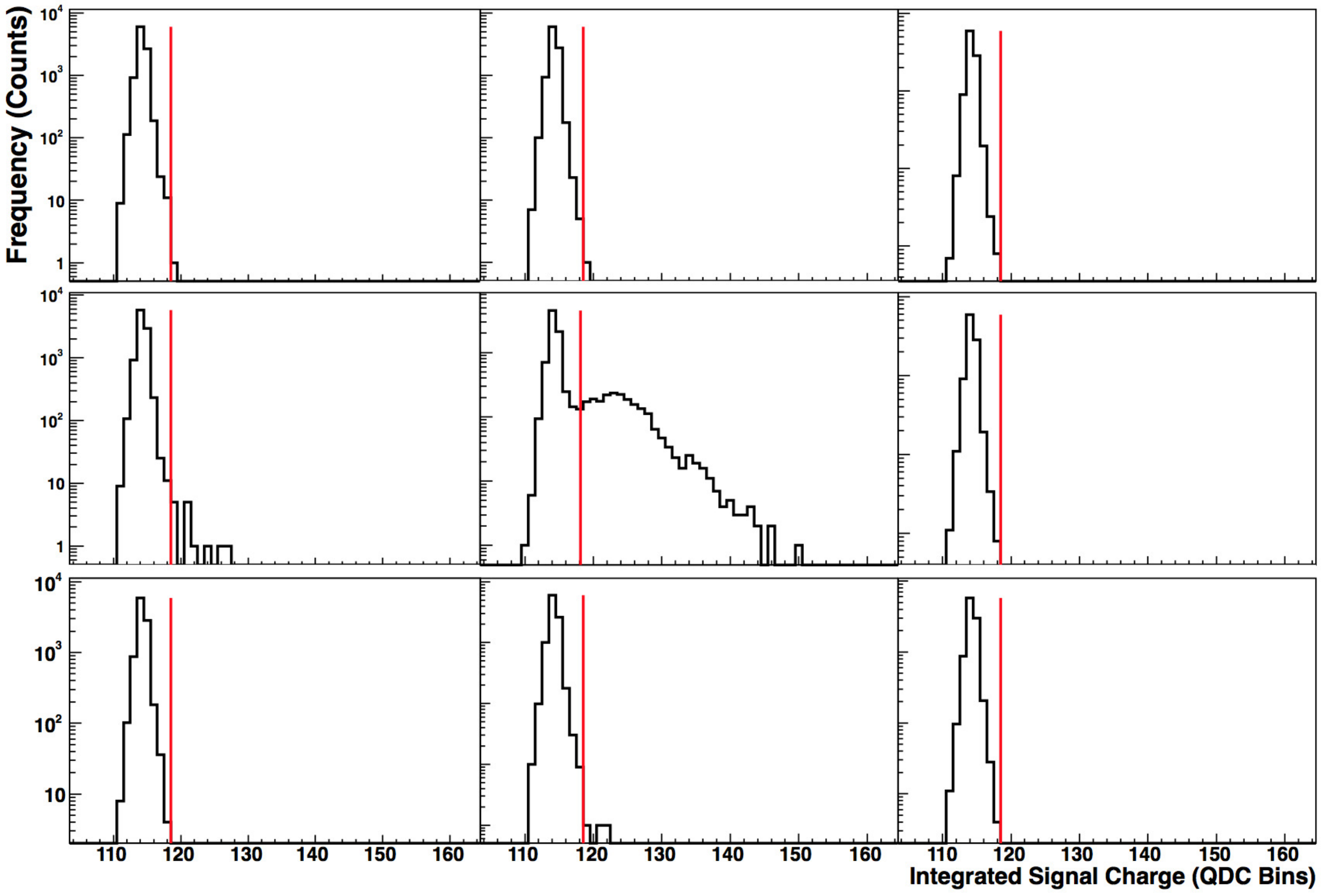}}}
}
\caption{\small \sf QDC distributions for one pixel of the CA4686 MAPMT when it is illuminated (central histogram) and when the laser strikes its eight neighbouring pixels (surrounding histograms), (a) without and (b) with the use of a mask to eliminate stray laser photon contributions to the measured cross talk. As for Fig.~\ref{fig:ca4658_Xtalk_34_1075}, the y-axes denote the frequency (counts) and the x-axes denote the integrated signal charge (QDC bins, 100fC/bin).}
\label{fig:h8500_Xtalk_Mask}
\end{center}
\end{figure}

We started by looking at the QDC spectrum of the illuminated pixel (central plot in Fig.~\ref{fig:ca4658_Xtalk_34_1075}) and we calculated the number of events above the pedestal threshold cut. Here, a pedestal threshold cut of 5\,$\sigma$ was used to reduce the contribution from electronic noise to below 10$^{-5}$.
Then, we looked at the QDC spectrum of the same pixel when it was not illuminated, but when the laser struck one of its 8 adjacent pixels, the 4 side-sharing and the 4 cornering neighbours. Again for each case we counted the number of events above the threshold cut in the spectra. As visible in Fig.~\ref{fig:ca4658_Xtalk_34_1075}, a small fraction of events above the threshold cut is recorded when the pixel is not illuminated, especially when the laser strikes the side-sharing neighbours. Such events are not due to electronic noise (compare with the pedestal shown in Fig.~\ref{fig:DAQ_noise}), but can be caused by cross talk between adjacent pixels or stray photons from the laser (as will be shown later in Fig.~\ref{fig:h8500_Xtalk_Mask}). Normalising the number of such events to the number of events above threshold when the pixel itself is illuminated, we calculated the cross talk probability for that given pixel from its adjacent ones. This is shown in Fig.~\ref{fig:ca4658_Xtalk_1075} for the CA4658 MAPMT for each of its 64 pixels for the 8, 5 or 3 neighbouring pixels, respectively, depending on whether it is a centre, edge or corner pixel. The results have been divided to show the cross talk contributions from either side-sharing or cornering neighbours separately.
As visible, the cross talk contribution from each individual side-sharing neighbour (Fig.~\ref{fig:ca4658_Xtalk_1075}\,(a)) typically is 1\,-\,1.5\,\% and from each cornering neighbour (Fig.~\ref{fig:ca4658_Xtalk_1075}\,(b)) it is typically less than 0.5\,\%. From Fig.~\ref{fig:ca4658_Xtalk_1075}, no strong dependence upon cross talk direction within the side-sharing or cornering neighbour sub-sets is observed.

\begin{figure}[h!]
\begin{center}
\mbox{
\subfigure[]{\scalebox{1.0}{\includegraphics[width=0.49\linewidth]{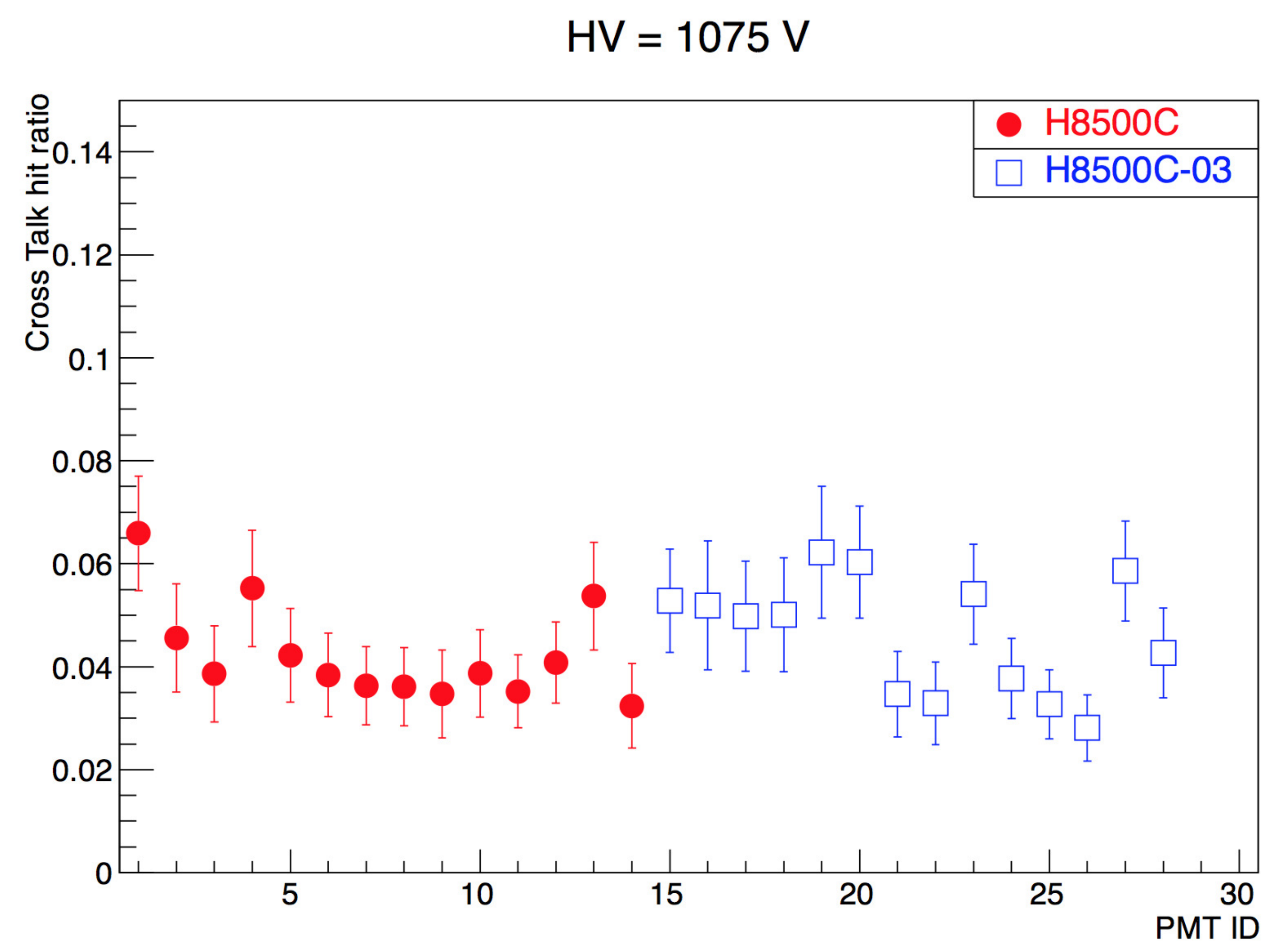}}}
\subfigure[]{\scalebox{1.0}{\includegraphics[width=0.49\linewidth]{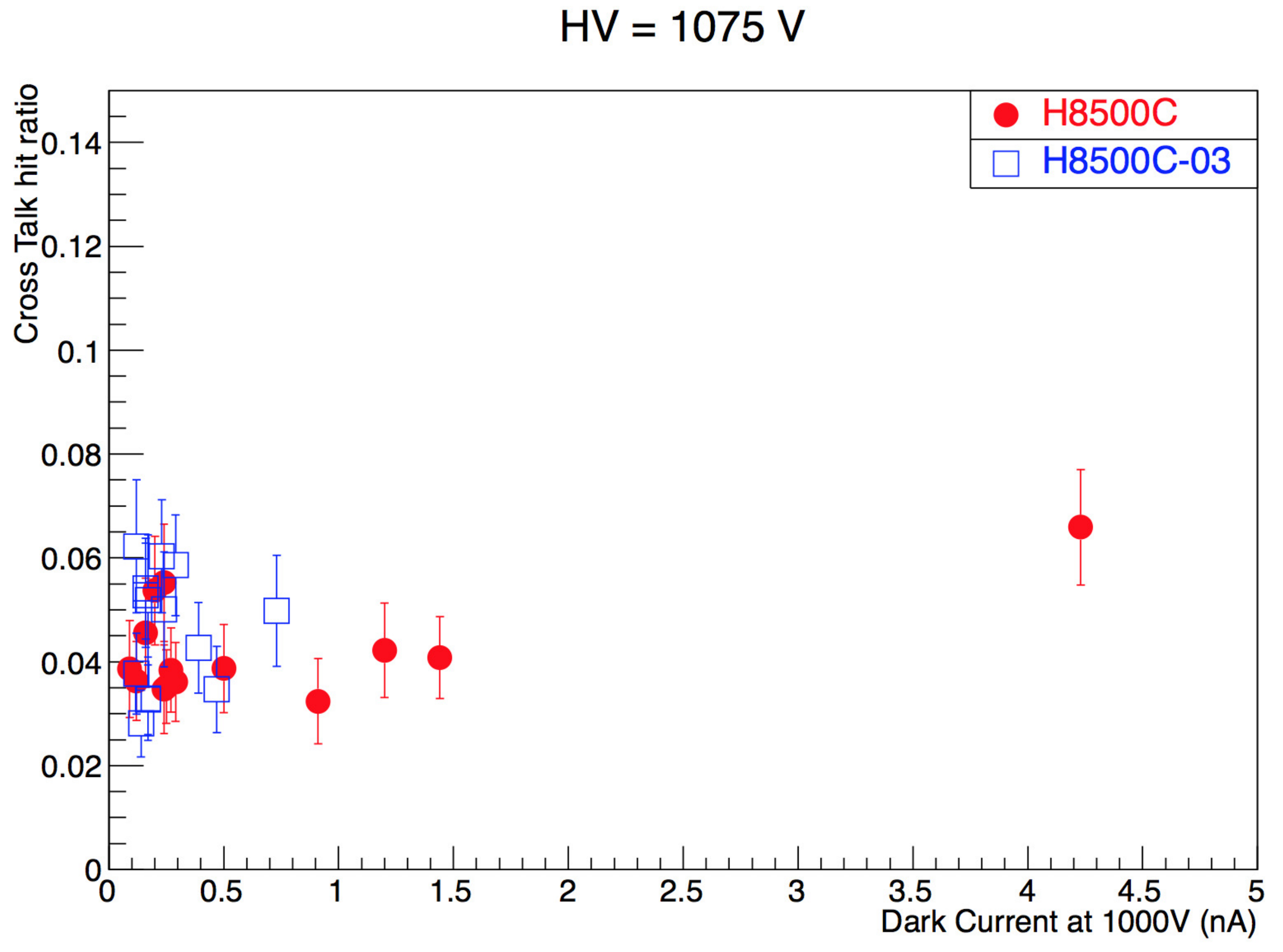}}}
}
\caption{\small \sf (a) Average fraction of cross talk hits for the 14 H8500C MAPMTs (red circles) and the 14 H8500C-03 MAPMTs (empty squares, with the error bars representing the RMS values of the 64 pixels. (b) Average fraction of cross talk hits as a function of MAPMT dark current (from Tab.~\ref{tab:pmt_sheets}).}
\label{fig:h8500_Xtalk_1075}
\end{center}
\end{figure}

\begin{figure}[h!]
\begin{center}
\mbox{
\subfigure[Side-sharing neighbours]{\scalebox{1.0}{\includegraphics[width=0.49\linewidth]{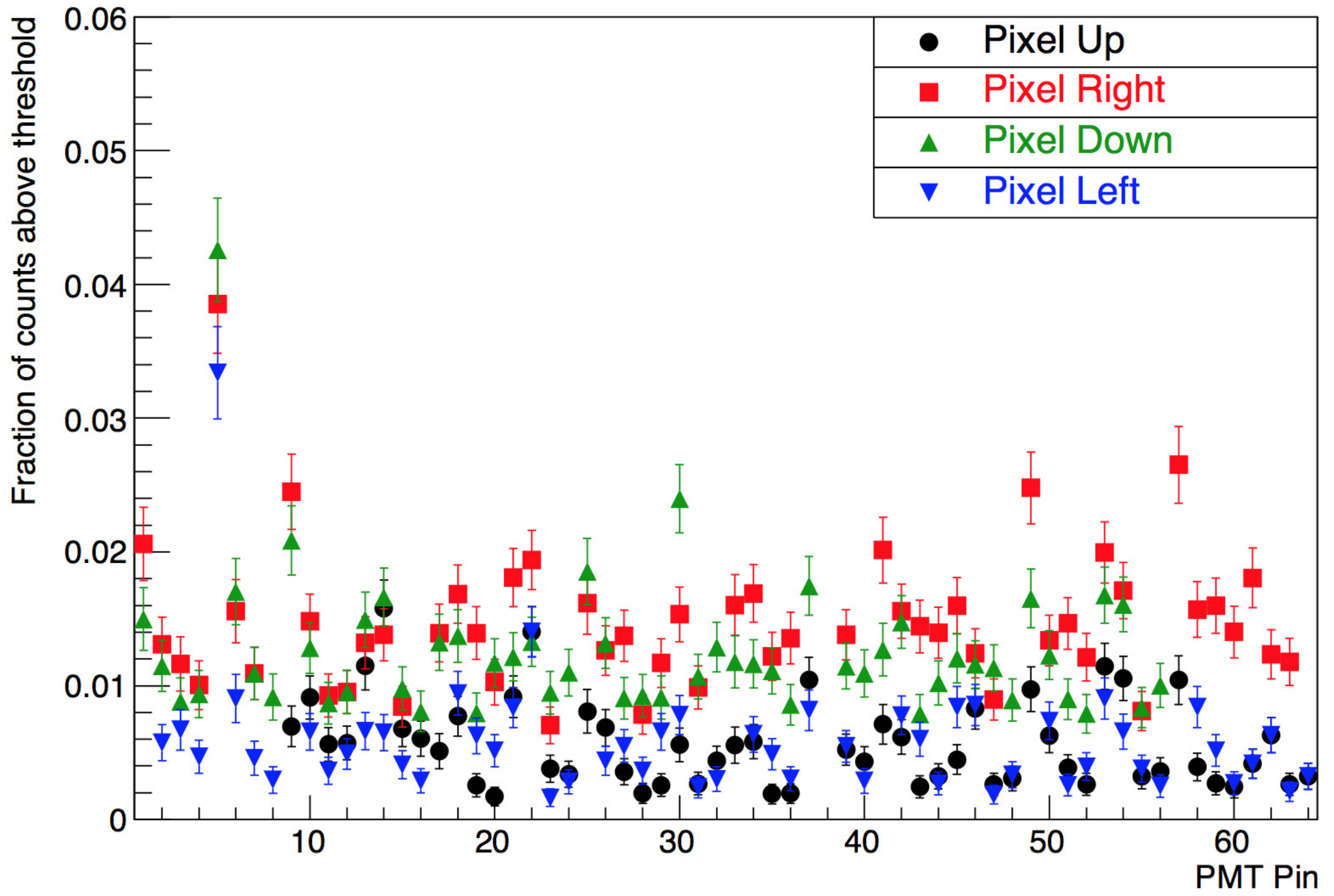}}}
\subfigure[Cornering neighbours]{\scalebox{1.0}{\includegraphics[width=0.49\linewidth]{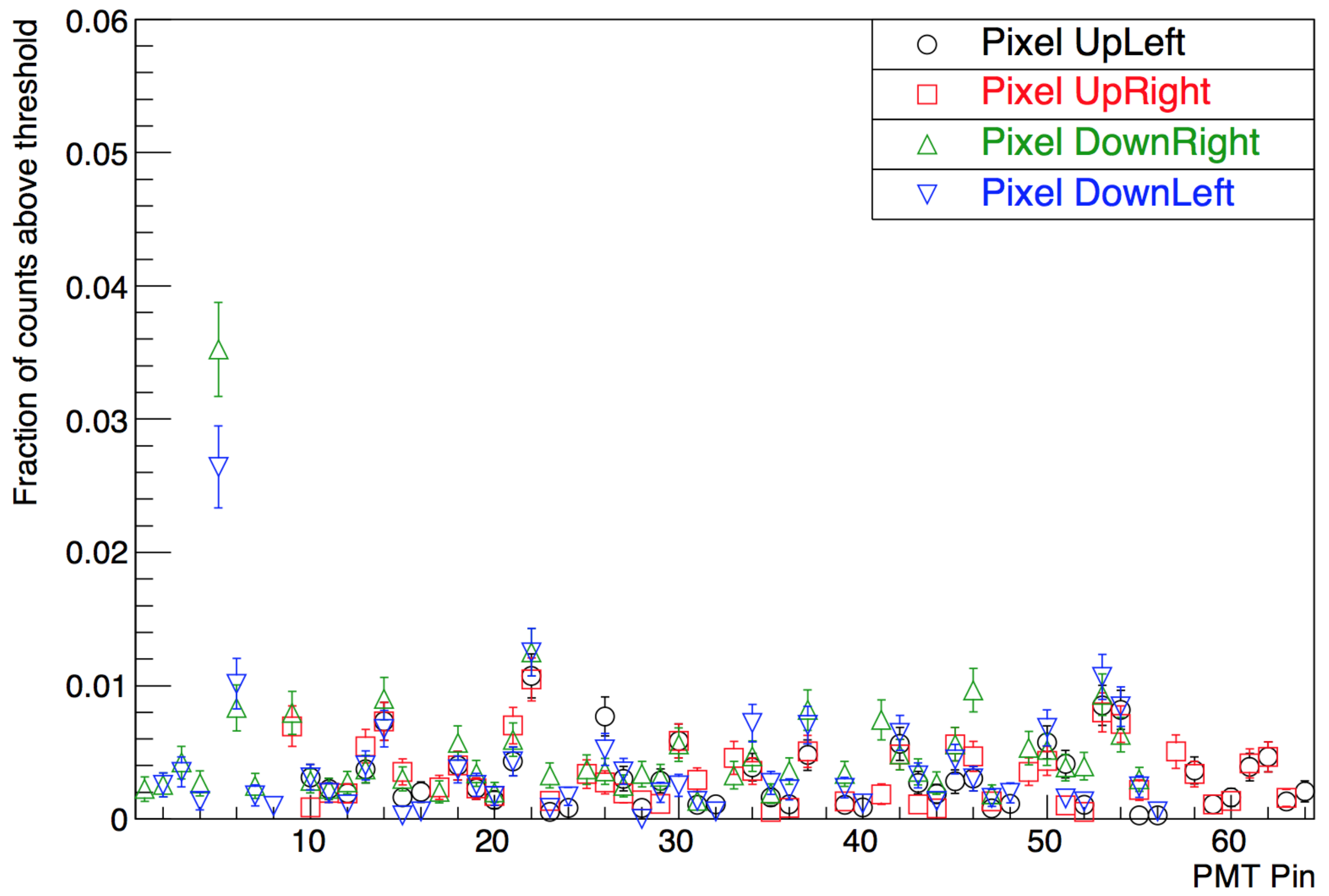}}}
}
\caption{\small \sf Fraction of cross talk hits for the 64 pixels of the CA4686 MAPMT when each of the eight adjacent pixels ((a) up, down, left and right side-sharing neighbours and (b) the 4 cornering neighbours) are illuminated without use of a mask.}
\label{fig:h8500_Xtalk_NoMask_AllPixels}
\end{center}
\end{figure}

\begin{figure}[h!]
\begin{center}
\mbox{
\subfigure[Side-sharing neighbours]{\scalebox{1.0}{\includegraphics[width=0.49\linewidth]{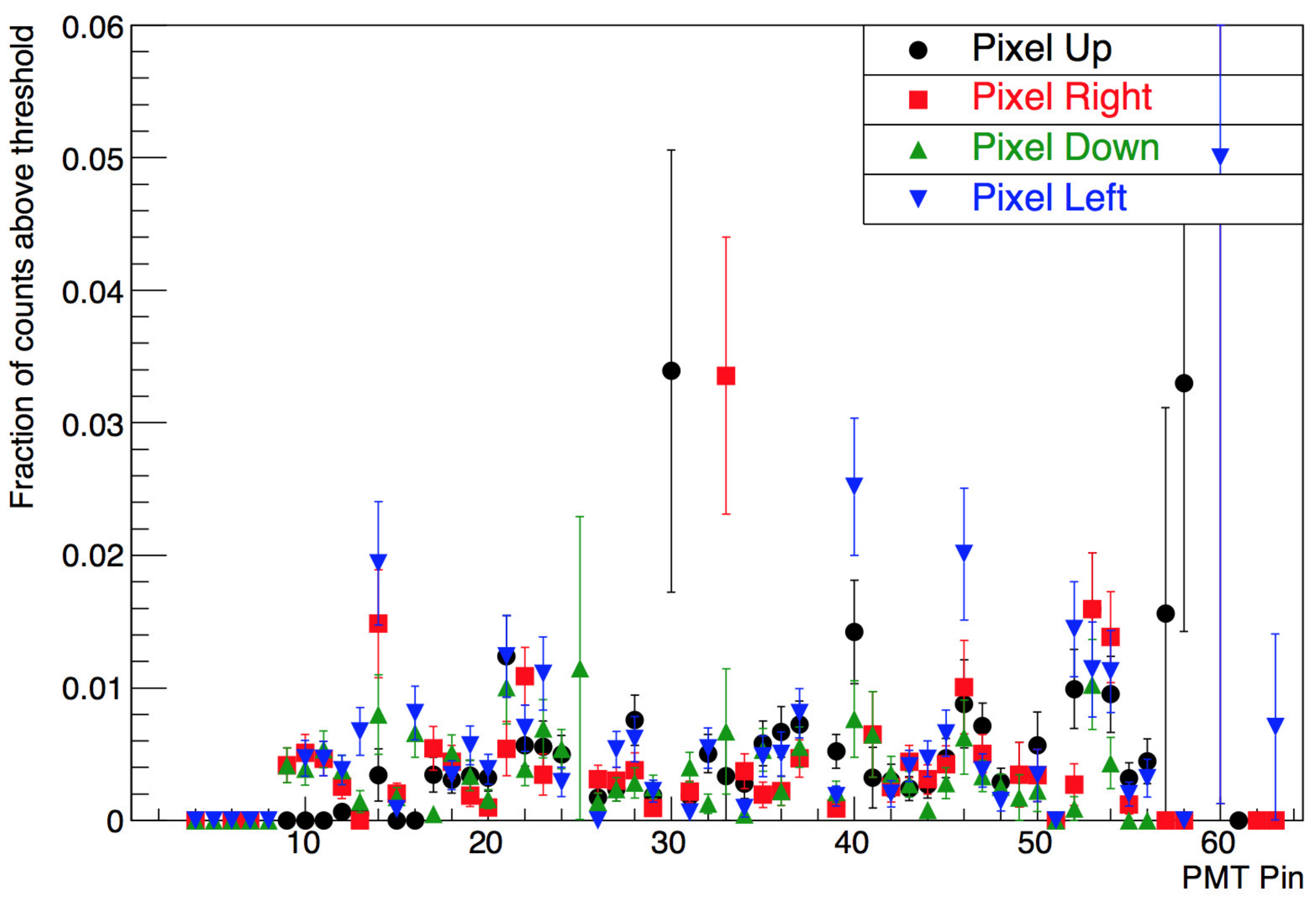}}}
\subfigure[Cornering neighbours]{\scalebox{1.0}{\includegraphics[width=0.49\linewidth]{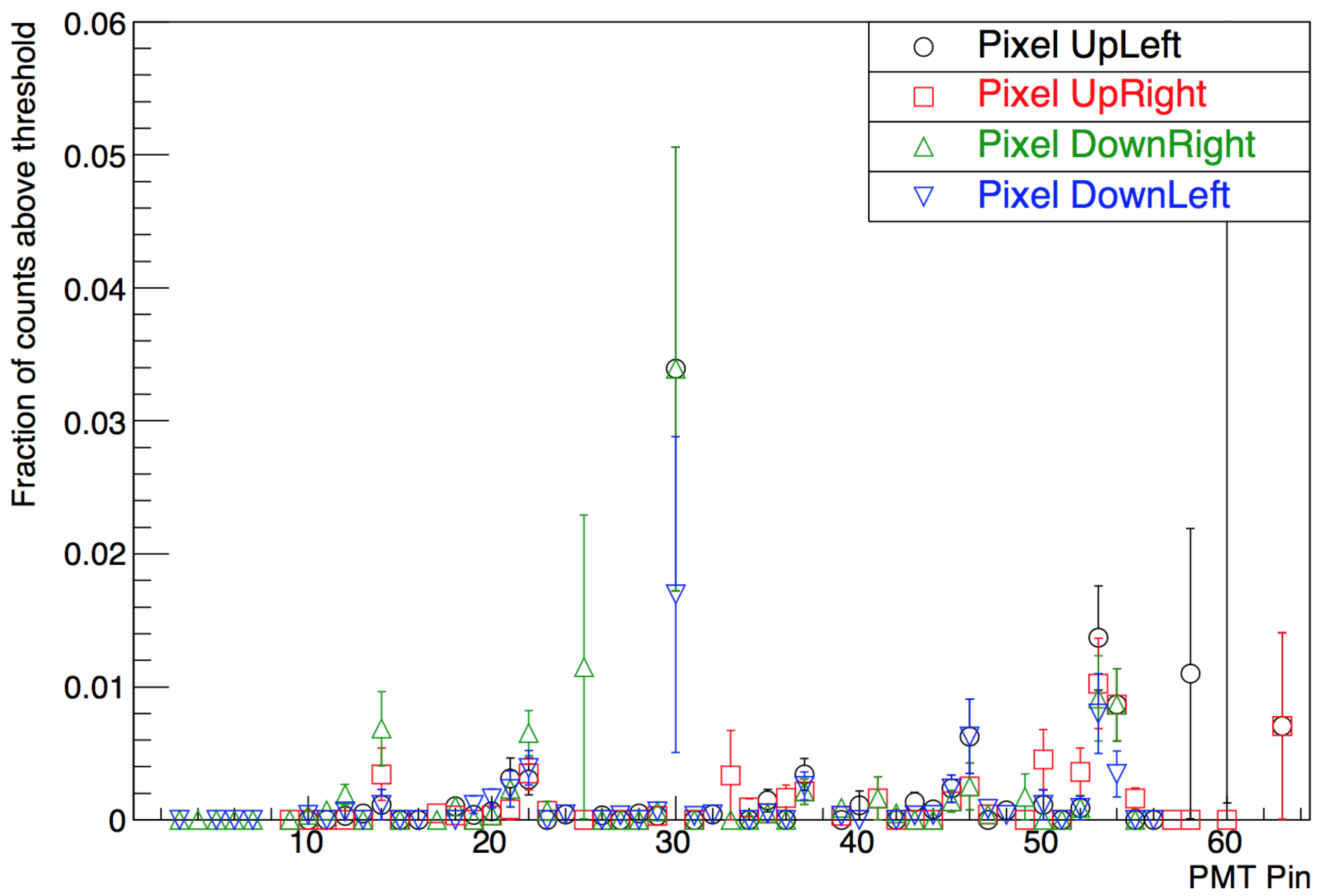}}}
}
\caption{\small \sf Fraction of cross talk hits for the 64 pixels of the CA4686 MAPMT when each of the eight adjacent pixels ((a) up, down, left and right side-sharing neighbours and (b) the 4 cornering neighbours) are illuminated with use of a mask.}
\label{fig:h8500_Xtalk_Mask_AllPixels}
\end{center}
\end{figure}

Summing the cross talk contributions from the eight adjacent pixels, we obtain average cross talk levels of typically 4\,-\,6\,\% for all the tested MAPMTs, matching well the Hamamatsu specifications. This is reported in Fig.\ref{fig:h8500_Xtalk_1075}\,(a), which shows the average values of the cross talk for all the MAPMTs tested, with the error bars again representing the RMS values amongst the 64 pixels of each MAPMT.
In Fig.~\ref{fig:h8500_Xtalk_1075}\,(b), we show the average values of the cross talk as a function of the MAPMT dark current levels provided by the Hamamatsu test sheets (and as reported in Tab.~\ref{tab:pmt_sheets}). As seen, no dependence of the cross talk on the dark current appears to emerge, which again confirms that the dark currents of the H8500 MAPMTs are not a concern for their application to the RICH detector. The cross talk values obtained in this study represent an upper limit on the magnitudes, since the measurements were performed without use of a mask covering all MAPMT pixels, except from the illuminated one. Subsequent studies have shown that the incorporation of such a mask reduces the cross talk observed, due to the elimination of contributions from stray laser light.
For example, Fig.~\ref{fig:h8500_Xtalk_Mask}\,(a) shows the spectra of one pixel from the CA4686 MAPMT, like in Fig.~\ref{fig:ca4658_25_hv_QDC}, when illuminated directly with the collimated light source or when its neighbours are illuminated. In Fig.~\ref{fig:h8500_Xtalk_Mask}\,(b) we show the equivalent spectra for the same pixel obtained with a mask put in front of the MAPMT, which blocks any illumination of neighbouring pixels. The mask consists of a black PVC square with outer dimensions matching those of the MAPMT and a thickness of 0.5\,mm. The mask is perforated with a matrix of 1\,mm diameter holes which are positioned at the centres of each of the 64 pixel positions. In the second case the number of signals above threshold seen when the neighbouring pixels are illuminated is significantly reduced. For the pixel presented in Fig.~\ref{fig:h8500_Xtalk_Mask} the cross talk integrated over all neighbours went down from $\sim$4\,\% without mask to $\sim$1\,\% with mask. This represents one of the strongest reductions we have seen. The cross talk probabilities for all pixels of the CA4686 MAPMT at 1075\,V  without and with the mask in place are shown in Figs.~\ref{fig:h8500_Xtalk_NoMask_AllPixels} and \ref{fig:h8500_Xtalk_Mask_AllPixels} respectively. As expected, the cross talk levels are mostly reduced when a mask is used, with the exception of some spurious points.
It is concluded that, even based upon the slightly higher cross talk values measured without use of a mask, the MAPMTs are low cross talk devices and this noise source is not problematic for their use in the CLAS12 RICH detector.


\section{Magnetic Field Studies}
\label{sec:MagField}
\begin{figure}[h]
\centering
\includegraphics[width=0.6\textwidth]{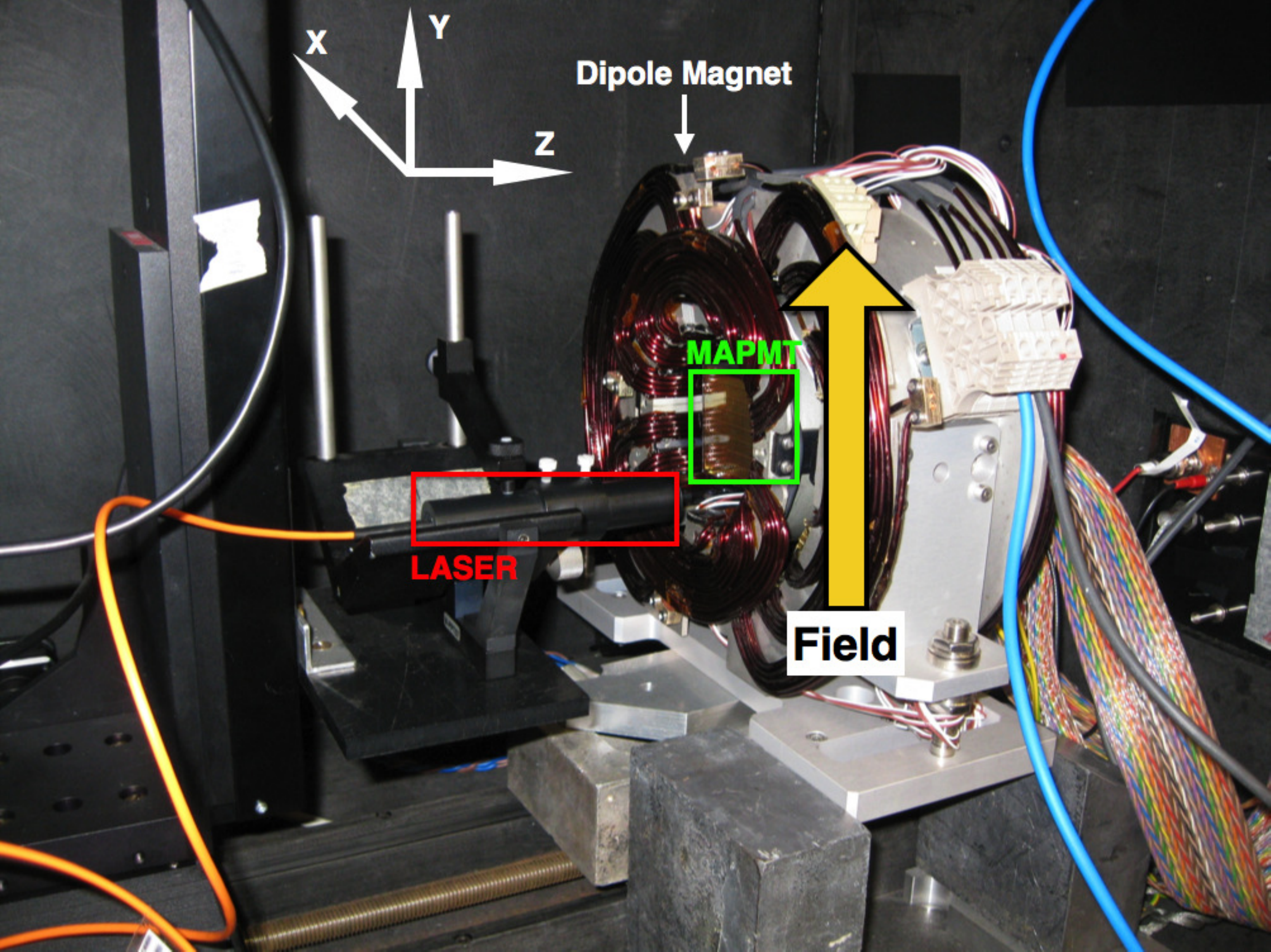}
\caption{\small \sf Set-up for the magnetic field tests. The MAPMT was placed completely inside a dipole magnet, with the orientation of the magnetic field transverse to the MAPMT axis.}
\label{fig:MagneticFieldSetup}
\end{figure}

The positions of the RICH detector MAPMTs in the CLAS12 spectrometer are located within the fringe field of the torus magnet. From simulations only a very small maximum field strength of 3.5\,Gauss is expected as the worst case, for a few of the MAPMTs, mainly the transverse direction. We studied the performance of an example H8500 MAPMT (CA4655) when placed within a magnetic field, ranging from strengths of 5\,Gauss to 50\,Gauss, to evaluate any degradation of its signal properties.  The HV of the MAPMT was set to -1040\,V for the duration of the tests, and laser scans were performed with and without the presence of the magnetic field.  For this, the experimental set-up was modified to include a dipole magnet within the light-tight box. The arrangement and orientation of the dipole magnet used is shown in Fig.~\ref{fig:MagneticFieldSetup}.
The magnet was cylindrical in shape, with an outer diameter of $\sim$\,25.0\,cm and an inner diameter of $\sim$\,8.5\,cm, allowing the MAPMT to be fully inserted within the magnet. The magnetic field was roughly mapped before use using an external probe. It was uniform in the vertical (Y) direction within 10\,\%, transverse to the MAPMT axis and along the direction from pixel 57 to pixel 1 (according to the map in Fig.~\ref{fig:PixelMap}), and negligible in the horizontal (X) and longitudinal (Z) directions. Within the 10\,\% uniformity, the field in the Y direction slightly increased towards the surfaces of the dipole coils. The upper limit for the field strength of the magnet was $B_{Y}=50$\,Gauss, where $B_{Y}$ denotes the magnetic field strength in the vertical Y direction. Four field strengths in total were tested: 5, 10, 25 and 50\,Gauss.

\begin{figure}[h!]
\begin{center}
\mbox{
\subfigure[\scriptsize No Field: gain = (8.89\,$\pm$\,0.20)\,Bins]{\scalebox{1.0}{\includegraphics[height=4cm]{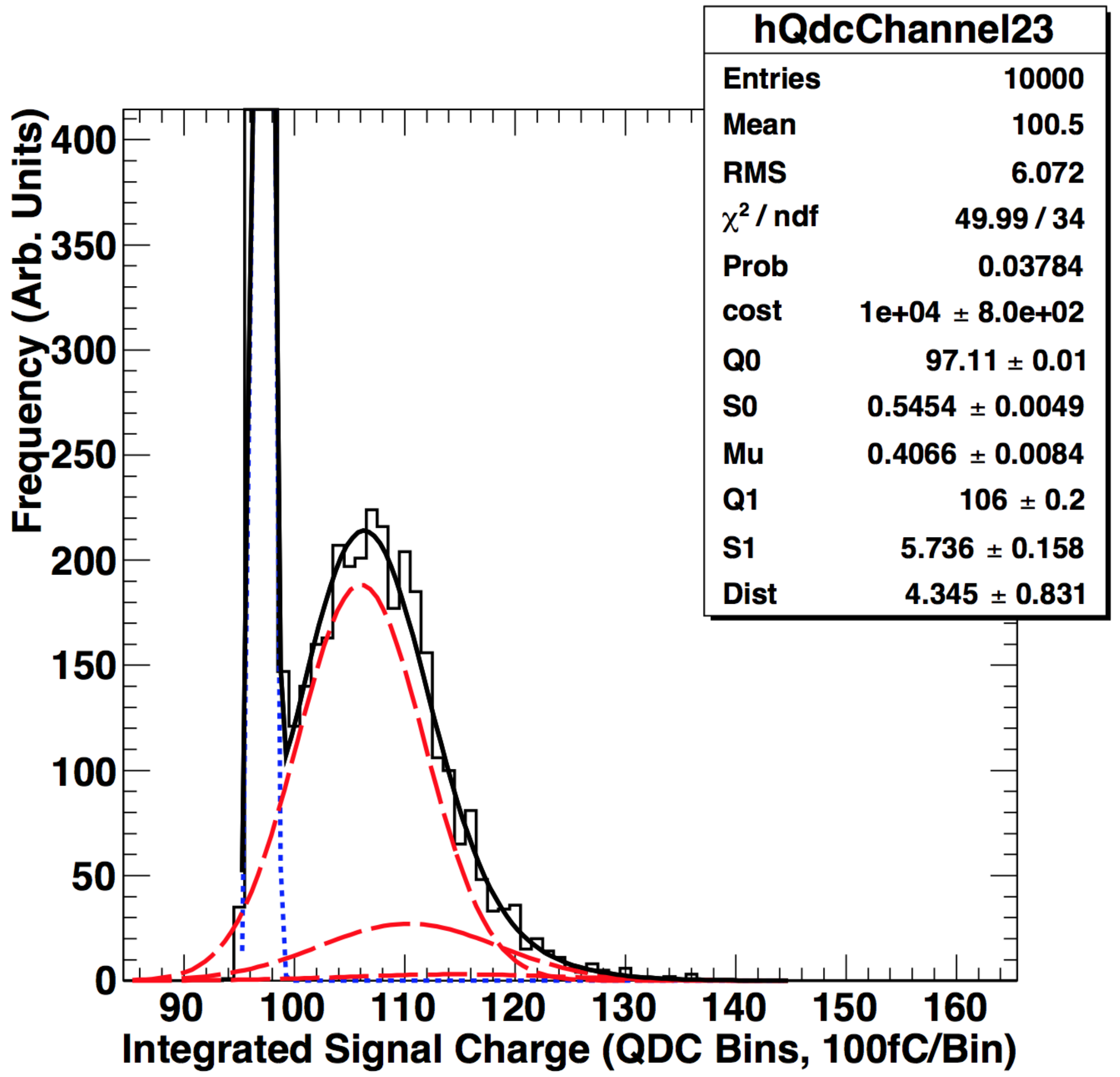}}} \quad
\subfigure[\scriptsize 5\,Gauss: gain = (8.70\,$\pm$\,0.10)\,Bins]{\scalebox{1.0}{\includegraphics[height=4cm]{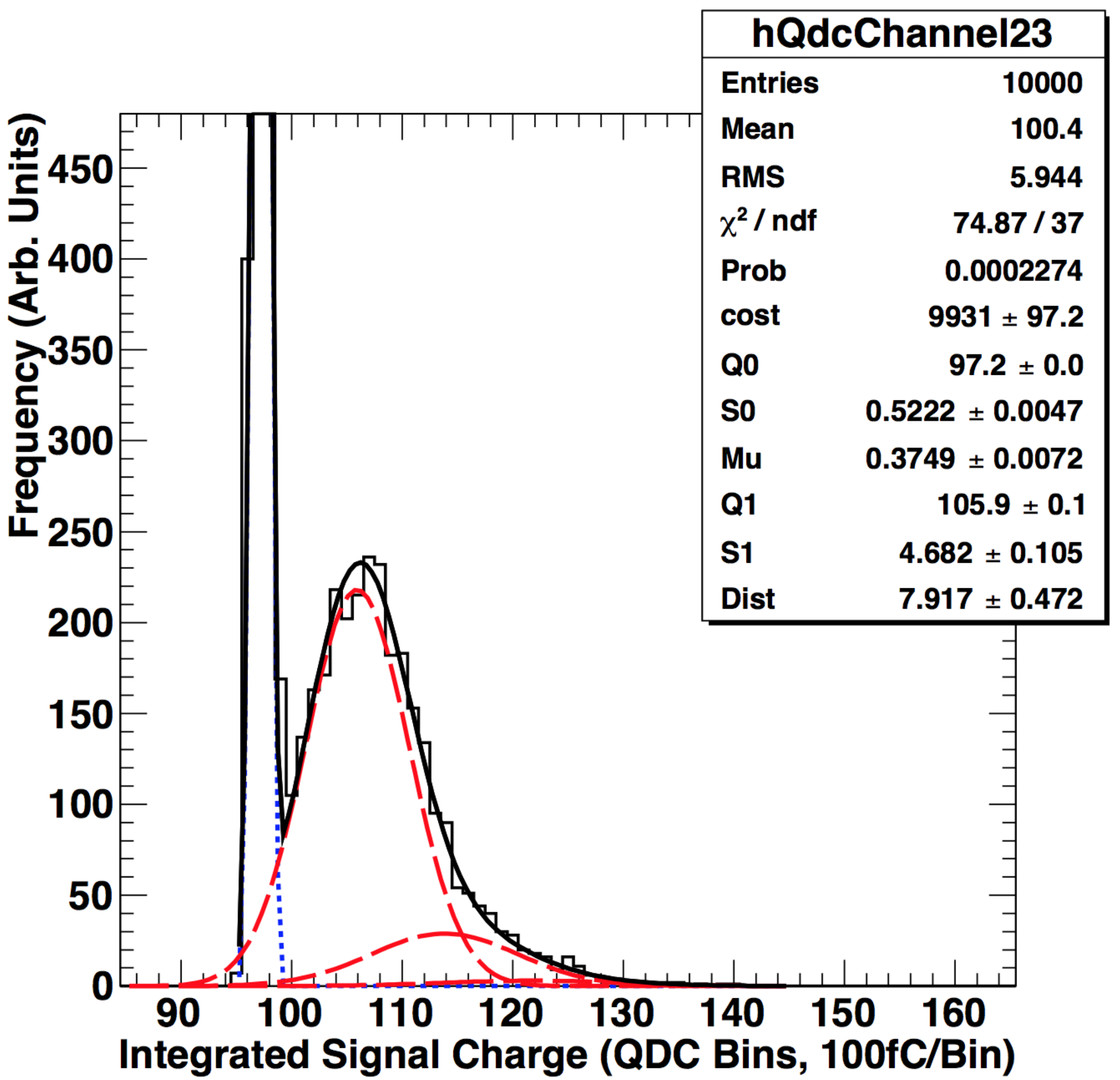}}}
\subfigure[\scriptsize 10\,Gauss: gain = (9.06\,$\pm$\,0.10)\,Bins]{\scalebox{1.0}{\includegraphics[height=4cm]{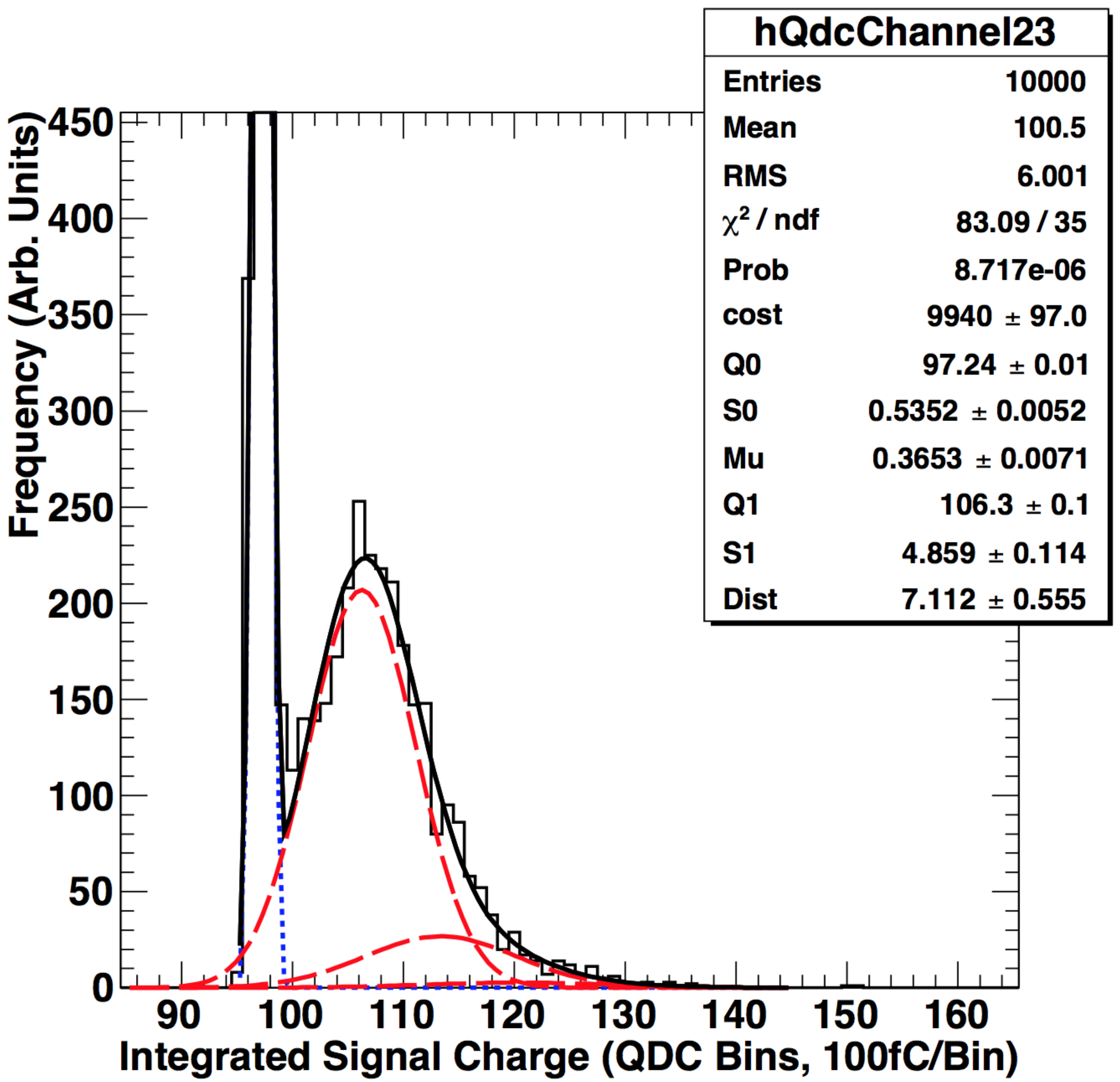}}}
}\\
\mbox{
\subfigure[\scriptsize 25\,Gauss: gain = (8.52\,$\pm$\,0.20)\,Bins]{\scalebox{1.0}{\includegraphics[height=4cm]{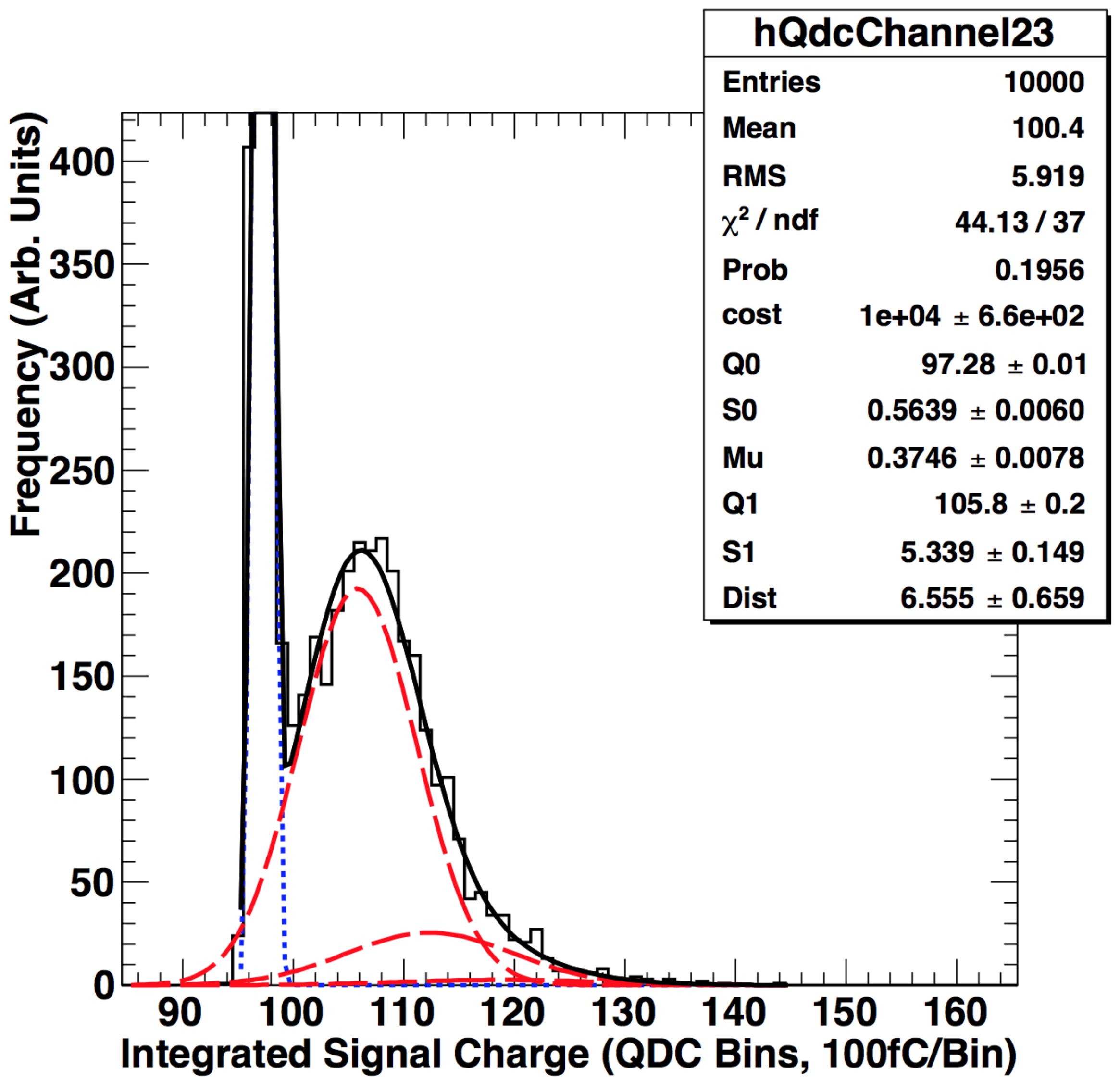}}} \quad
\subfigure[\scriptsize 50\,Gauss: gain = (7.78\,$\pm$\,0.20)\,Bins]{\scalebox{1.0}{\includegraphics[height=4cm]{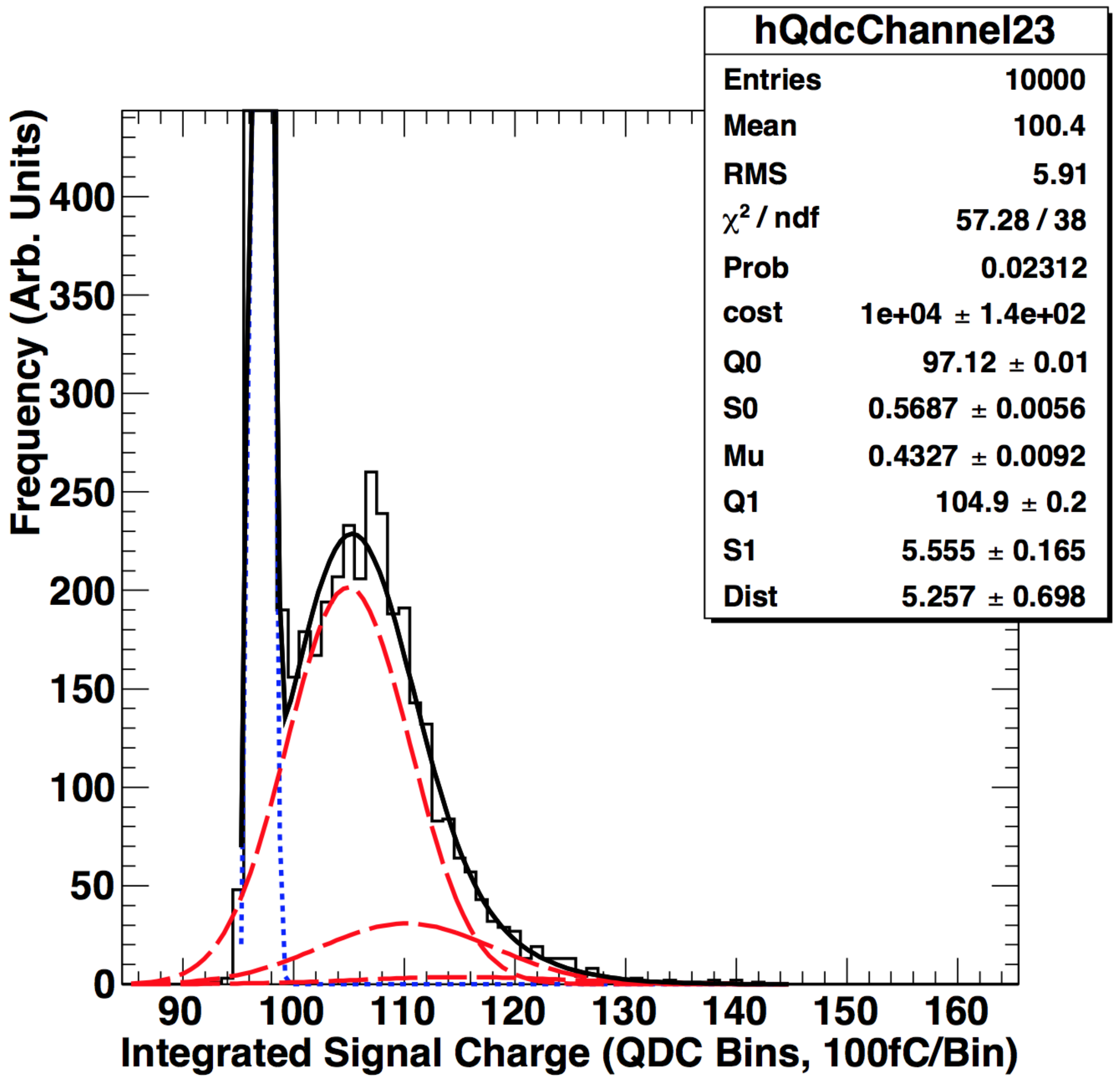}}}
}
\caption{\small \sf Fitted spectra obtained at the different magnetic field strengths for an example inner pixel of the CA4655 MAPMT (pixel 28, which mapped to QDC channel 23). The first p.e. gain, $Q_{1} - Q_{0}$, is also given for each field setting.}
\label{fig:MagFieldStudy_Spectra_P28}
\end{center}
\end{figure}

\begin{figure}[h!]
\begin{center}
\mbox{
\subfigure[\scriptsize No Field: gain = (8.47\,$\pm$\,0.11)\,Bins]{\scalebox{1.0}{\includegraphics[height=4cm]{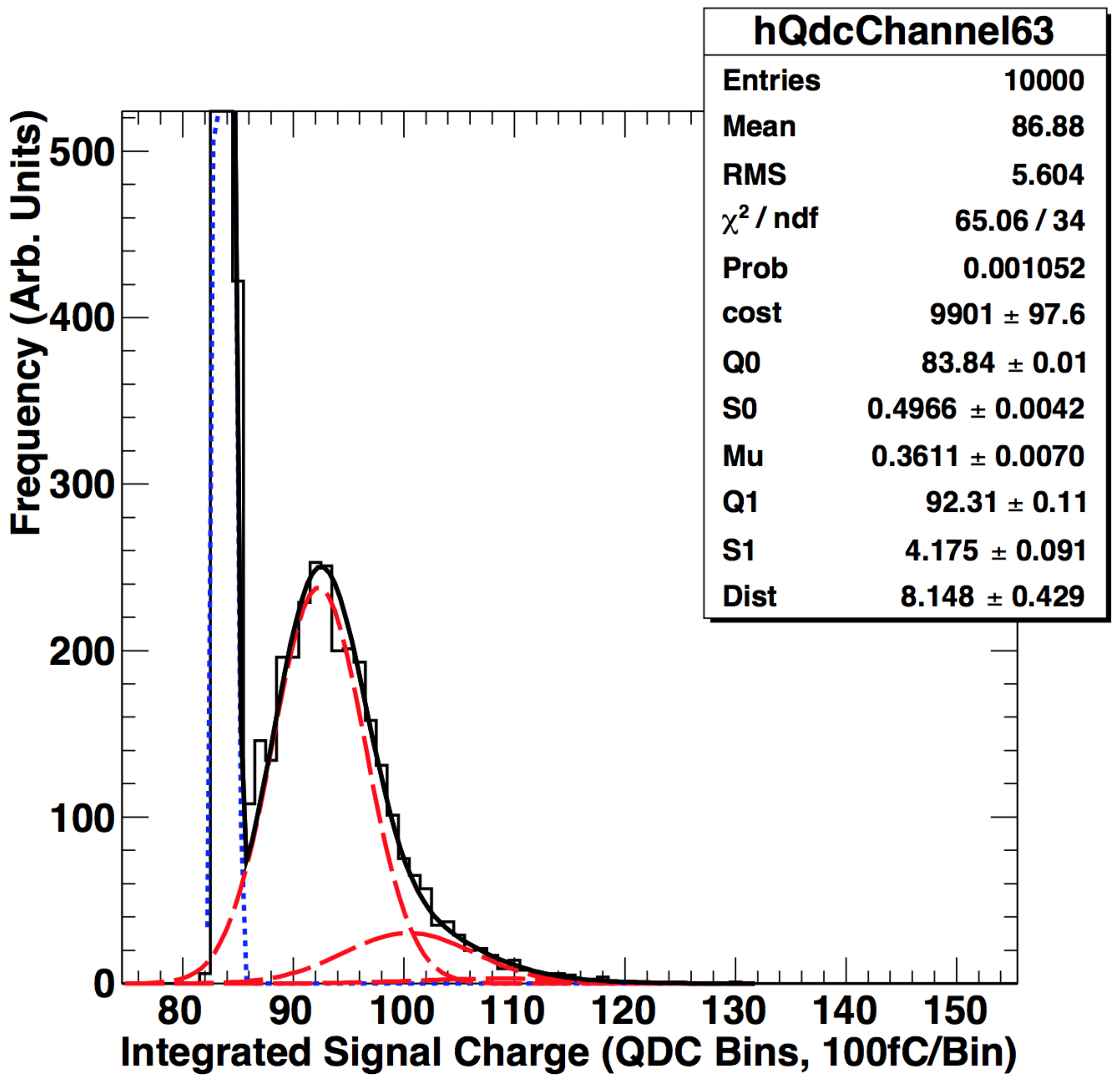}}} \quad
\subfigure[\scriptsize 5\,Gauss: gain = (8.29\,$\pm$\,0.12)\,Bins]{\scalebox{1.0}{\includegraphics[height=4cm]{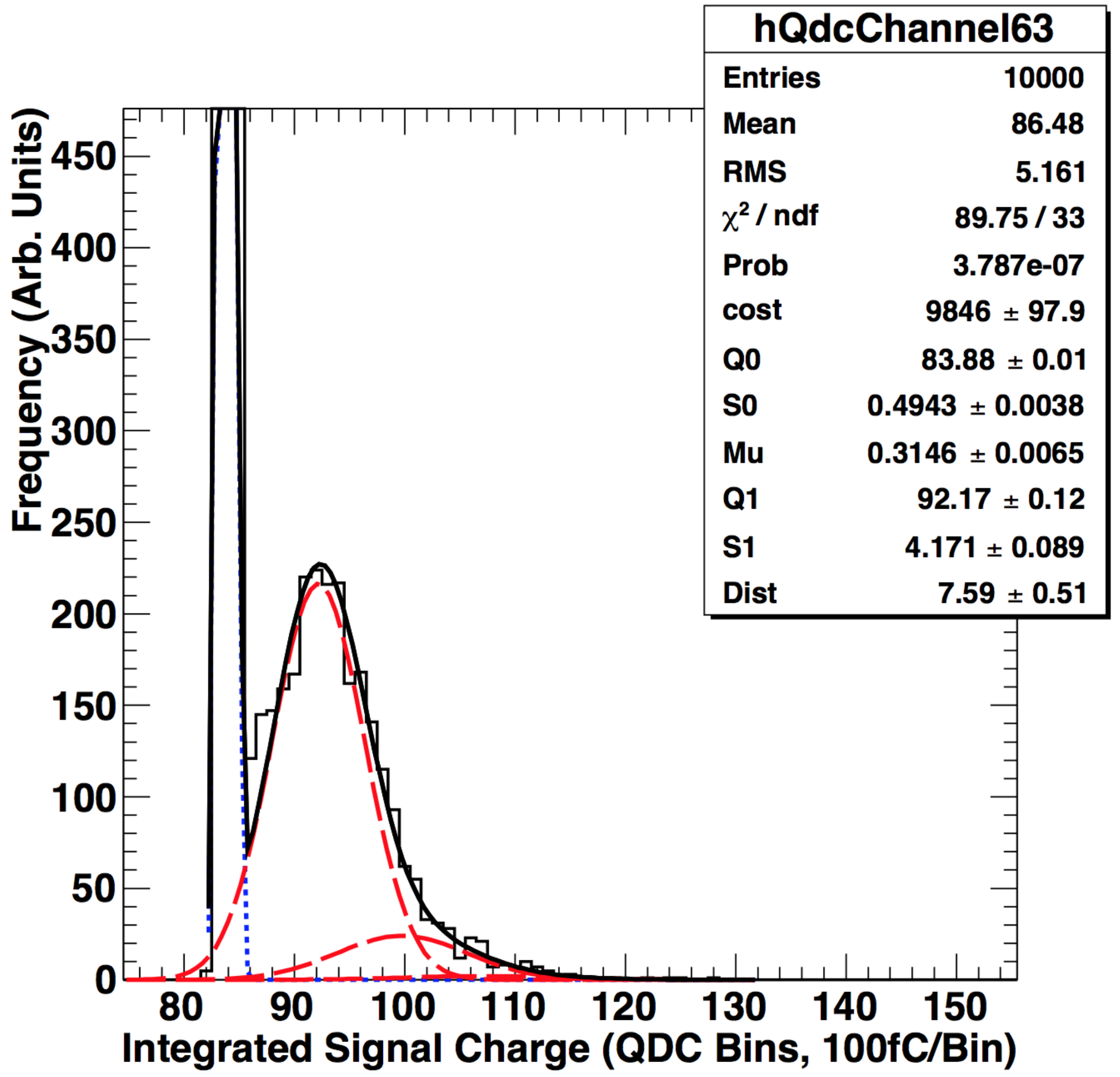}}}
\subfigure[\scriptsize 10\,Gauss: gain = (8.39\,$\pm$\,0.12)\,Bins]{\scalebox{1.0}{\includegraphics[height=4cm]{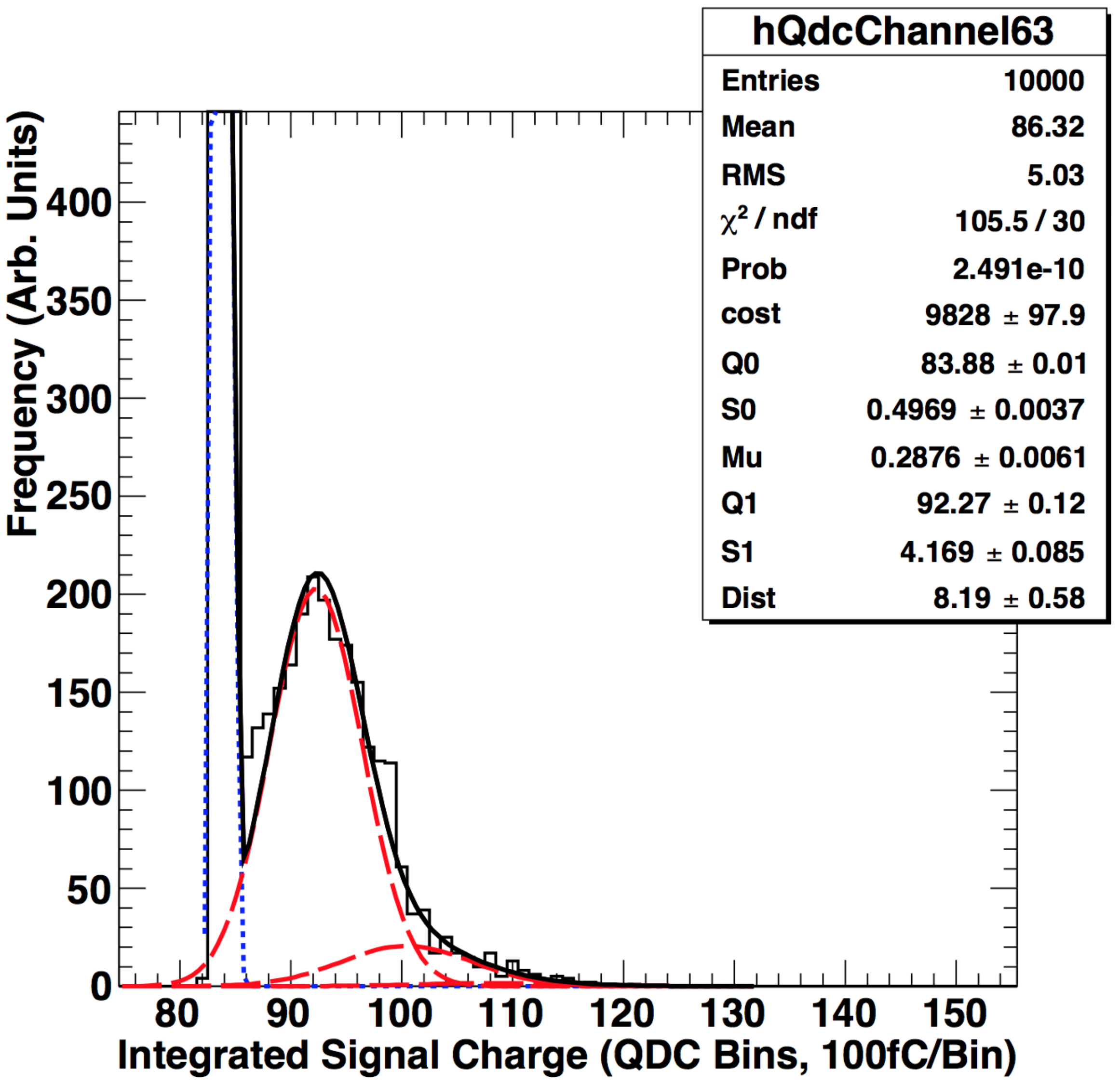}}}
}\\
\mbox{
\subfigure[\scriptsize 25\,Gauss: gain = (7.71\,$\pm$\,0.13)\,Bins]{\scalebox{1.0}{\includegraphics[height=4cm]{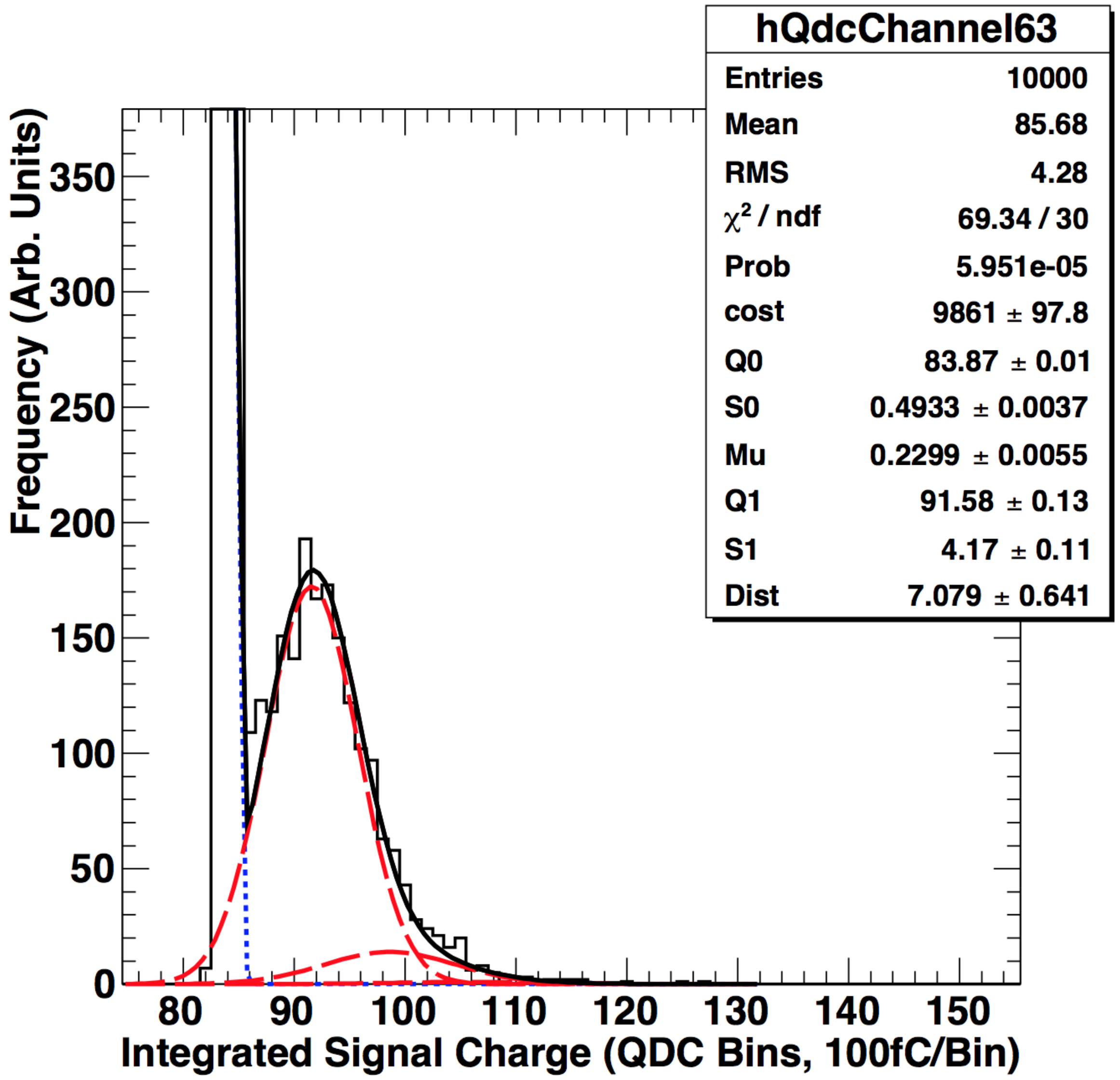}}} \quad
\subfigure[\scriptsize 50\,Gauss: gain = (6.57\,$\pm$\,0.13)\,Bins]{\scalebox{1.0}{\includegraphics[height=4cm]{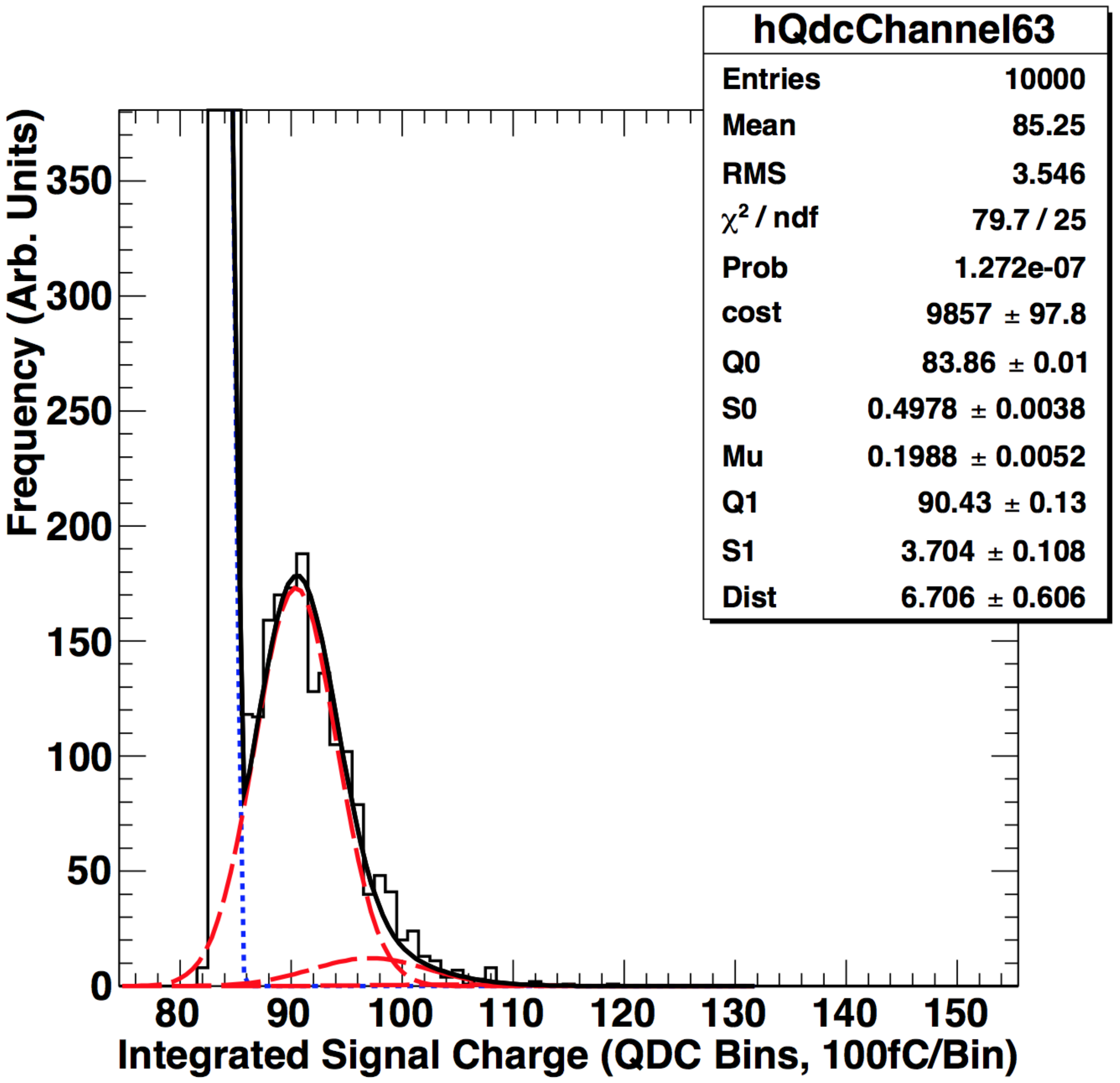}}}
}
\caption{\small \sf Fitted spectra obtained at different magnetic field strengths for an example corner pixel of the CA4655 MAPMT (pixel 64, which mapped to QDC channel 63). The first p.e. gain, $Q_{1} - Q_{0}$, is also given for each field setting.}
\label{fig:MagFieldStudy_CornerSpectra}
\end{center}
\end{figure}

The standard fit function (Equation~\ref{eq:QDCfit}) was used to describe the spectra. Some examples of fitted spectra obtained from a typical inner pixel (pixel 28) at the different field strengths are shown in Fig.~\ref{fig:MagFieldStudy_Spectra_P28}. 
As shown, the spectra were unaffected by the magnetic field from 5\,Gauss up to 25\,Gauss. The pedestal and first p.e. peak fit parameter, $Q_{1}$, were stable consistent with the fit error. For a field of 50Gauss a small increase in the number of events filling the valley region separating pedestal caused the fit to be marginally less successful in describing the amplitude of the first p.e. peak. This resulted in a reduction of the fitted gain ($Q_{1}-Q_{0}$) which was on the order of only one QDC bin (100\,fC). A similar behaviour was observed for the other inner pixels of the MAPMT and it was concluded that no major degradations in the signals were visible. A priori it was expected that corner and edge channels of the MAPMT may be affected worse by the magnetic fields than centre pixels. This is demonstrated in Fig.~\ref{fig:MagFieldStudy_CornerSpectra} for a corner pixel of the MAPMT (pixel 64). For field strengths 5 and 10\,Gauss no significant change was observed in the signal spectra, but starting from 25\,Gauss a reduction in the signal gain, $Q_{1}-Q_{0}$, is observable. At 50\,Gauss this reduction is $\sim$200\,fC or about 20\,\% of the signal gain. A similar behaviour was observed in the remaining corner and edge pixels.

\begin{figure}[h!]
\begin{center}
\mbox{
\subfigure[]{\scalebox{1.0}{\includegraphics[height=6.3cm]{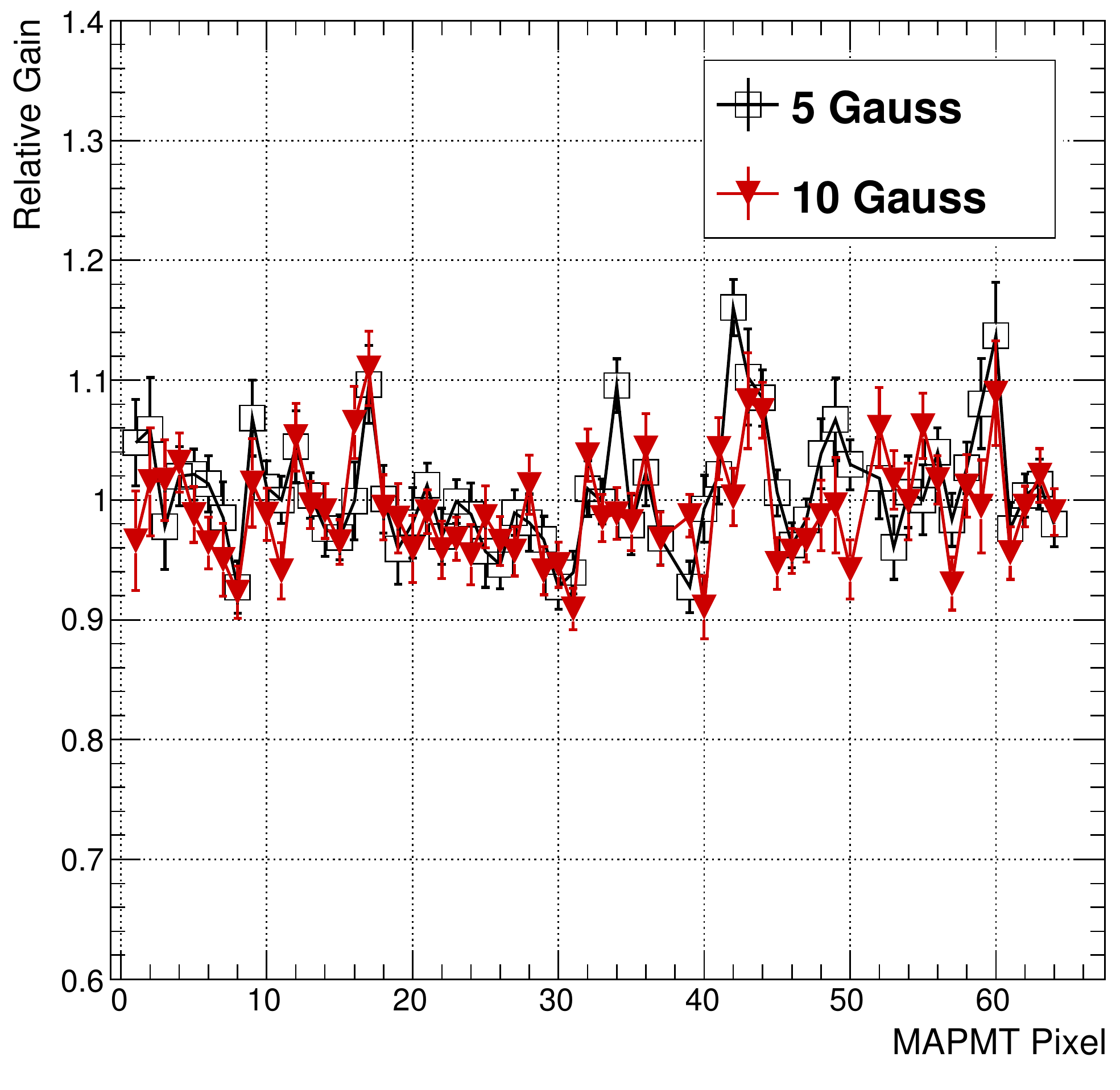}}}
\subfigure[]{\scalebox{1.0}{\includegraphics[height=6.3cm]{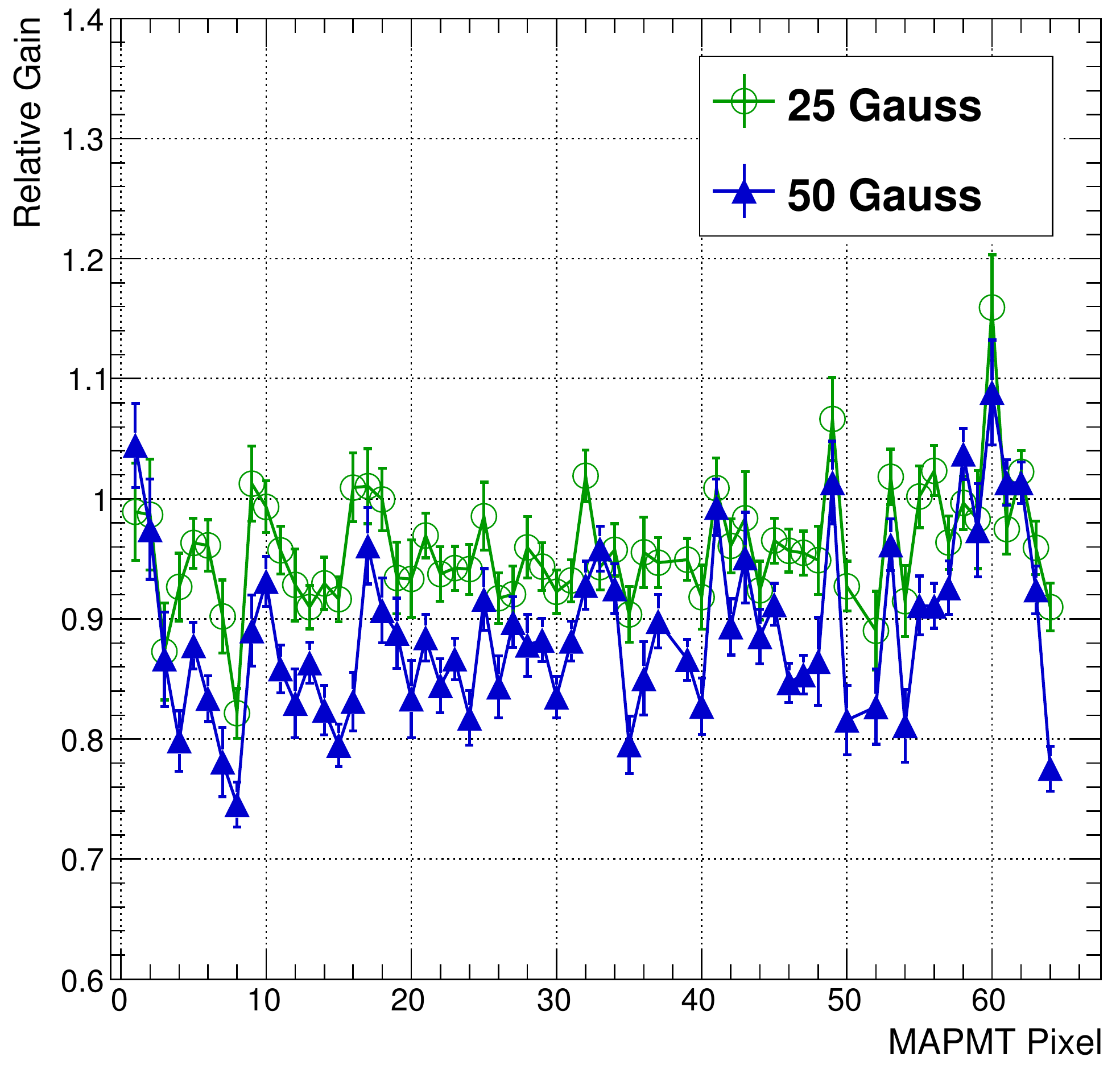}}}
}
\caption{\small \sf First p.e. peak gains, $Q_{1}-Q_{0}$, calculated relative to the values for no magnetic field and for all pixels of the CA4655 MAPMT. Results are grouped into magnetic field strengths of: (a) 5 and 10\,Gauss; (b) 25 and 50\,Gauss.}
\label{fig:MagFieldStudy_RelGainsVPixel}
\end{center}
\end{figure}

\begin{figure}[h!]
\begin{center}
\mbox{
\subfigure[]{\scalebox{1.0}{\includegraphics[height=6.3cm]{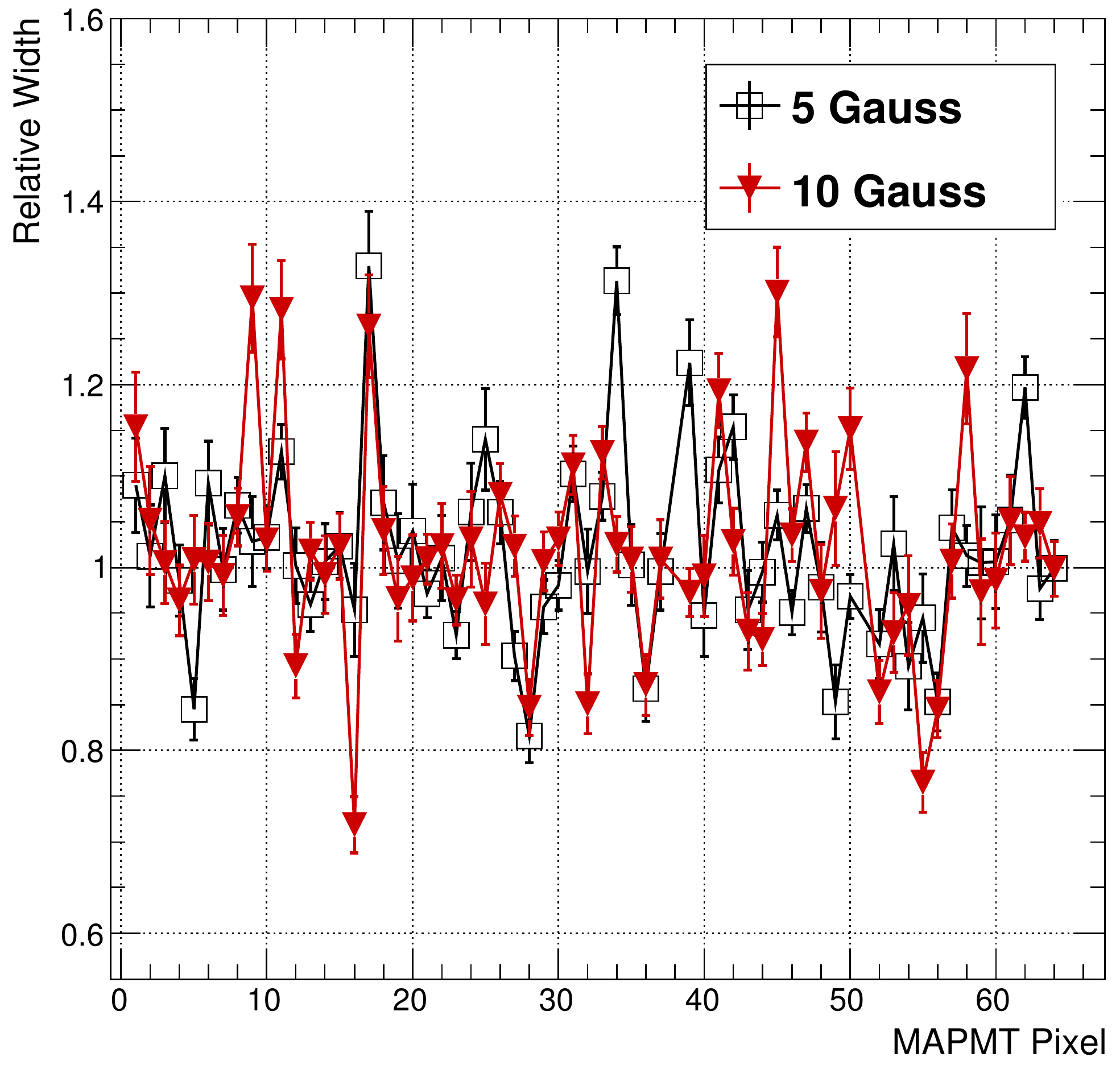}}}
\subfigure[]{\scalebox{1.0}{\includegraphics[height=6.3cm]{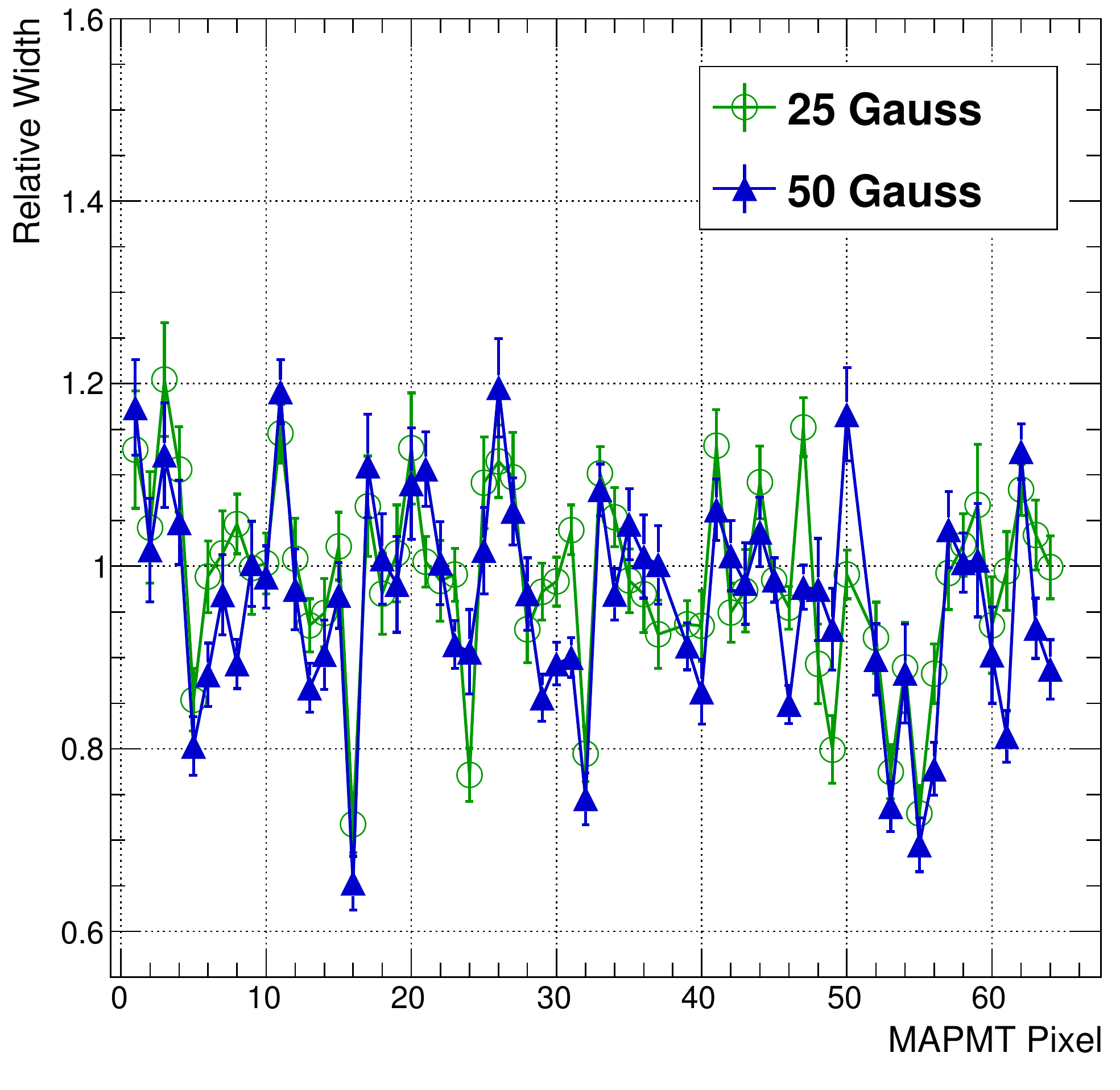}}}
}
\caption{\small \sf First p.e. peak widths, $S_{1}$, calculated relative to the values for no magnetic field and for all pixels of the CA4655 MAPMT. Results are grouped into magnetic field strengths of: (a) 5 and 10\,Gauss; (b) 25 and 50\,Gauss.}
\label{fig:MagFieldStudy_RelWidthsVPixel}
\end{center}
\end{figure}

In Figs.~\ref{fig:MagFieldStudy_RelGainsVPixel}\,(a) and (b) we show the gains extracted from fits ($Q_{1}-Q_{0}$) to the spectra of all pixels of the MAPMT relative to the values obtained without magnetic field, at magnetic field settings of 5 and 10\,Gauss, and 25 and 50\,Gauss respectively.
For a few data points, typically for pixels on the edge of the MAPMT, an increase of the gain $Q_{1}-Q_{0}$ with magnetic field was deducted. This is regarded as an artefact, caused by combination of the low resolution of the spectra and the lower relative intrinsic gains of the affected channels. In such circumstances the applied fit may not represent the signal gain well for one of the data points in the ratio presented. For the majority of pixels, the relative $Q_{1}-Q_{0}$ gain values were largely unaffected by magnetic fields of 5 and 10\,Gauss (Fig.~\ref{fig:MagFieldStudy_RelGainsVPixel}\,(a)), when fitting errors are considered. Stronger systematic drops in the $Q_{1}-Q_{0}$ gain were more visible across all pixels for the 25 and 50\,Gauss results (Fig.~\ref{fig:MagFieldStudy_RelGainsVPixel}\,(b)), with relative reductions of around 5\,-\,10\,\% and 10\,-\,20\,\% extracted on average for the two field settings respectively. A relative reduction of $\sim$\,10\,-\,20\,\% corresponds to a difference on the order of 1 or 2\,QDC bins. The results therefore do not indicate a significant problem for single photon counting with this MAPMT in the CLAS12 RICH application. The fit results also do not show any strong distortion in the shapes of the spectra, as a function of magnetic field strength. Fig.~\ref{fig:MagFieldStudy_RelWidthsVPixel} gives the width of the first p.e. peak,$S_{1}$, for all pixels relative to the values without field, again for all field settings.
Overall the shape of the signals were mostly consistent, up to a field strength of 50\,Gauss. Only a few data points with poor fit results, already discussed for Fig.~\ref{fig:MagFieldStudy_RelGainsVPixel}, deviated from this general behaviour. The resolution limit of this method is of the order of 1\,QDC channel, or about 30\,\% of the typical signal width. But this is sufficient to deduct that the requirements for operation in the CLAS12 RICH detector are met.

\begin{figure}[h!]
\begin{center}
\subfigure[First p.e. peak gain.]{\scalebox{1.0}{\includegraphics[height=6.3cm]{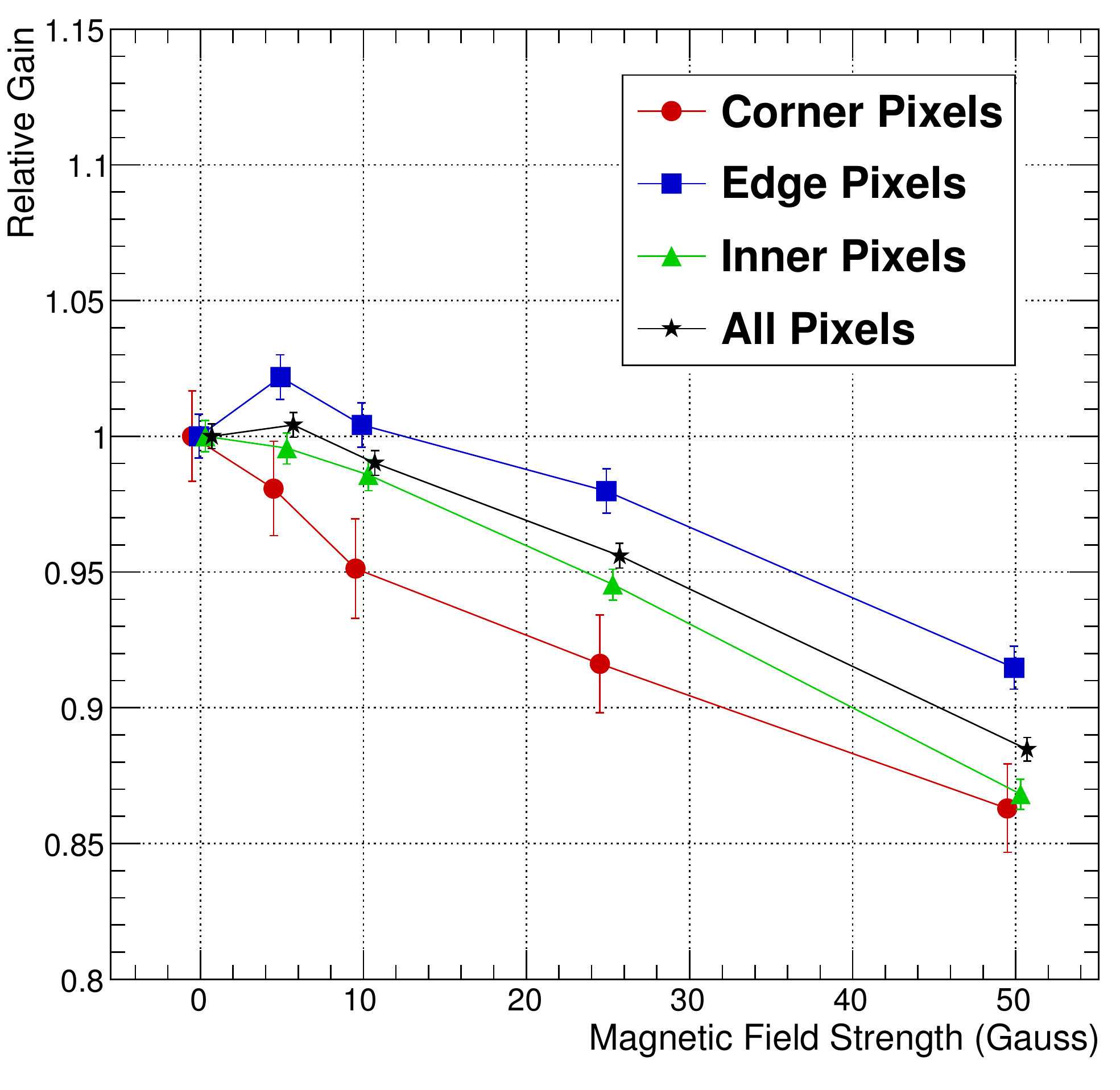}}} 
\subfigure[First p.e. peak width.]{\scalebox{1.0}{\includegraphics[height=6.3cm]{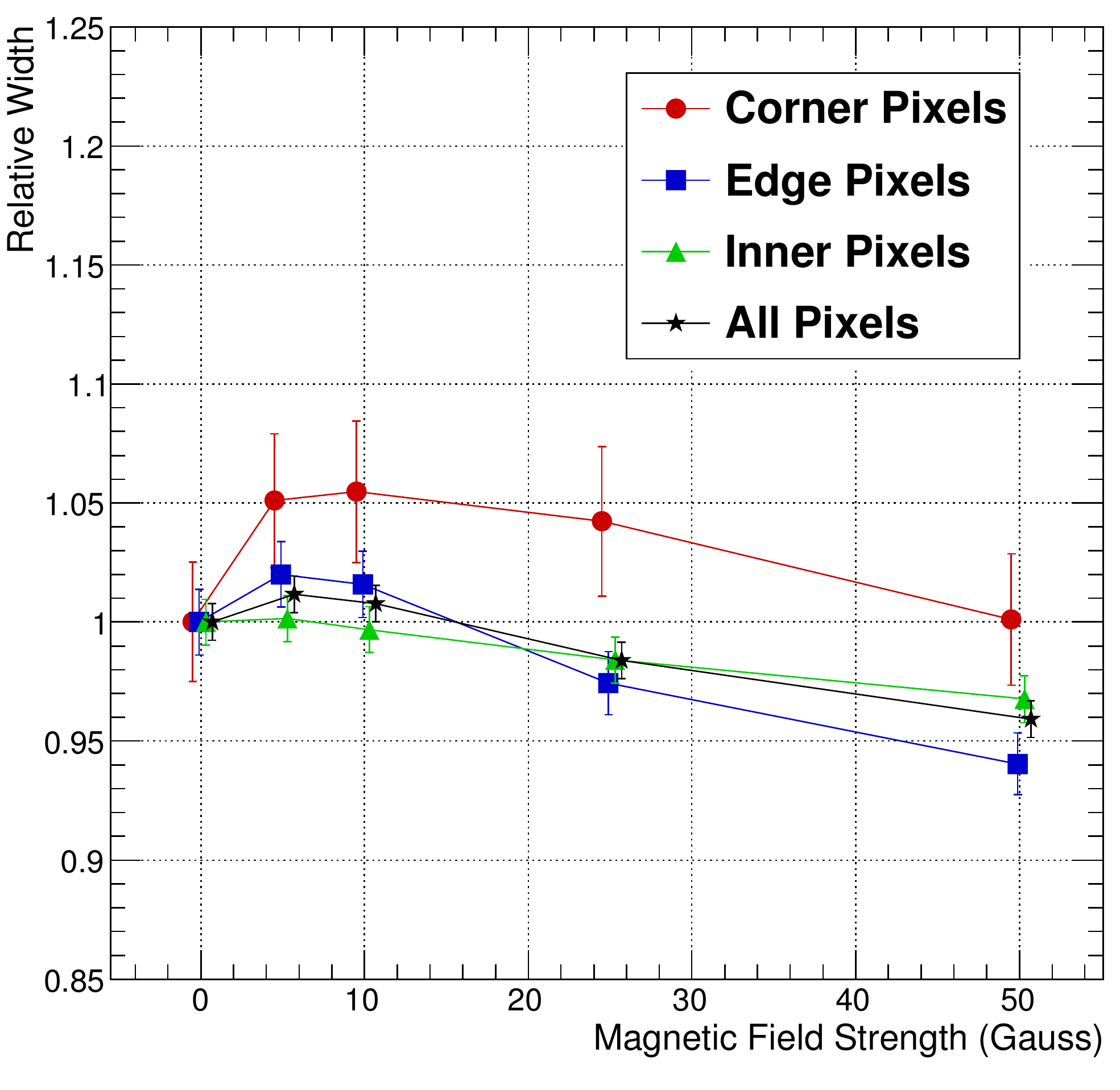}}}
\caption{\small \sf Average fitted first p.e. peak gain,$Q_{1}-Q_{0}$, and width, $S_{1}$, obtained at the tested magnetic field strengths. The results averaged amongst all pixels simultaneously are shown (black stars), in addition to sub-sets which have been averaged according to the geometrical positions of the pixels on the MAPMT face. The sub-sets are: 4 corner pixels (red circles); 24 edge pixels (blue squares) and 36 inner pixels (green triangles).}
\label{fig:MagFieldStudy_MeanResultsGainSigma}
\end{center}
\end{figure}

The first p.e. peak gain ($Q_{1}-Q_{0}$) and width ($S_{1}$) results averaged over all MAPMT pixels are shown in Figs.~\ref{fig:MagFieldStudy_MeanResultsGainSigma}\,(a) and (b), where the results have been separated into sub-sets depending upon the pixel locations on the MAPMT face. These averaged results illustrate more clearly a dependence upon pixel position, with edge and corner pixels being the most affected by the magnetic field. Even though in the magnetic field survey a 10\,\% variation in the field strength has been measured towards the edges, the data significantly characterises the behaviour of the MAPMT.
For comparison the averages over all pixels are given as well. Overall there is a relatively significant reduction in the averaged gain, $Q_{1}-Q_{0}$, of the first p.e. peak of $\sim$1\,QDC bin or 100\,fC observed, corresponding to about $\sim$12\,\% reduction in the signal gain for an external field of 50\,Gauss in the studied transverse direction. This yields no appreciable deterioration for the application in the CLAS12 RICH detectors. The drop in the averaged signal width, $S_{1}$, is also relatively significant and corresponds to roughly 4\,\% for a field of 50\,Gauss. This reduction is not as strong as may be expected, and the increases in $S_{1}$ widths for edge and corner pixels at field settings of 5 and 10 Gauss are also unexpected. However, these effects are likely caused by the success of the fit function in combination with the low resolution spectra and the absence of any signal amplification, resulting in the parameter $S_{1}$ to be over-estimated if the first p.e. peak is too close to the pedestal.

It was also planned to study the effects of longitudinal magnetic fields, with respect to the direction of the MAPMT axis, since from previous studies of the H8500 MAPMT these fields are expected to have the strongest effect on its performance~\cite{JLabH8500Paper}, however these studies were not possible with the set-up due to space restrictions. Moreover, the RICH MAPMTs will not be subjected to any longitudinal fields in CLAS12. Only small components in this direction are expected, which could arise from non-uniformities in the torus magnet and would not be large enough to affect the behaviour of the MAPMTs. To summarise, at small field strengths, even which are larger than those expected in the CLAS12 RICH, no effect was observed in the MAPMT response within the sensitivity of the set-up. Significant differences in the signals only occurred at the largest field setting of 50\,Gauss. Within an equivalent scope, the results are in agreement with previous studies of the H8500 behaviour in magnetic field environments~\cite{JLabH8500Paper}.


\section{Conclusions}
For the first time, a significant sample of two different types of Hamamatsu H8500 MAPMTs (14\,$+$\,14) has been tested, pixel by pixel, using a dedicated laser beam, at different HV values and also within a (weak) magnetic field. The overall aim of these studies was to assess the performances of the H8500  MAPMTs as single photon detectors, in view of their use in the CLAS12 RICH detector.

The results show that the Hamamatsu H8500 MAPMTs can be used successfully for such purposes, even if a voltage slighter higher than the recommended one (-1000\,V) significantly improves their performances for single photon detection. The fraction of the single photoelectron spectrum lost below the pedestal peak is important for the application of MAPMTs to a RICH detector and, for the majority of the tested H8500 MAPMTs a loss fraction of less than 20\,\% was measured at an HV value of 1075\,V. Additionally,  the performances of the H8500 MAPMTs are expected to be practically unaffected in the weak magnetic fringe field foreseen in the CLAS12 spectrometer.

These results, which agree with and expand upon what has similarly been reported for a smaller sample of the H8500 MAPMT~\cite{JLabH8500Paper}, confirm their selection and suitability for use in a RICH detector for CLAS12.

\bibliographystyle{elsarticle-num}


\end{document}